\documentclass[]{aastex631}

\usepackage{color}
\usepackage{soul}
\usepackage{comment}

\shortauthors{Manek et al.}

\graphicspath{{./}{figures/}}

\begin{document}

\title{The Rise of Buoyant Magnetic Structures through Convection with a Background Magnetic Field}

\correspondingauthor{Bhishek Manek}
\email{bmanek@ucsc.edu}

\author[0000-0002-2244-5436]{Bhishek Manek}
\affiliation{Department of Applied Mathematics, Jack Baskin School of Engineering, University of California Santa Cruz, \\ 1156 High Street, Santa Cruz, California 95064, USA}

\author{Christina Pontin}
\affiliation{Department of Applied Mathematics, \\ 
School of Mathematics, University of Leeds, \\ 
Leeds, LS2 9JT, UK}

\author[0000-0003-4350-5183]{Nicholas Brummell}
\affiliation{Department of Applied Mathematics, Jack Baskin School of Engineering, University of California Santa Cruz, \\ 1156 High Street, Santa Cruz, California 95064, USA}

\begin{abstract}

Inspired by observations of sunspots embedded in active regions, it is often assumed that large-scale, strong magnetic flux emerges from the Sun’s deep interior in the form of arched, cylindrical structures, colloquially known as flux tubes.  Here, we continue to  examine the different dynamics encountered when these structures are considered as concentrations in a volume-filling magnetic field rather than as isolated entities in a field-free background.   Via 2.5D numerical simulations, we consider the buoyant rise of magnetic flux concentrations from a radiative zone through an overshooting convection zone that self-consistently (via magnetic pumping) arranges a volume-filling large-scale background field.  This work extends earlier papers that considered the evolution of such structures  in a purely adiabatic stratification with an assumed form of the background field.  This earlier work established the existence of a bias that created an increased likelihood of successful rise for magnetic structures with one (relative) orientation of twist and a decreased likelihood for the other.  When applied to the solar context, this bias is commensurate with the solar hemispherical helicity rules (SHHR).  This paper establishes the robustness of this selection mechanism in a model incorporating a more realistic background state,
consisting of overshooting  convection and a turbulently-pumped mean magnetic field. 
Ultimately, convection only weakly influences the selection mechanism, since it is enacted at the initiation of the rise, at the edge of the overshoot zone.  Convection does however add another layer of statistical fluctuations to the bias, which we investigate in order to explain variations in the SHHR.

\end{abstract}

\section{Introduction} \label{sec:intro}

Observations of active regions and their sunspots have arguably had the largest influence on our understanding of the operation of the global dynamo that powers the magnetic activity of the Sun. That these objects, first associated with the magnetic field by  \cite{Hale:1908}, emerge from the highly turbulent plasma and yet obey a set of strict rules, is quite remarkable.  Sunspots occur in pairs, with a leading polarity and a trailing polarity that switches cyclically with a period of about 11 years (Hale's Polarity Laws), and with a definite tilt of the leading object towards the equator (Joy's Law;  \cite{Hale:Ellerman:Ferdinand:Nicholson:1919}).   So-called ``butterfly diagrams'' of tracers of this cyclic activity exhibiting the regular regeneration of magnetic field strongly suggest that a dynamo is likely responsible for this behaviour.

The prevailing theory behind these observations attributes active regions and sunspots to the emergence of elements of strong toroidal magnetic field through the visible surface of the Sun \citep{Parker:1975}.  If cylindrical or tube-like structures of toroidal flux were to arch up through the visible solar surface, then the opposite polarity of the sunspot pairs would be explained by the oppositely-directed field in the legs of these arches where they pierce the solar surface. Furthermore, it is plausible that some interaction of this emerging flux loop with the background solar rotation could lead to a writhe of the structure that accounts for the Joy's Law tilt.  The cyclic change in polarity would have to be imposed during the dynamo origination of the field, requiring a reversal of the toroidal field.

It has long been thought that strong toroidal field is most likely generated deep in the solar interior, towards the base of the convection zone or in the tachocline, where strong shear in the form of differential rotation can amplify fields significantly.  Since these processes are far removed from the surface, magnetic buoyancy is invoked as a transport mechanism between the two regions \citep{Parker:1975}.
The concept of magnetic buoyancy can loosely be described as the situation where a strong concentration of magnetic field contributes significant magnetic pressure to the total pressure. Under the simple assumptions that the total pressure and the temperature equilibrate quickly, this leads to a decrease in the density associated with the magnetic concentration that results in an upwards buoyant force.
This process can be cast more formally as an instability \citep{Acheson:1979}, and a great success of simulations of magnetic buoyancy instabilities has been that they naturally exhibit the creation of rising, arching, tube-like structures of magnetic field, akin to what was envisaged for the solar case \citep[e.g.][]{Matthews:Hughes:Proctor:1995,Vasil:Brummell:2008}.

Some of the more recent observations of magnetically-active areas of the sun have focused on a relatively new signature of the activity, namely, the magnetic helicity of these regions \citep[see e.g.][for a complete review]{Pevtsov:etal:2014}.  Magnetic helicity is defined as $H_m=\int_V \mathbf{A} \cdot (\mathbf{\nabla} \times \mathbf{A})~\mathrm{d}V$.  Here, $\mathbf{B}$ is the magnetic field, and $\mathbf{B}= \mathbf{\nabla} \times \mathbf{A}$, so that $\mathbf{A}$ is a vector potential defining the divergenceless magnetic field. Since $\mathbf{A}$ is not uniquely defined, nor can it be directly measured in observations, a more commonly-used measure is current helicity, $H_c=\int_V \mathbf{B} \cdot (\mathbf{\nabla} \times \mathbf{B})~\mathrm{d}V$. 
The current helicity is a measure of the twist, writhe and connectedness of the magnetic field \citep[see e.g.][]{Moffatt:1969}.  These  quantities are important for two main reasons.  Firstly, $H_m$ is an invariant of the ideal MHD equations and thereby provides strong constraints on those dynamics.  Secondly, the release of energy from twisted magnetic field is commonly cited as one of the major sources powering eruptive events in the solar atmosphere \citep[e.g.][]{Low:1996, Amari:Luciani:Aly:Mikic:2003, Nindos:Andrews:2004}.  The origin of such twist is therefore important.  Since $H_c$ is more measurable, it is often used as a proxy for $H_m$, although it is unclear under what circumstances this is useful \citep{Seehafer:1990, Pevtsov:Canfield:Metcalf:1995, Abramenko:Wang:Yurchishin:1997, Bao:Zhang:1998}.

The most striking result from observations of the current helicity in active regions is that, once again, such dynamics seem to be remarkably ordered, considering the turbulent environment through which the magnetic structures emerge.  It has been found that the sign of the current helicity, when averaged over active region areas, has a strong dependence on which solar hemisphere it belongs to: in the Northern hemisphere, in general the helicity is negative, whereas in the Southern hemisphere, it is positive \citep[see e.g.][]{Pevtsov:etal:2014}. This is a trend and not a strict rule, only being obeyed approximately  60-80\% of the time.  However, this trend is independent of which half of the 22-year magnetic activity cycle is examined; that is, it is independent of the polarity of the sunspots pairs, or the direction of the toroidal field making up the sunspot structures.  This set of rules is known collectively as the Solar Hemispherical Helicity Rule (SHHR).  

The origin of this rule is clearly of interest, and there have been a number of theories proposed.  This paper presents the most complete and dynamically-realistic version to date of the dynamics postulated by one of these potential theories, previously developed in \cite{Manek:Brummell:Lee:2018} and \cite{Manek:Brummell:2021} (hereinafter known as Papers 1 and 2).  There are indeed other reasonable theories apart from the one pursued here \citep[see e.g.][]{Longcope:etal:1996, Choudhuri:Chatterjee:Nandy:2004}, and it is entirely plausible, if not perhaps even  likely, that the processes conceived in each of these theories all contribute to the overall helicity budget of a rising magnetic structure to some degree.  The introduction of Paper 2 contains more details about the pros and cons of each theory, so they are not repeated again here.  Instead, we follow the theory of Papers 1 and 2 and examine the  robustness of these ideas under conditions that are more representative of solar conditions than were examined in those earlier works.

The theory proposed in Papers 1 and 2 arises from adopting a slightly different perspective from those used previously in the modelling of the rise of structures by magnetic buoyancy.  In general, earlier work done in this context has proceeded mainly along three fronts. 
Firstly, there have been many studies of the rise of preconceived, {\it isolated}, tube-like magnetic structures.  Here, the key point is that the structures are isolated in the sense that they are embedded in an environment that is free of other magnetic field.  Such studies have been performed under the thin flux tube approximation \citep[e.g.][]{Spruit:1981} and, more realistically, where the tube has a finite cross-section \citep[e.g.][]{ Moreno:1983, Moreno:1986, Choudhuri:1989, Dsilva:Choudhuri:1993, Fan:Fisher:Deluca:1993, Fan:Fisher:Mcclymont:1994, Caligari:Moreno:Schussler:1995, Longcope:Klapper:1997}. An interesting discovery here was that locally-azimuthal field in the tube is required in order to create twisted field lines that supply a centrally-directed tension that provides coherence to the tube during its rise
\citep{Moreno-Insertis:Emonet:1996}.  Without sufficient twist, the tubes are ripped apart by the trailing vortices they generate in their wake \citep{Longcope:Fisher:Arendt:1996}.  The second class of modelling efforts concerns true instabilities of layers of magnetic field \citep[see e.g.][]{Acheson:1979, Cattaneo:Hughes:1988,Matthews:Hughes:Proctor:1995, Vasil:Brummell:2008}.
Here, magnetic structures are not initially present, but horizontal layers of magnetic field containing sufficiently strong vertical gradients are  unstable to magnetic buoyancy instabilities and therefore evolve to produce such structures.  Typically, tube-like structures with a mushroom-like cross-section are created, that can arch and kink in simulations if they are three-dimensional.  The final category is that of global dynamo simulations.  Some spherical shell simulations at certain parameters show the production of strong bands of toroidal field amid the convection, including rising segments that again look similar to the concept of emerging flux tubes.

It should be noted that both of the latter two categories can end up with the existence of concentrated strong magnetic structures embedded in a large-sale background field (usually the field from which the structures were created).  This is contrary to many of the earlier studies in the first category.  The work performed in Papers 1 and 2 was a simple attempt at examining the different dynamics of magnetic concentrations embedded in a volume-filling large-scale background field without specifically having to be concerned about the process that created the magnetic structures.  To that end, Papers 1 and 2 examined the dynamics of a preconceived flux concentration (often still colloquially referred to as a ``flux tube'' for convenience) embedded in a chosen background field in an initially quiescent, adiabatically-stratified fluid layer.  Those numerical studies, and the work of the current paper, were performed in a Cartesian 2.5D domain, where all three components of vector fields are kept, but are assumed to be independent of one horizontal direction (say, $y$), so that only a two-dimensional domain (in the other horizontal direction and the vertical, say $x$ and $z$ respectively) is actually computed.  This allows magnetic structures to have twist in a two dimensional calculation.  That is, the structures are typically initially confined to a circular region and contain both axial ($y$) and locally-azimuthal field (in $x$ and $z$).  Papers 1 and 2 then also imposed a horizontal (in $x$) magnetic field that had an assumed variation in the vertical ($z$) so that the flux tube was a structure embedded in this larger-scale field.  The use of 2.5D simulations rather than full 3D simulations provides a huge computational advantage, but clearly restricts and affects the problem.  For example, despite the tendency of magnetic field to be much closer to invariant along fieldlines than across, thanks to the stiffening tendency of magnetic tension, flux tubes can arch and writhe in an interesting manner in fully 3D simulations.  Furthermore, turbulent convection (incorporated in the current studies but not in Papers 1 and 2) in reality is clearly 3D, and not 2.5D as in these models.  We discuss the expected differences in the resultant physics due to three-dimensionality in detail in the discussion of \S~\ref{sec:discussion_conclusions}.  However, we ultimately conclude that the main results of this series of papers are likely robust to this change of dimensionality.  This will be verified in another paper in this series, soon to appear.

Papers 1 and 2 remarkably found that, for a certain region of parameter space (termed the Selective Rise Regime or SRR), there was a selection mechanism that dictated different dynamics for differently twisted initial tubes (relative to a fixed background field direction).  It was found that tubes that had twist such that the locally azimuthal field at the bottom of the tube aligned with the background field had a tendency to be more likely to rise than tubes whose twist was such that the azimuthal field was aligned at the top.  These selective dynamics only occur for intermediate relative strengths of the tube and the background field.  If the tubes are relatively weak, both signs of twist cannot rise;  if the tubes are relatively strong, both signs of tubes rise.  If the background field strength is between $\sim 5-15$\% of the tube strength (roughly, for the other parameters considered in Papers 1 and 2), then one sign of twist rises and the other does not.  The reason for the existence of the selection effect is the differential influence of the background field on internal tension forces in the tube.  Where the background field and the locally-azimuthal field of the tube align, the tension forces are increased, and they are reduced where the two components are anti-aligned.  When the enhancement is at the bottom of the structure and the reduction is at the top, a net tension force acts upwards, in concert with the buoyancy forces, enhancing the opportunity for rise.  When the alignment is the other way around, the net tension force in the tube acts downwards and opposes buoyancy, reducing the likelihood of rise.  

Most remarkably, if translated into the solar context, the preferred signs of twist translate into the correct helicities to obey the SHHR.
Paper 2 explored this selection mechanism and the SRR in great detail, and included an explanation of the dynamics via a simple mathematical model.  An intriguing point is that the existence of the SRR suggests that violations to the rule are entirely plausible, if twist is assumed to be created initially with some distribution.  Paper 2 confirmed this via synthetic SHHR maps generated from Monte Carlo simulations of multiple tubes with random twist strengths and locations.

The biggest difference between this model and other theories is that all the other theories essentially require a rising tube to \emph{acquire} the correct twist (by some method) as it rises.  In the models of Papers 1 and 2, instead, all signs of twist are assumed to be created randomly (as seems to be the case in simulations of magnetic buoyancy instabilities, see e.g.  \cite{Matthews:Hughes:Proctor:1995, Vasil:Brummell:2008}) and then the selection mechanism \emph{acts as a filter}, allowing a certain sign of twist to rise preferentially (but with significant violations to the rule).

The simulations examined in Papers 1 and 2 employed highly simplified models of the background dynamics.  The magnetic structures simply buoyantly rose in an initially quiescent, adiabatic fluid layer, and the large-scale background field was artificially imposed as an exponential function that decreased upwards.  These choices were supposed to represent the thermodynamic background state that would result from a well-mixed convective state and also the large-scale magnetic field that would result from transport by convection in such a state, even though the actual convection was not included.  This paper aims to relax these simplifications and study the fully  convective problem.  Prevailing theories for the origin of emerging active region flux assume that flux tubes are most likely formed in the tachocline, and therefore rising structures must traverse perhaps the upper radiative zone, the overshoot zone and the convection zone in their rise towards emergence at the surface.  Here, in this study, we model all of these zones, and examine rise through them.  However, we note that  we still make no attempt to model any process that creates the magnetic concentrations from the large-scale field since we exclude the shear of the tachocline.  

We do, however, pay great attention to modelling the overshooting convection that should be present at the base of the solar convection zone, and the magnetic fields that might be associated with it.  Overshooting convection has been much studied in the astrophysical context \citep[see e.g.][]{Hurlburt:etal:1989, Zahn:1991, Hurlburt:etal:1994,Singh:etal:1995, Brummell:etal:2002, Korre:etal:2019}.
When a convectively-unstable layer is abutted by a convectively-stable layer, the motions of the convection are not confined to the convective layer alone, and can ``overshoot'' into the adjacent stable layer, where they are buoyantly decelerated.  If the overshooting motions are sufficiently strong enough to mix the thermodynamic background close to adiabatic in the overshoot zone, the effect is termed ``penetration'' rather than overshoot.  There is much astrophysical interest in the degree of overshoot and penetration since such motions imply extra mixing, and much attention has been applied to the extent of such motions and mixing since it may have a significant impact on stellar evolution. 

A particularly intriguing magnetic effect also occurs in overshooting convection.  Magnetic field that exists on much larger scales than the convective turbulence can be expelled from the convective region to form a layer at the edge of the overshoot zone, in a process known generically as ``magnetic pumping'' \citep[see e.g.][]{Dorch:etal:2001,Tobias:etal:1998,Tobias:etal:2001, Korre:etal:2021}.  There are a number of effects that can contribute to such expulsion but likely the dominant one is the transport of large-scale field down a gradient of turbulent intensity.   This process has been   characterized in mean-field models by the $\gamma$-effect \citep{Raedler:1968}.  Clearly such a gradient exists between the convection zone and the radiative layer below the overshoot.  The comprehensive three-dimensional simulations of \cite{Tobias:etal:2001} showed that mean field would accumulate at the lower edge of the overshoot zone even though there was a constant circulation of smaller-scale magnetic field leaving and arriving at the layer by magnetic buoyancy and advection effects.  Such a layer was only slowly eroded by diffusion and boundary effects.

This paper therefore extends the work of Papers 1 and 2  substantially in two main ways.  Firstly, the  magnetic concentration evolves by magnetic buoyancy through two  vertically-stacked regions. The flux concentration begins in a deeper layer that is initially convectively-stable (representing a radiative zone, or a tachocline), and then, if it buoyantly rises, it transits through a convection zone.  This is in contrast to the previous work where the buoyant rise only transited through a single quiescent adiabatic layer. Secondly, the presence of this two-layer system of overshooting convection serves to generate a self-consistent distribution of any imposed large-scale background field by magnetic pumping.  In the previous works of Papers 1 and 2, the profile of the background field was merely imposed artificially.

The structure of this paper is as follows: in \S~\ref{sec:model}, we outline our two-layer magnetoconvection model, including the governing equations and the  numerical methods used; in \S~\ref{sec:conv_2D}, we exhibit results from simulations only incorporating  overshooting convection in 2.5D in a similar manner to the 3D  results of \cite{Tobias:etal:2001}; in \S~\ref{sec:fluxtube_conv}, we study under what conditions an \emph{isolated} rising flux concentration is able to transit the overshooting convection, as a pre-cursor to examining the effect of the existence of a background field; in \S~\ref{sec:Pumping_conv}, without any flux tubes present, we examine the self-consistent rearrangement of a  large-scale (mean) horizontal magnetic field by magnetic pumping in 2.5D in a similar manner to the 3D simulations of \cite{Tobias:etal:2001}; in  \S~\ref{sec:fluxtube_backgroundfield}, for a canonical set of parameters, we combine the acquired knowledge into simulations that study the rise of a flux concentration in a two-layer overshooting convective system incorporating a self-consistent large-scale (mean) background magnetic field; in \S~\ref{sec:fluxtube_backgroundfield_Bs} - \ref{sec:vary_S}, we examine the influence of various parameters of the model, check that the theoretical explanation of Papers 1 and 2 still holds, and check the statistical robustness of the results.  Ultimately, we draw conclusions in \S~\ref{sec:discussion_conclusions}, in particular, noting  that the selection mechanism found in Papers 1 and 2 still operates in this far more complex situation.

\section{Model and Methods} \label{sec:model}

We follow the formulation used in \cite{Tobias:etal:2001}  for our two-layer overshooting magnetoconvection simulations.  We consider a 2-D Cartesian domain in one horizontal ($x$) and one vertical ($z$) direction, that contains a fully compressible ideal gas confined between two horizontal, impenetrable, stress-free boundaries. We keep all three components of the velocity and magnetic vector fields, but all components are independent of the missing third direction (i.e. $\partial / \partial y = 0$).  This type of setup is often referred to as 2.5D. Note that $z$ is depth and varies downwards  from the top.  The Cartesian box is $x_{\rm m}d$ wide and $z_{\rm m}d$ deep, where $d$ is the depth of the convection zone in the two layer system. 
We use $x_m=6, z_m=2.5$ for all our simulations here.
We non-dimensionalize our system using $d$, $T_0$ (the temperature at the upper boundary), $\rho_0$ (the density at the upper boundary) and $B_0$ (some measure of the initial field strength) as our units of length, temperature, density and magnetic field strength. The thermal sound crossing time at the top of the domain, $({d^2}/({(c_p-c_v)T_0)})^{1/2}$, is our unit of time, where $c_p$ and $c_v$ are the specific heats of the fluid at constant pressure and constant volume respectively (and their ratio $\gamma = c_p/c_v$ will be used later). 
With these units, the governing non-dimensional equations (the conservation of mass, momentum and
energy, the equation of state for a perfect gas, the induction
equation and the divergence-free condition for magnetic fields) are \citep{Tobias:etal:2001}

\begin{equation} \label{eq:cont}
    \partial_t \rho + \nabla \cdot \left(\rho \mathbf{U} \right) =  0,
\end{equation}

\begin{equation} \label{eq:motion}
    \partial_t \left(\rho \mathbf{U} \right) + \nabla \cdot \left( \rho \mathbf{U} \mathbf{U} -
\alpha \mathbf{B} \mathbf{B} \right) = -\nabla p_t  + \sigma C_k\left[ \nabla^2 \mathbf{U} +
\frac{1}{3} \nabla(\nabla \cdot \mathbf{U}) \right]
+ \rho g \hat{z},
\end{equation}

\begin{equation} \label{eq:energy}
    \partial_t T + \nabla \cdot (\mathbf{U} T) + (\gamma - 2) T \nabla \cdot \mathbf{U} = \frac{\gamma C_k}{\rho} \nabla
\cdot \left( {\kappa_z} \nabla T \right) + \frac{\zeta C_k \alpha (\gamma-1)}{\rho}
| \nabla \times \mathbf{B} | ^2  + V_{\mu},
\end{equation}

\begin{equation} \label{eq:induction}
    \partial_t \mathbf{B} = \nabla \times (\mathbf{U} \times \mathbf{B}) + C_k \zeta \nabla^2 \mathbf{B},
\end{equation}

\begin{equation} \label{eq:solenoidal}
    \nabla \cdot \mathbf{B}  =  0,
\end{equation}

\begin{equation} \label{eq:pressurebalance}
    p_{t} = p_{g} + p_{m} = \rho T + \alpha \frac{|\mathbf{B}|^2}{2}.
\end{equation}
Here, in nondimensional form, $\mathbf{U} = (u,v,w)$ is the velocity, $\mathbf{B} = (B_x,B_y,B_z)$ is the magnetic field, $T$ is the temperature and $\rho$ is the density. 
The total pressure, $p_t$, is the sum of the gas pressure, $p_g$ ($= \rho T$ for an ideal gas), and the magnetic pressure, $p_m = \alpha {|\mathbf{B}|^2}/{2}$, where $\alpha = \sigma \zeta Q C_k^2$ (where the latter parameters will be explained shortly). The rate of viscous heating is $V_\mu = {(\gamma -1) {C_k \ \rho}} \sigma \partial_i u_j (\partial_i u_j + \partial_j u_i - {(2 / 3)} \nabla \cdot \mathbf{U} \delta_{ij} )$. 

\begin{figure}
    \centering
    \includegraphics[width=13cm, height=9cm]{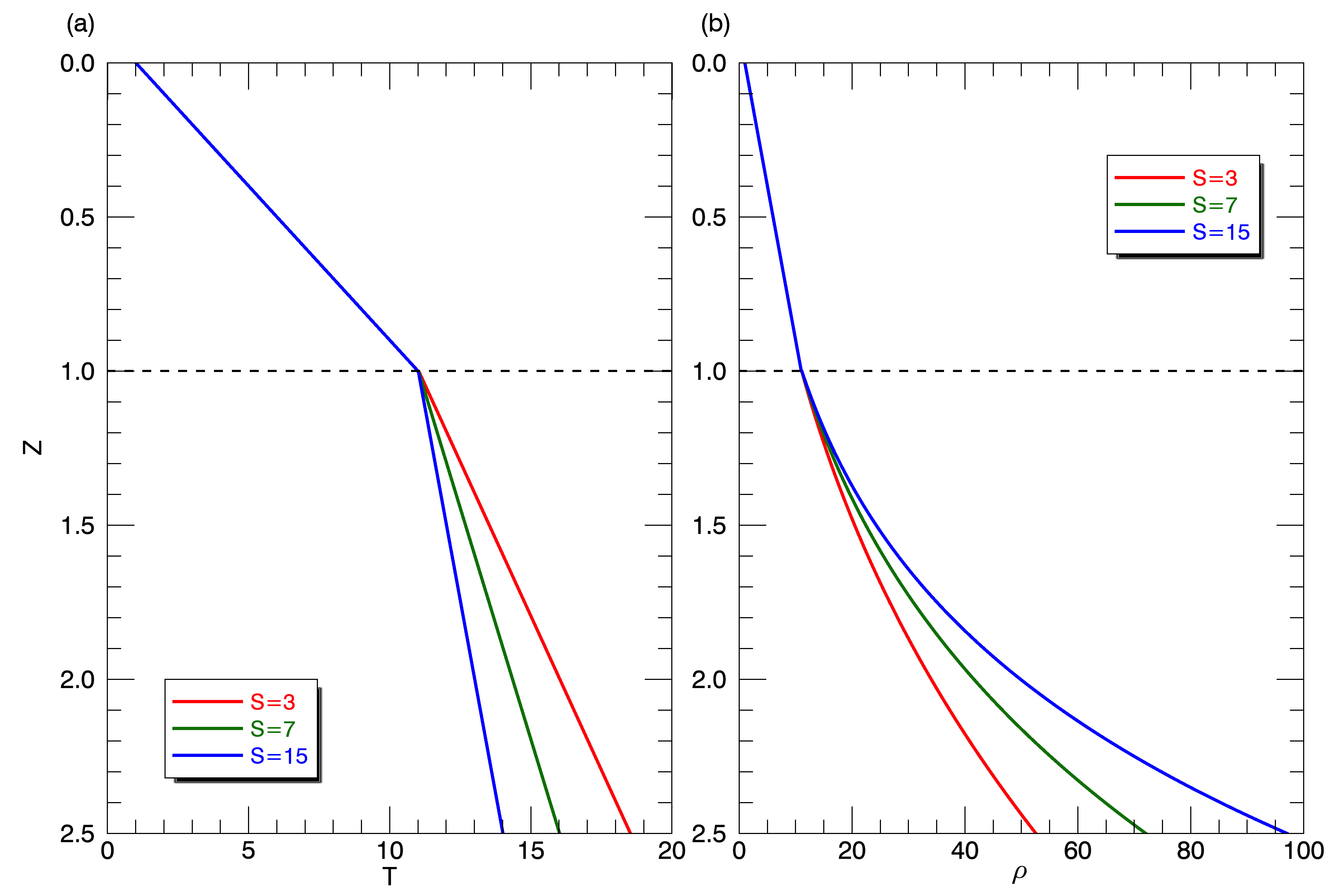}
    \caption{Profiles of the ($x$-independent) polytropic  thermodynamic initial conditions, a) $T$, and b) $\rho$, as a function of depth, $z$, for $S=3,~7$ and $15$. The dashed line marks the transition between the convection zone and the radiative zone.}
    \label{fig:thermo_init_cond}
\end{figure}

Our two layer system consists of a convective layer (layer $1$) overlying a convectively-stable layer (layer $2$).  This layering is enforced by a piecewise constant thermal conductivity (with a narrow smoothed junction between the two layers) that defines the two layers as being piecewise continuous polytropes when in a hydrostatic state. 
Since the total hydrostatic heat flux through the domain must remain the same at any depth, the ratio of the thermal conductivities in the two layers, $\kappa_{i}$, can be described by the parameter $S=(m_2-m_{\rm ad})/(m_{\rm ad}-m_1)$, related to the polytropic indices in the two layers, $m_i$, and the adiabatic index $m_{\rm ad}=1/(\gamma-1)$ ($=1.5$ for an ideal monatomic gas where $\gamma=5/3$):
\begin{equation}
\frac{\kappa_2}{\kappa_1} = \frac{m_2+1}{m_1+1} = \frac{S (m_{\rm ad}-m_1) +m_{\rm ad}+1} 
{m_1+1}.
\end{equation}
We choose $m_1=1$ always, and specify the temperature gradient in the hydrostatic upper layer as $\theta$.  The relative stability of the two domains is then measured by $S$, often referred to as the stiffness parameter (\cite{Hurlburt:etal:1994,Brummell:etal:2002}).  Increasing $S$ increases the relative stability (``stiffness") of the lower layer.
Figure \ref{fig:thermo_init_cond} shows the polytropic temperature and density profiles versus depth in the two layers for the three values of $S$ that we use in these simulations, $S=3,7$ and $15$. 

The other dimensionless parameters that govern the system are as follows.  The Rayleigh number,
\label{eqn_rayleighno}
\begin{equation}
Ra(z) = {\theta^2 (m_i+1) \over \sigma C_{k_z}^2}
\left( 1 - {(m_i + 1)(\gamma-1) \over \gamma}
\right) (1 + \theta z)^{2m_i-1} ,
\end{equation}
is a derived measure that evaluates the competition between buoyancy driving (given in terms of the stratification inputs $\theta$ and $m_i$) and diffusive effects. This is therefore a measure of the supercriticality and vigor of the convection.  The $Ra$ 
involves the non-dimensional thermal conductivity $C_{k_z} = C_k \kappa_z$,
where $\kappa_z =  {\kappa_i}/{\kappa_1}$ and $C_k = \kappa_1/\{d\rho_0 c_p
[(c_p - c_v)T_0]^{1/2}\}$. which is different in the two layers. The $Ra$ is a function of depth, and, in this study, the quoted value of $Ra$ is evaluated at the middle of the unstable upper layer ($z=0.5$) under the conditions of the initial polytrope.

The Prandtl number at a given depth is 
\label{eqn_prandtlno}
\begin{equation}
\sigma_z = \frac{\mu c_p}{\kappa_z},
\end{equation}
where $\mu$ is the (constant) dynamic viscosity.  The Prandtl number is depth-dependent, but   $C_{k_z} \sigma_z = C_k \sigma$ is independent of $\kappa_z$ and therefore is independent of depth. Any quoted Prandtl number here is $\sigma$, the value in the upper layer. Similarly, the non-dimensional version of the  magnetic resistivity $\eta$ is a depth-dependent  Schmidt number 
\label{eqn_schmidt}
\begin{equation}
\zeta_z = \frac{\eta c_p}{\kappa_z}, 
\end{equation}
but $C_{k_z} \zeta_z = C_k \zeta$ is also  independent of depth, and any quoted value is $\zeta$, the value in the upper layer.

The Chandrasekhar number,
\label{eqn_chandra}
\begin{equation}
Q = \frac{B_0^2 d^2}{\mu_0 \mu \eta},
\end{equation}
where $\mu_0$ is the magnetic permeability, is a non-dimensional measure of the strength of the imposed magnetic field $B_0$, comparing the associated Alfven timescale (squared) to the timescales of viscous and resistive effects.  
Note that the parameter truly governing magnetic effects in the equations is $\alpha = \sigma \zeta Q C_k^2$. This parameter determines the relative importance of the magnetic pressure compared to the gas pressure, and is the inverse of the quantity often referred to in other contexts as ``the plasma $\beta$''.  Increasing $Q$ (for fixed diffusivities) increases the influence of all components of the Lorentz forces and therefore the dynamical back-reaction of any magnetic field on the flow field.

At the upper and lower boundaries of our domain, we apply impenetrable and stress-free boundary conditions
\begin{eqnarray}
&w = \partial_z u = \partial_z v = 0\  {\rm at}\ z = 0,z_m,
\label{eqn_boundaryconditions}
\end{eqnarray}
which ensure that the mass flux and mechanical energy flux vanish on the boundaries conserving the total mass. The boundary conditions on temperature are
\begin{eqnarray}
&T = 1 \ {\rm at}\ z=0,\ \partial_z T= \frac{\kappa_2}{\kappa_1}\theta \ {\rm at}\  z=z_m,
\end{eqnarray}
and this imposed heat flux at the lower boundary is the only flux of energy into and out of the system.  The magnetic boundary conditions are specified as
\begin{equation}
\label{mag_bc_1}
B_x = B_y = 0\  {\rm at}\ z = 0,z_m.
\end{equation}
Note, that it is sufficient to impose boundary conditions only on the horizontal
components of the magnetic field due to the solenoidality of $\mathbf{B}$. We note that non-zero vertical gradients of the horizontal field can be present at the boundaries so that the magnetic energy can decrease with time, i.e. these can be ``run-down" systems. 

The domain is periodic in the horizontal in all variables and therefore the system can be solved numerically using the same high-performance hybrid pseudo-spectral code as in \cite{Tobias:etal:2001}.  The resolution typically used is $512 \times 600$ for the $6 \times 2.5$ aspect ratio 2D domain.

\section{Results} \label{sec:Results}

In order to achieve our ultimate goal of examining the effect of a self-consistent large-scale background field on the rise of a magnetic concentration through a convection zone, we must first complete some preparatory steps.  First of all, we need to establish overshooting convection in our two layer system representing the base of the convection zone.  Secondly, we must examine what is the necessary strength of field in an \emph{isolated} flux tube for it to have enough buoyancy to rise successfully through the overshooting convection established in the first step.  Since we are interested in bias effects on \emph{emerging} flux tubes, we need to eliminate flux tubes that would not normally emerge.   Thirdly, we must establish a vertical profile of large-scale (mean) horizontal field within the overshooting convection self-consistently via turbulent pumping, in the absence of the flux tube.  Finally, we can then evolve the (formerly emergent, when isolated) flux tube from step 2 as a concentration amongst the  volume-filling, self-consistently organized background field from step 3 in order to examine the effect of that background field on the different signs of twist in the tube/concentration.  We now describe results from each of these steps in detail.

\subsection{Convection in 2D} \label{sec:conv_2D}

For purely hydrodynamic convection, we effectively solve the system of  equations (\ref{eq:cont}-\ref{eq:pressurebalance}) with ${\bf B}=0$ (although practically, of course, equations (\ref{eq:induction}-\ref{eq:solenoidal}) and the Lorentz terms in equation (\ref{eq:motion}) are omitted completely).
We choose some canonical parameters for our study as follows.  Our two-dimensional Cartesian box is of size $x_{m}=6, z_{m}=2.5$ so that our convection zone (of depth unity) has an aspect ratio of 6.  These dimensions allow plenty of horizontal space for naturally-sized  convective cells to form, and plenty of vertical room for overshooting.    Throughout this study, we canonically use  $Ra=4\times 10^{4}$ and $Pr=0.1$, in order to achieve reasonably supercritical dynamics at a Prandtl number less than unity. These values are in the correct regime for  astrophysical purposes, but are orders of magnitude away from their true astrophysical values, due to numerical limitations.  We set the stratification using $m_1=1$ and $\theta=10$ and choose $S$ from the values $S=3,7,15$.  These choices of $S$ provide a range of overshooting dynamics from deep to fairly confined, as we shall see shortly.  For much of this paper, we use $S=7$ as the canonical value, but all $S$ are investigated and described in detail.

\begin{figure}
    \centering
    \includegraphics[width=13cm,height=9cm]{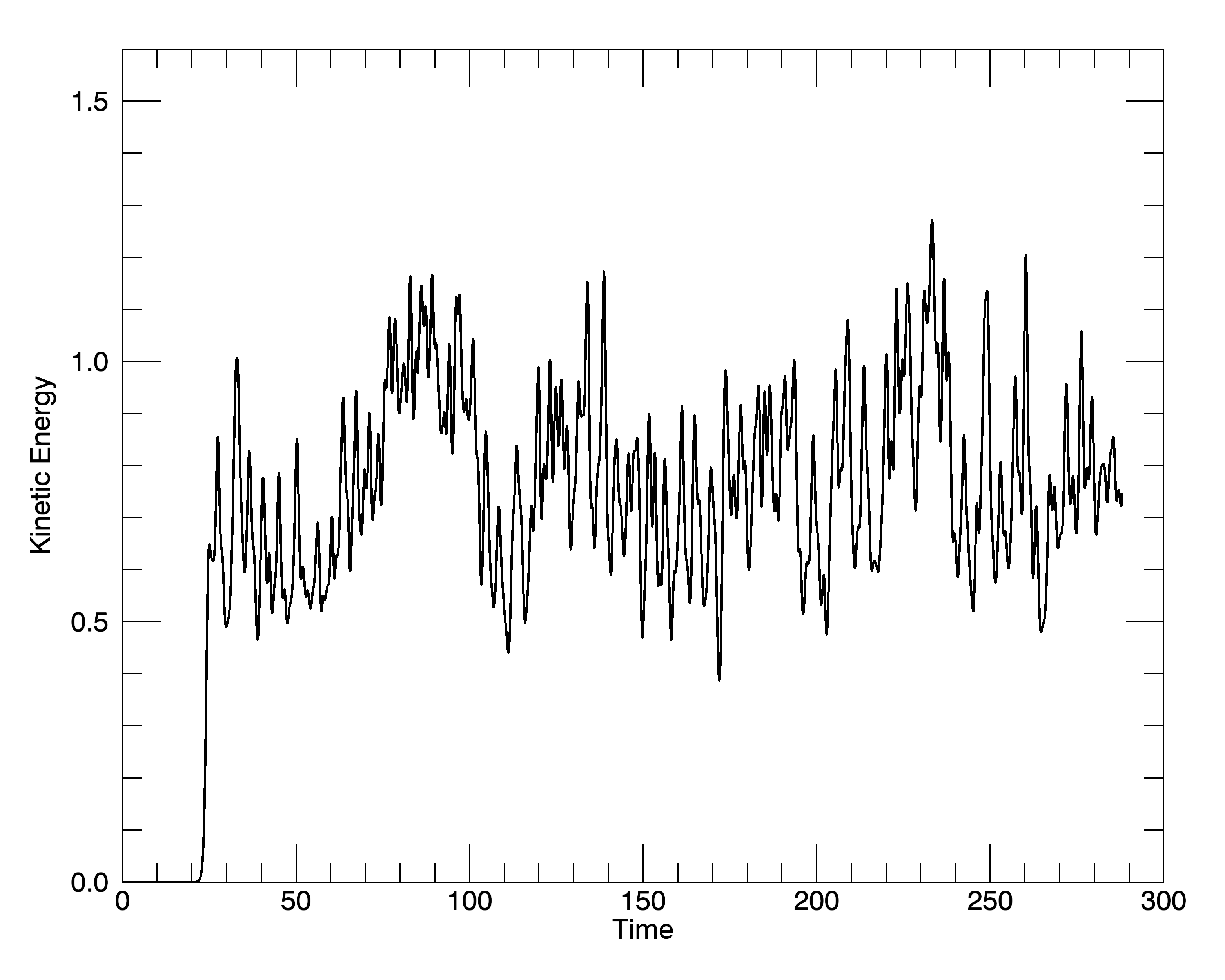}
    \caption{Kinetic energy as a function of time for the hydrodynamic case with $S=7,~Ra=4\times 10^{4}$ and $Pr=0.1$.}
    \label{fig:ke_S7_highreso}
\end{figure}

\begin{figure}
    \centering
    \includegraphics[width=\columnwidth,height=10cm]{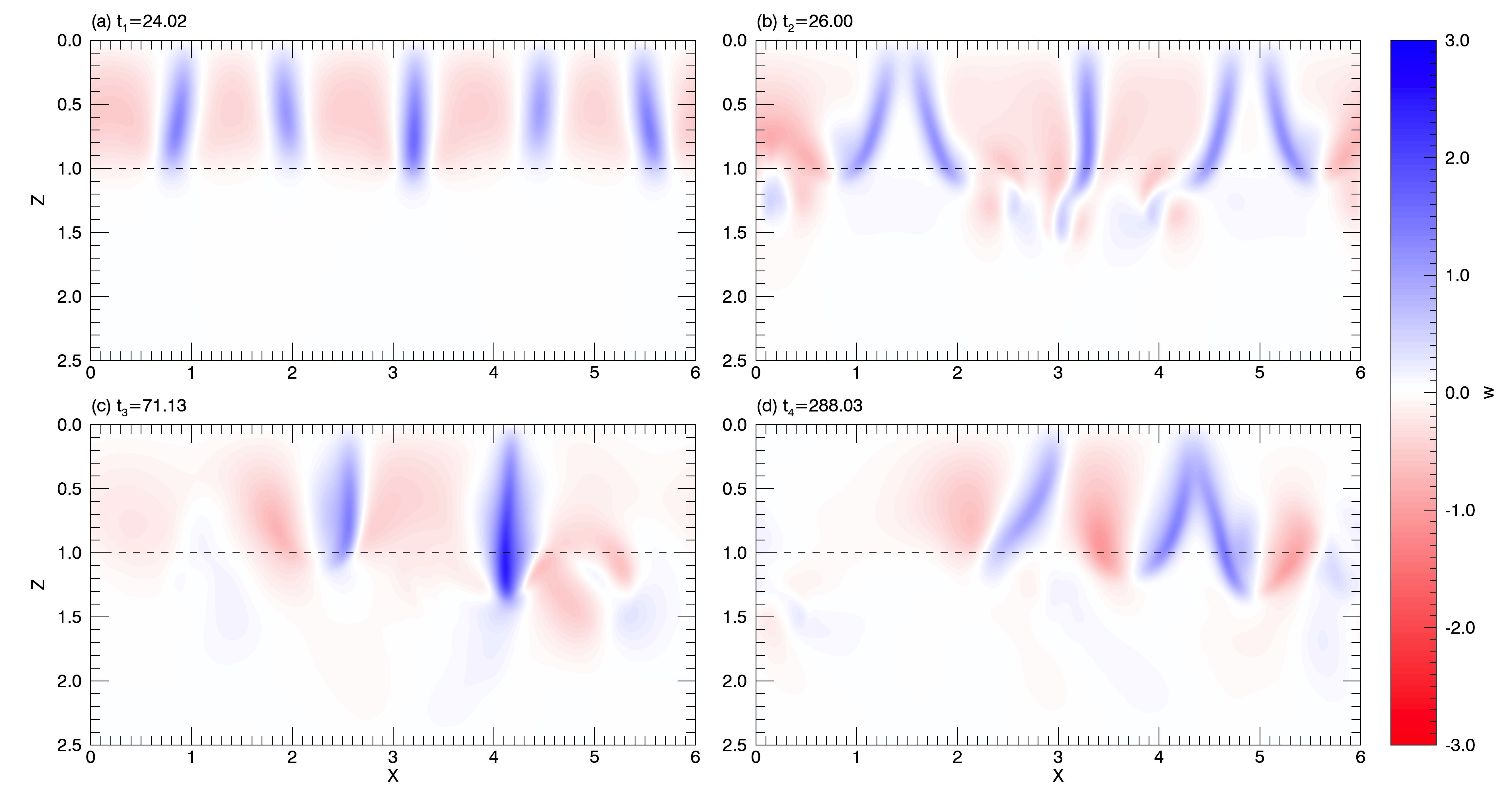}
    \caption{Snapshots of the Vertical velocity, $w$,  plotted at four different times (as marked) for the hydrodynamic case with $S=7,~Ra=4\times 10^{4}$ and $Pr=0.1$. Blue tones indicate downflows whereas upflows are colored red. The dashed line marks the transition between the convection zone and the radiative zone.}
    \label{fig:snapshow_w_S7}
\end{figure}

The time evolution of these equations from initial conditions consisting of a small amount of thermal noise in the convectively unstable layer leads to convection in that layer which can then overshoot into the lower convectively-stable layer.  As an example, the time evolution of total kinetic energy in the simulation domain for the case with $S=7$ is shown in Figure \ref{fig:ke_S7_highreso}. An initial increase in the kinetic energy characterizes the onset of the convective instability in the unstable layer, and, eventually, the kinetic energy saturates nonlinearly to a statistically-steady (stationary) state. 
Figure \ref{fig:snapshow_w_S7} shows snapshots of the vertical velocity ($w$) from the simulation domain at four different times during this evolution ($t \sim 24,26,71,288$). Figure \ref{fig:snapshow_w_S7}a shows that the initial evolution consists of the formation of a cellular pattern of roughly five cells of convective motions, characterized by narrow downflows (blue) and broader upflows (red). As the convection proceeds, these regular cellular patterns quickly evolve into turbulent nonlinear  dynamics, consisting of  time-dependent plumes that begin to  overshoot into the stable layer below, as seen in Figure \ref{fig:snapshow_w_S7}b.  During the later, more characteristic, evolution, as shown in Figures \ref{fig:snapshow_w_S7}c and d, plumes can form, migrate, split and merge, and the flows   establish a strong identity in the stable layer.  Flow structures in the stable layer remain  connected to those in the convective layer as if there were no boundary between the two regions; such dynamics is known as  ``overshooting".

The depth to which this overshooting occurs is a topic of great interest in astrophysics and has been studied in detail in both two-  and three-dimensional simulations  \citep[e.g.][]{Hurlburt:etal:1989,Hurlburt:etal:1994, Brummell:etal:2002}.
This overshooting depth 
depends on many of the parameters, such as $Ra$ and $Pr$, but here we concentrate on the effect of the relative convective stability of the two layers, quantified by the parameter $S$ (introduced by \cite{Hurlburt:etal:1994}, defined here in \S \ref{sec:model}).
Increasing $S$ increases the relative stability of the lower layer since we have fixed the stratification of the upper layer.  This increases the  ``stiffness" of the lower layer, in the sense that the resistance to downward convective motions entering the layer is greater, due to enhanced buoyant deceleration.  For greater $S$, due to the more rapid increase in density in the stable region (as can be seen in Figure \ref{fig:thermo_init_cond}), a plume entering this region feels a larger (negative) density perturbation and therefore decelerates more rapidly than it would at lower $S$.

\begin{figure}
    \centering
    \includegraphics[width=11cm,height=11cm]{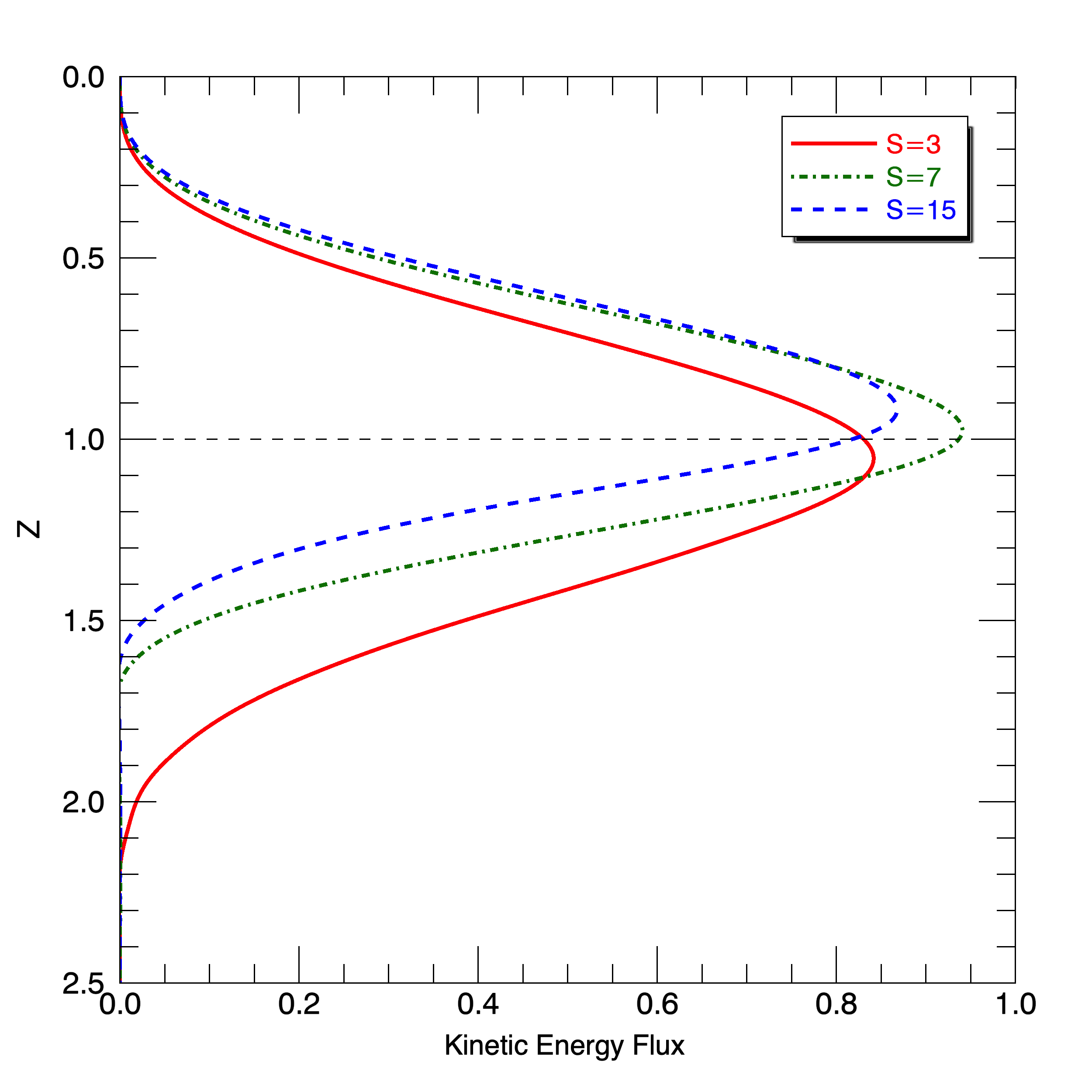}
    \caption{Time- and horizontally-averaged kinetic energy flux $\langle F_k \rangle$ as a function of $z$ for three different cases with $S=3,~7$, and $15$ at $Ra=4\times 10^{4}$ and $Pr=0.1$. The horizontal dashed line at $z=1$ marks the transition between the convection zone and the radiative zone.}
    \label{fig:ke_avg_svar}
\end{figure}

\begin{figure}
    \centering
    \includegraphics[width=\textwidth,height=5cm]{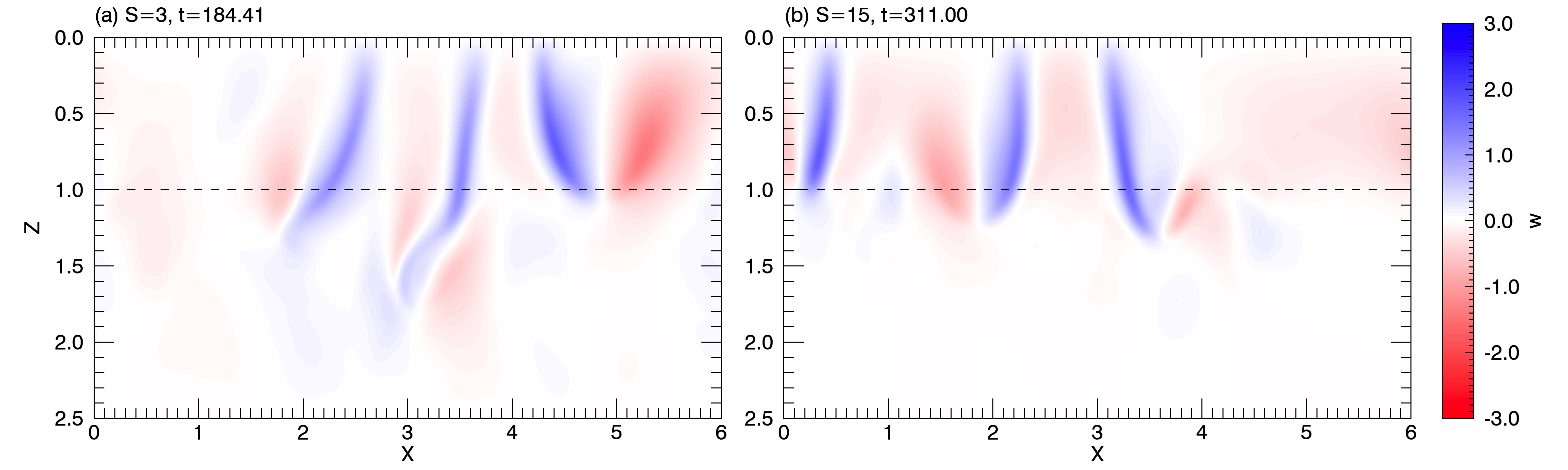}
    \caption{Vertical velocity ($w$) snapshots at a representative time in simulations at $Ra=4\times 10^{4}$ and $Pr=0.1$ with (a) $S=3$ and (b) $S=15$. }
    \label{fig:w_conv_s3_15_combined_highreso}
\end{figure}

The level to which the convection overshoots into the stable region can be quantified by examining the time- and horizontally-averaged kinetic energy flux, 
$<{F}_k>$ (where $F_k = (1/2) \rho w |{\bf u}|^2$, and $<\cdot>$ denotes a horizontal average over the whole vertical domain and a time-average over a representative time in the stationary state of the convection), as shown in Figure \ref{fig:ke_avg_svar}.  
The overshoot depth is often taken to be the depth at which the kinetic energy flux falls to a certain fraction of its maximum value, typically $1\%$ \citep[e.g.][]{Hurlburt:etal:1994, Brummell:etal:2002}.  This value reflects where, on average (both in time and space), the convective motions die out. 
Figure \ref{fig:ke_avg_svar} shows this average kinetic energy flux profile for the three different values of $S=3,7,15$. We can clearly see that the extent of overshoot decreases as $S$ increases, as has been found many times before \citep[e.g.][]{Hurlburt:etal:1994}.  The measured overshoot depths here are $z_o=2.08, 1.63, 1.55$ for $S=3,7,15$ respectively. This result is an average over many instantaneous realizations, such as those shown as examples in Figure \ref{fig:w_conv_s3_15_combined_highreso}, which shows the vertical velocity at a representative time in each of the $S=3$ and $S=15$ cases (for comparison with the case for $S=7$ in Figure \ref{fig:snapshow_w_S7}) .  Between these figures, it can easily be confirmed that the dynamics are indeed active to greater depths in the stable layer for lower $S$ and are less extensive for higher $S$.
The scaling of the overshoot depth with $S$ is a topic of some debate, but here we are more concerned with creating initial conditions for the later parts of our study.

\begin{figure}
    \centering
    \includegraphics[width=11cm, height=11cm]{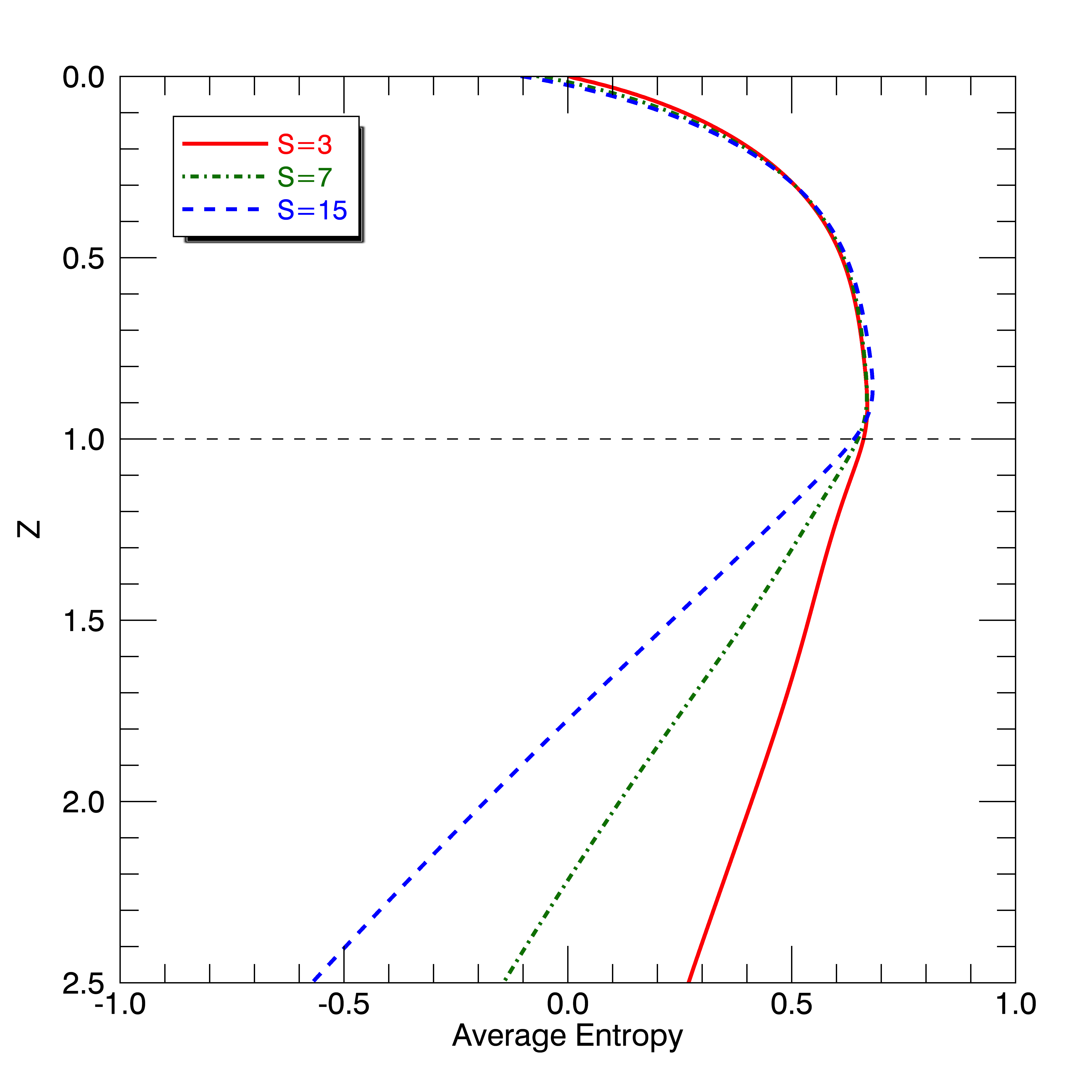}
    \caption{Time- and horizontally-averaged entropy as a function of height for $S=3,~7,$ and $15$.}
    \label{fig:avg_entropy_Svariation}
\end{figure}

We note that if the dynamics in the overshoot layer below the convection zone are sufficiently energetic to mix the thermodynamics of the stable region strongly, driving it towards a well-mixed adiabatic system with a constant entropy, then the overshooting convection becomes known as ``penetrative convection" \citep{Zahn:1991}.  All of our cases here exhibit overshooting convection but not penetrative convection. Figure \ref{fig:avg_entropy_Svariation} confirms this by exhibiting the time- and horizontally-averaged entropy for the  three cases at different $S$.  We can see that any  mixing that has  occurred below the convection zone was insufficient to reduce the strong entropy gradient there.

\subsection{Rise of Flux tubes in Pure Convection} \label{sec:fluxtube_conv}

Having established a purely convective background state, we now wish to understand under what conditions a flux tube is able to rise through such a state {\it in the absence of any large-scale volume-filling field}.  This is necessary since we are interested later in processes that affect flux tubes that would otherwise emerge buoyantly through  the convection zone.  We therefore now turn to MHD simulations where we add an {\it isolated} flux tube 
to established convection.  Again, similar  simulations have been performed before, for example, in 3D for fully compressible and anelastic convection  \cite{Cline:thesis:2003,Abbett:etal:2004}. A significant discovery of this past body of work was that magnetic structures need to possess magnetic energy greater than the peak kinetic energy of the flow to survive transit of the convection zone.  Here, we need to establish exact values for our particular setup to inform later, more complete and complex simulations that also include a large-scale background field.

\begin{figure}
    \centering
    \includegraphics[width=\columnwidth,height=10cm]{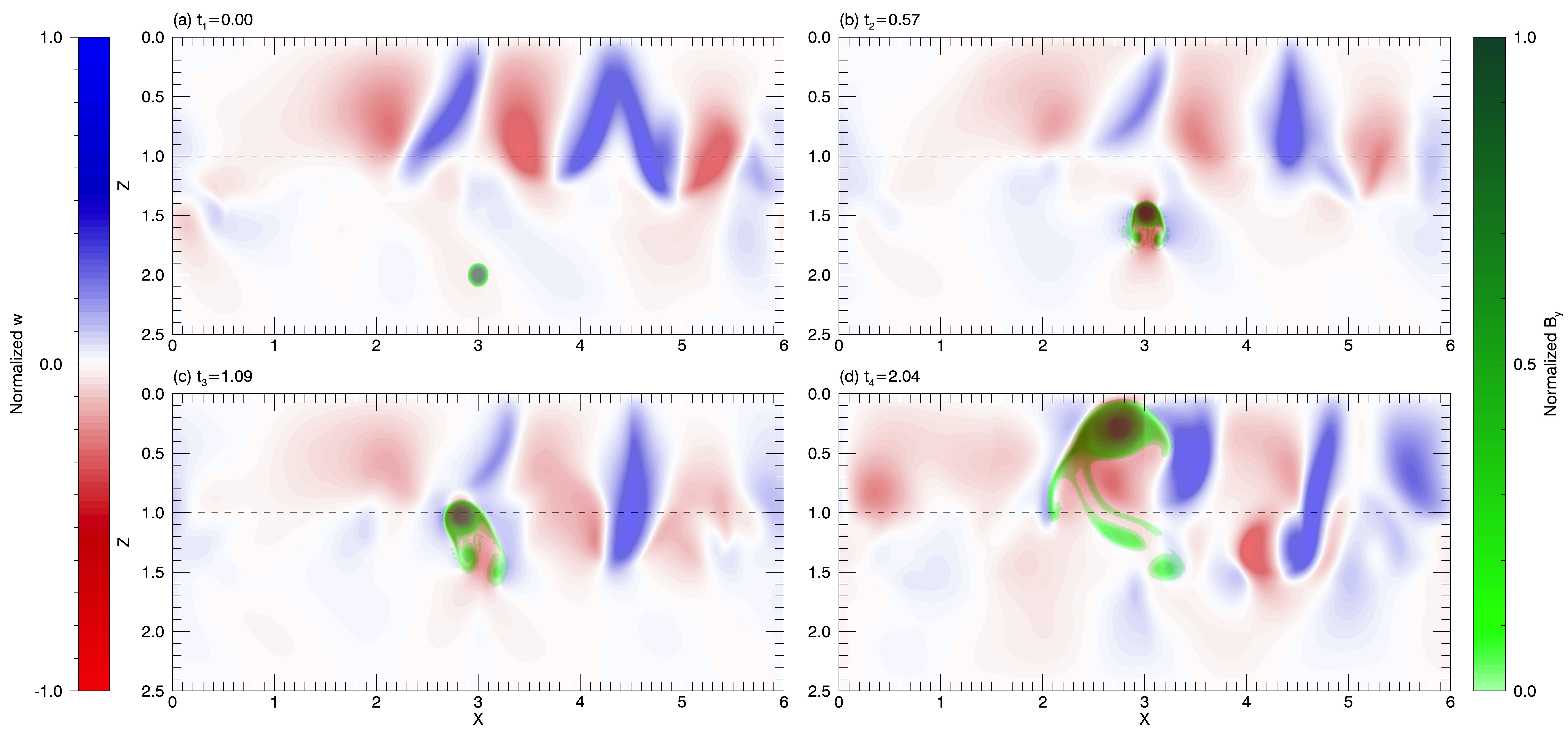}
    \caption{Intensity plots of normalized $B_{y}$, overplotted on the normalized vertical velocity, $w$, at four different times (as shown) in the evolution for the case with $S=7$, $Ra=4\times 10^{4}$, $Q=2 \times 10^{8}$, $Pr=0.1$ and $Pm=100$. The initial location of the flux tube is $(x_{c},z_{c}) = (3.0,2.0)$. The dashed horizontal line indicates the transition between the convection zone and the radiative zone.}
    \label{fig:tube_convection_s7}
\end{figure}

\begin{figure}
    \centering
    \includegraphics[width=\columnwidth,height=10cm]{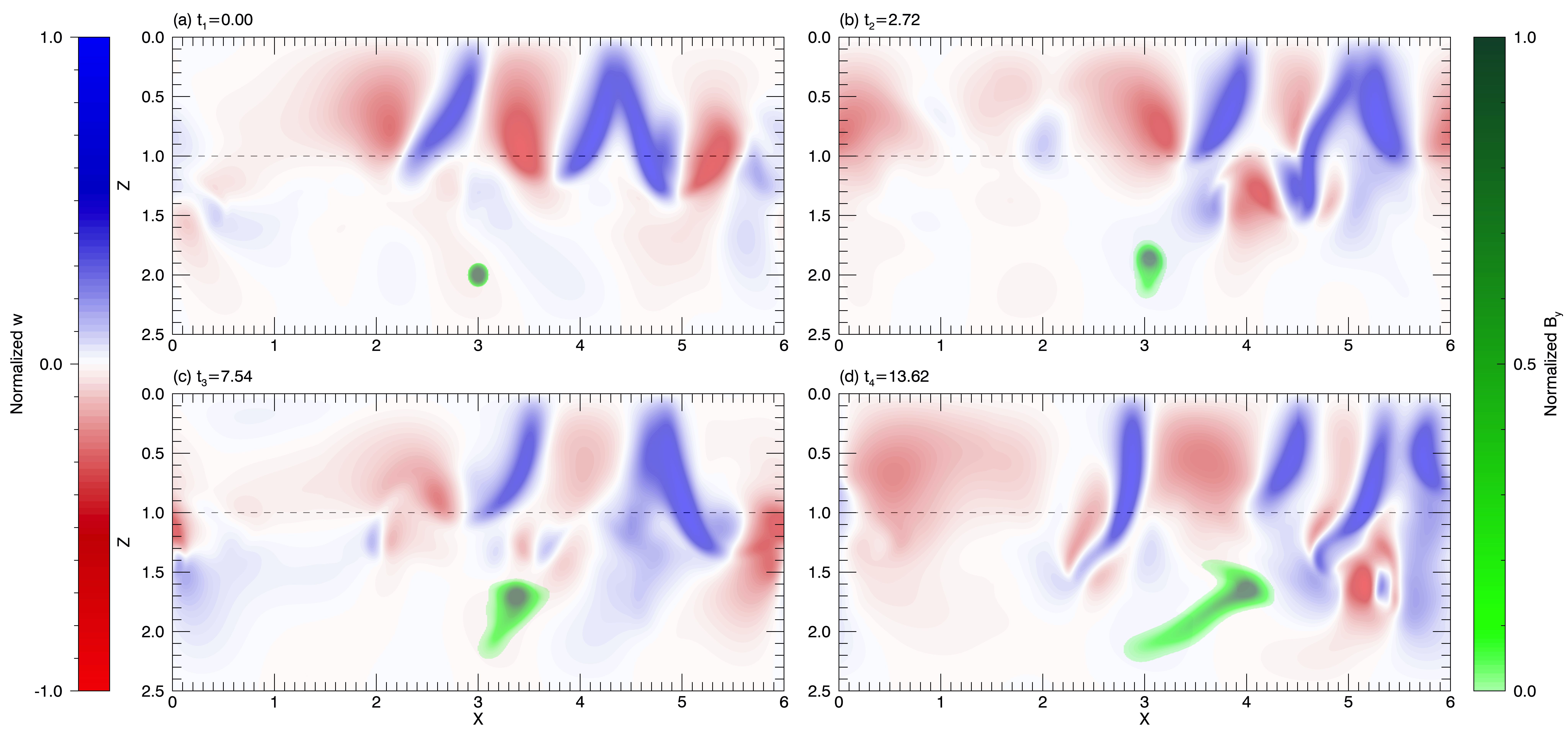}
    \caption{As for Figure \ref{fig:tube_convection_s7} but for $Q=5\times 10^{6}$.}
    \label{fig:tube_convection_s7_Q5e6}
\end{figure}

The initial setup for these cases consists of an instance in time from the stationary state of one of the overshooting convection simulations (such as those shown in Figures \ref{fig:snapshow_w_S7}c, d or Figure \ref{fig:w_conv_s3_15_combined_highreso}).  To this state, we add a twisted magnetic flux tube embedded at a chosen location in the stable zone (see, for example, Figure \ref{fig:tube_convection_s7}a or \ref{fig:tube_convection_s7_Q5e6}a, described later). The total magnetic field of the flux tube is given by

\begin{equation} 
    \mathbf{B} = 
    (B_x,B_y,B_z)
    = \mathbf{B_{\rm tube}}
    = \Bigg( -2q\frac{(z_{c}-z)}{r_{o}}, 1 - \frac{r^{2}}{r_{o}^{2}} ,-2q\frac{(x-x_{c})}{r_{o}} \Bigg) ~\text{for}~ r \le r_{o}
    \label{eq:fluxtube}
\end{equation}
where $q$ is a measure of the initial twist of the field, $x_{c}$ and $z_{c}$ are the horizontal and vertical locations of the center of the flux tube, $r_{o}$ is the outer radius of the structure, and $r=({(x-x_c)^2+(z-z_c)^2})^{1/2}$ is the cylindrical radius measured from the center of the tube.  We choose $x_c=3$ (the horizontal middle) and $z_c=2$ (solidly in the stable layer) as the standard initial location. Note that the field is purely axial at the center of the flux tube and the amplitude there is unity.  With $|q|=0.5$, the azimuthal field in the tube is also unity at $r=r_0$.  We use $|q|=0.5$ exclusively throughout this study, and in this portion of the work, the sign of $q$ does not matter, since there is no background field against which its orientation can be judged.   

We adopt $\zeta=0.001$ for all magnetic simulations from now on.  This value is small so that tubes, which can have strong gradients at the edges, do not diffuse faster than other dynamical times of interest, such as the buoyant rise time.  This value renders the magnetic Prandtl number, $Pm = \sigma/\zeta = 100$, which is not representative of astrophysical values (that are typically considerably less than one).  This fact is a statement that our simulations are too viscous and therefore not as turbulent as they should be, but this is a numerical limitation.  

At this point, we only examine $S=7$ and leave the study of the effects of varying $S$ until later (see \S  \ref{sec:vary_S}).  
The canonical set of governing parameters is therefore now  $S=7,Ra=4\times 10^{4},~Pr=0.1,~\zeta=0.001$. We are left to vary the Chandrasekhar number, $Q$, as the key parameter of interest in this section.   With the maximum amplitudes of both the axial and azimuthal fields of the flux tube being unity, the Chandrasekhar number entirely determines the initial strength of the magnetic field (in this case, the flux tube) and determines the dynamical influence (via $\alpha$) of the Lorentz force  (see equation \ref{eq:motion}). 
When we add the flux tube to the existing convective fields, we do so assuming that the total pressure and the temperature equilibrate quickly, so that the increase in magnetic pressure is entirely compensated for by a drop in density.  This means that the initial condition is not in equilibrium and the tube experiences an initial upwards buoyancy force.

Figure \ref{fig:tube_convection_s7} shows the evolution of the full set of equations (\ref{eq:cont}-\ref{eq:pressurebalance}) for such an initial state.  This figure shows intensity plots of the normalized axial field, $B_{y}$, which is a good indicator of the location of the tube, overplotted with the normalized vertical velocity, $w$, representing the convective flows. We use the absolute maximum values of $B_{y}$ and $w$ as the normalizing factors for the axial field and vertical velocity, respectively. Red and blue colors indicate upflows and downflows.  The case shown is at the canonical parameters, for a positively-twisted ($q>0$) flux tube  with $Q=2\times 10^{8}$, and the figure displays four different times.  
Figure \ref{fig:tube_convection_s7}a shows the initial location of the cylindrical flux tube at ($x_{c}, z_{c}$) = ($3.0, 2.0$) deep in the stable region, along with the initial overshooting convective motions. As the flux tube starts rising due to its magnetic buoyancy (Fig. \ref{fig:tube_convection_s7}b), the initial motion is reasonably symmetric, and quickly becomes accompanied by the formation of vortices, a characteristic observed in  non-convective flux tube rise. The impact of overshooting convection on the dynamics of the flux tube is minimal at this stage. As the flux tube enters the overshooting region ($1.0 \le z \le 1.5$) and gets closer to the transition between the stable and convective layer, the effect of convective motions becomes evident on the flux tube. Figure \ref{fig:tube_convection_s7}c shows the flux tube entering the convection zone from the overshoot zone directly beneath a downflow. The symmetric rise of the flux tube is broken, as both buoyant and advective forces act on the tube, and the driving vortices are distorted. However, due to sufficient initial twist ($|q|=0.5$) and sufficient buoyancy ($\propto Q$), the flux tube maintains coherency and navigates its way into an upflow and thereby to the top of the convective simulation domain (Fig. \ref{fig:tube_convection_s7}d). We consider these dynamics to be characteristic of a successful rise of the flux tube (but will develop a more quantitative measure shortly).

Figure \ref{fig:tube_convection_s7_Q5e6} represents a case identical to Figure \ref{fig:tube_convection_s7} but computed at a lower $Q$. The simulation starts from the same initial steady-state convection and initial magnetic profile, albeit now with $Q=5\times 10^{6}$ (Figure \ref{fig:tube_convection_s7_Q5e6}a).  With a weaker buoyancy perturbation due to the lower $Q$, the initial rise of the flux tube is slower. Figure \ref{fig:tube_convection_s7_Q5e6}b shows that the flux tube barely rises in the stable zone and no distinct vortices form in the wake. The slow rise brings the tube to the edge of the overshooting region ($z \sim 1.5$) where even these weak flows affect the cylindrical flux structure (Fig. \ref{fig:tube_convection_s7_Q5e6}c).  Eventually, other forces dominate over buoyancy, the flux tube gets drastically distorted, and its rise becomes halted (Figure \ref{fig:tube_convection_s7_Q5e6}d). We consider this as an example of an unsuccessful rise of the flux tube. Even with minimal interaction with the overshooting, the buoyancy was insufficient to avoid destruction of the coherence of the structure.

From these two examples, it seems reasonable to conclude that the rise or lack of rise of the flux tube is determined by the amount of buoyancy imparted to the flux tube (controlled by $Q$, for fixed other parameters). We now explore a wider range of $Q$ to understand this dependence in more detail.  To determine the ultimate fate of a flux tube more concisely than the visual plots of Figures \ref{fig:tube_convection_s7} and \ref{fig:tube_convection_s7_Q5e6}, we use a more quantitative and compact measure of the rise characteristics, denoted by $z_{ft}(t)$.  This measure tracks the vertical ($z$) location of the maximum of the axial field, $B_{y}(x,z)$, as a function of time.
It is clear from the earlier figures (Figures  \ref{fig:tube_convection_s7} and \ref{fig:tube_convection_s7_Q5e6}) that this quantity tracks the progress of a tube reasonably well, at least in those cases.

\begin{figure}
    \centering
    \includegraphics[width=8cm,height=9cm]{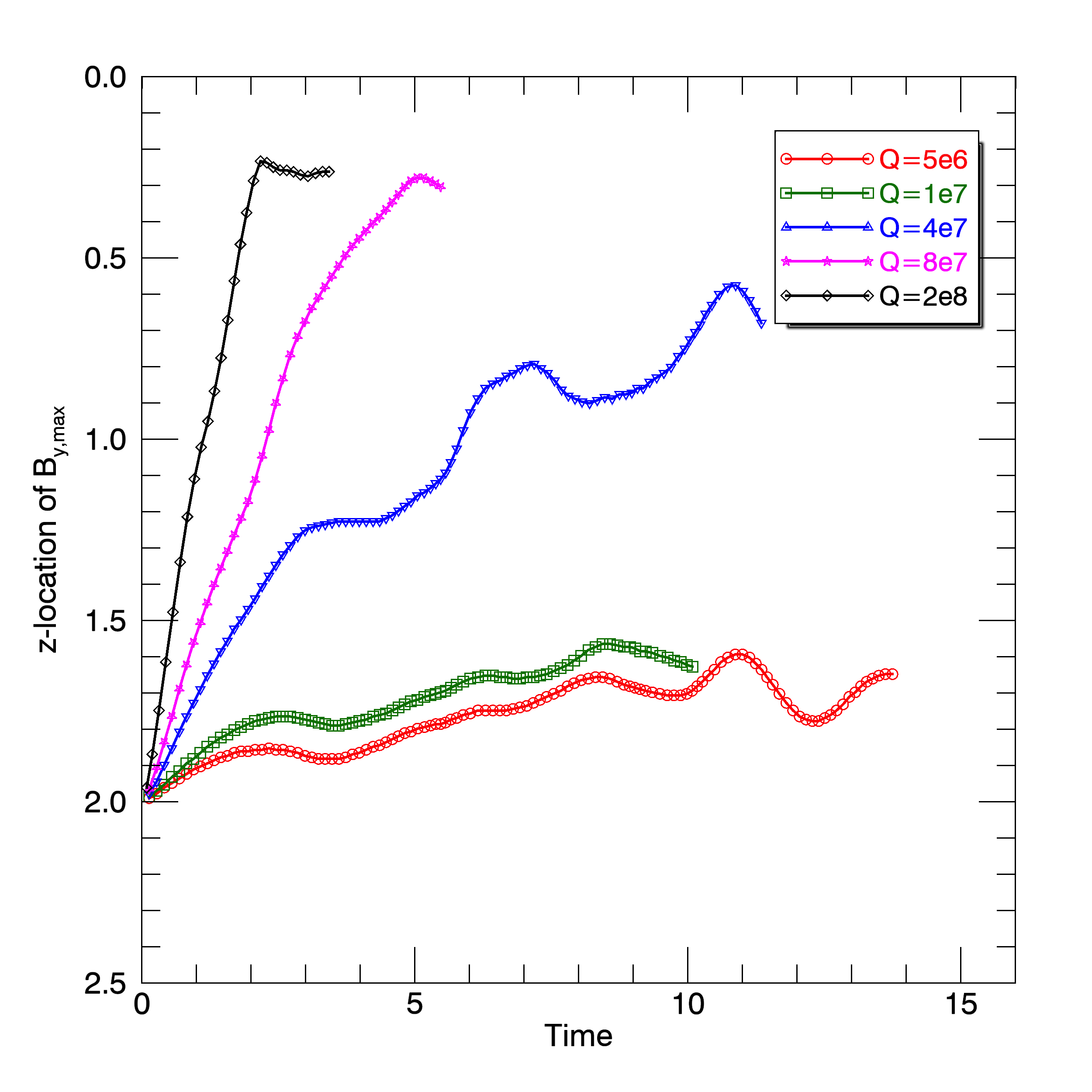}
    \caption{The vertical ($z$) location of the maximum of $B_{y}$, $z_{ft}$, plotted as a function of time for different $Q$. Other parameters are $Ra=4\times 10^{4},~Pr=0.1,~\zeta=0.001,~q=0.5,$ and $S=7$.}
    \label{fig:S7_TP05_NoBG}
\end{figure}

Figure \ref{fig:S7_TP05_NoBG} shows $z_{ft}(t)$ for a range of simulations that were initiated identically (as above) but for varying $Q$. It can be seen that a lower value Q (e.g. $Q=5 \times 10^6,1 \times 10^7$; red and green markers) leads to an almost neutrally buoyant flux tube that only rises very slowly initially and then ceases to rise, ultimately remaining embedded deep in the stable zone. A middling value of Q (e.g. $Q=4\times 10^{7}$; blue markers) leads to a moderately quick rise of the flux tube through the stable zone and into the convection zone, where it experiences some substantial variations due to interactions with upflows and downflows, but nevertheless rises. At higher $Q$ values (e.g. $Q=8 \times 10^7,2 \times 10^8$; magenta and black markers), the flux tube rises quickly and steadily through both the stable zone and the convection zone to the top of the domain.  The black markers in this plot correspond to the case with  $Q=2\times 10^{8}$ shown in Figure \ref{fig:tube_convection_s7} and the red markers correspond to the case at $Q=5\times10^6$ shown in Figure \ref{fig:tube_convection_s7_Q5e6}. Note that we  curtail the traces of $z_{ft}$ in such plots either when the trace  reaches the top of the simulation domain or when it becomes clear that a tube has stopped rising. 

From this information, we choose to use $Q=2\times 10^{8}$ as the value of $Q$ for our canonical set of parameters used in the next sections.  Our ultimate aim is to evaluate the effect of the sign of twist in conjunction with a large-scale background field on emergent tubes.  This case provides a clear and definite emergence of the flux tube through the convection zone in the absence of such effects and therefore is a solid basis for our canonical investigations.  It is, of course, interesting to examine cases at parameters other than this canonical set, and we will do this later in the paper.

\begin{figure}
    \centering
    \includegraphics[width=\textwidth,height=12cm]{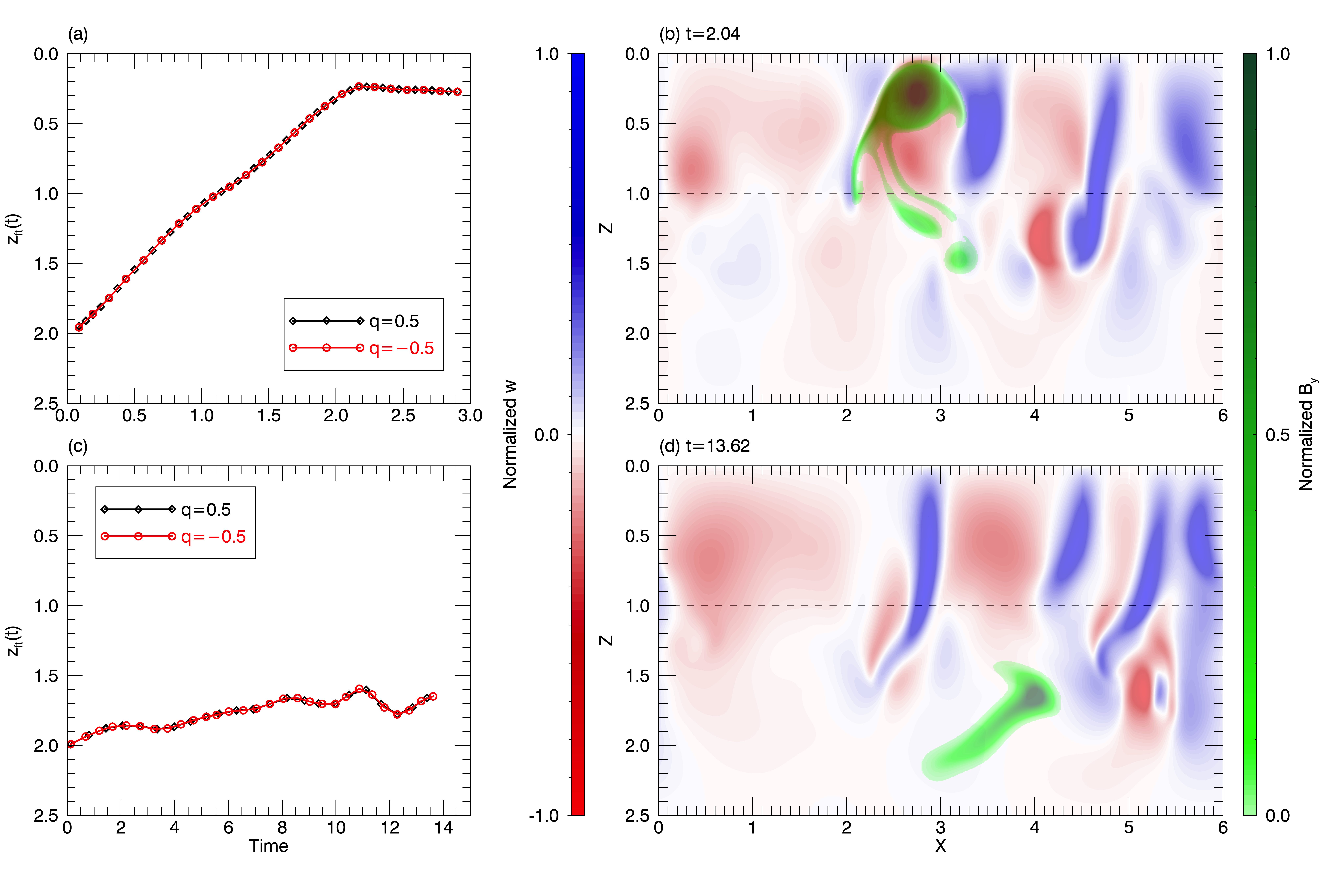}
    \caption{Comparison of the dynamics of positively- and negatively-twisted tubes for $Q=2\times 10^{8}$ (a and b) and $Q=5\times 10^{6}$ (c and d). (a) $z_{ft}$ plotted as a function of time for both $q=0.5$ and $q=-0.5$ at $Q=2\times 10^{8}$. (b) The equivalent of Figure \ref{fig:tube_convection_s7}d but for $q=-0.5$. (c) $z_{ft}$ plotted as a function of time for both $q=0.5$ and $q=-0.5$ at $Q=5\times 10^{6}$. (d) The equivalent of Figure \ref{fig:tube_convection_s7_Q5e6}d but for $q=-0.5$.}
    \label{fig:NegativeTwist_PureConvection_Zft_IntensityPlot}
\end{figure}

At this point in the discussion, it is important to realize that, at this stage of the simulation hierarchy, the sign of the twist of the flux tube does not make a significant difference to the dynamics.  The evolution and fate of a positively-twisted flux tube and a negatively-twisted flux tube inserted into convection without any background field is essentially the same.
This is confirmed by Figure \ref{fig:NegativeTwist_PureConvection_Zft_IntensityPlot} which shows results for  $Q=2\times 10^{8}$ in panels a and b and for $Q=5\times 10^{6}$ in panels c and d.  Each of the left panels (a and c) show time traces of $z_ft$ for {\it two tubes of opposite signs}.  The right panels (c and d) show snapshots at a late stage of the respective  simulations {\it for a negatively-twisted tube}, which can be compared directly with those shown for the positively-twisted tubes in Figures \ref{fig:tube_convection_s7}d and \ref{fig:tube_convection_s7_Q5e6}d respectively.  Clearly, the dynamics of positively- and negatively-twisted tubes are extremely similar.  Without a volume-filling background field, the twist serves mainly to maintain the coherence of the tube.  We will show that this symmetry between the dynamics of the differently-signed tubes is broken by the inclusion of a background field in later sections of this paper.

\subsection{Dynamical formation of Background Magnetic Field} \label{sec:Pumping_conv}

Papers 1 and 2 explored the effect of a background magnetic field on the rise of flux tubes by artificially imposing the background field as a function that decreased exponentially with height. These papers argued that the turbulent convective pumping of large-scale magnetic fields \citep[see e.g][]{Nordlund:etal:1992,Brandenburg:etal:1996, Tobias:etal:1998,Tobias:etal:2001} would likely achieve some profile of the background field where the majority of the field was confined below the convection zone.  The exponential profile used in Papers 1 and 2 was a simple model of part of this expected pumped profile.
In this section of this paper, as part of the setup of our initial state for the ultimate simulations, we explore the self-consistent formation of a large-scale background magnetic layer by magnetic pumping due to the presence of the overshooting turbulent convection, instead of assuming its existence, as was the case in the previous work of Papers 1 and 2.  Again, detailed simulations of this type have been performed before \citep[see e.g][]{Tobias:etal:2001} and transport of magnetic field out of a convection zone into a stable layer down a gradient of turbulent intensity is fairly well understood.  Our aim here is not to study this phenomenon comprehensively, but rather to create self-consistent initial conditions for the large-scale background field for later cases, to replace the artificial ones used previously in Papers 1 and 2.
Since the selection mechanism discovered in Papers 1 and 2 depends on the relative strengths of the tube and the background field, one of our main goals here is to be able to control the amplitude of the profile of the pumped field to explore this relationship.

In this section, we start with a steady-state pure convection solution from \S \ref{sec:conv_2D} and impose a thin horizontal magnetic layer concentrated in the convection zone. Note that in this section of work there is no flux tube imposed.  The magnetic layer is given by
\begin{equation} 
    \mathbf{B} = 
    (B_x,B_y,B_z) =
    \mathbf{B_{\rm layer}} =\Bigg( \text{tanh} \Big(\frac{z-z_{bot}}{\delta} \Big)\text{tanh} \Big(\frac{z_{top}-z}{\delta} \Big) , 0 ,0 \Bigg) 
    \label{eq:magneticlayer}
\end{equation}
so that $z_{top}$ and $z_{bot}$ are the top and bottom locations of the layer respectively, and $\delta$ is the width of a narrow smoothing region at the edges of the layer. For a standard initial condition, we use a layer where $z_{bot}=0.80, z_{top}=0.75$ and $\delta=0.01$. The other parameters remain at the canonical values ($Ra=4\times 10^{4}$, $Pr=0.1$, and $\zeta=0.001$) and we can again choose $Q$ to determine the initial strength of the magnetic layer.  This, in turn, affects the initial magnetic buoyancy of the layer, since we again adjust the background thermodynamic state to maintain total pressure and temperature equilibrium, as in the previous section.  We have experimented with omitting the thermodynamic adjustment to avoid certain numerical issues, and found that this makes little difference to the overall dynamics, since total pressure balance is adjusted very quickly.

Choosing the value of $Q$ here is a little complicated for the following reasons.
In Section \ref{sec:fluxtube_conv}, we established that a flux tube with $Q=2\times 10^{8}$ rises through the convection and therefore we will need to run our ultimate simulations (involving both flux tubes and background field) at this $Q$ value.  The work of Paper 2  showed that the key factor in the selection mechanism was the relative values (and orientation) of the azimuthal field in the tube and the background field strength.  The former is initially dictated by $q$ and can be different from the axial field strength of unity.  However, our choice of $|q|=0.5$ makes the peak azimuthal field also conveniently equal to unity, and therefore, the strength of both components of the tube field are explicitly governed solely by the value of $Q$.
However, we wish to be able to control the relative strengths of the azimuthal field in the tube and the background field in order to explore the regimes of Paper 2.  From that previous work, we expect the background field strength to be roughly between 5\% -- 20\% of the tube strength for the selection mechanism of Papers 1 and 2 to manifest.  We therefore need to be able to change either the resultant azimuthal field strength or the resultant background field strength (at the fixed input $Q$ of the whole simulation) to be able to realize this ratio.  We could achieve this by varying $|q|$, but instead we choose to control the resultant strength of the evolved background field (in order to avoid having to repeat the simulations of the previous section many times). This is not particularly artificial, since any pumping calculation with the magnetic boundary conditions specified (Equation \ref{eqn_boundaryconditions}) runs down (i.e. loses flux), so that the magnetic field amplitude decreases from unity to some significantly lower value as the simulation progresses.  Changing the initial layer amplitude allows any chosen final amplitude to be realized.  In essence, there is an ``effective'' $Q$ of this run-down state that we can control, given by the product of the square of the field amplitude and  the original $Q$: $Q_{\rm eff} = Q|{\bf B}|^2$.

With this in mind, we purposely start our pumping simulations here with $Q=4\times 10^{6}$, which is lower than the target $Q$ of the ultimate simulations ($Q=2\times 10^{8}$).  This allows a two-fold flexibility.  Firstly, at large $Q$, strong field can quickly accumulate at the upper boundary causing numerical problems; lower initial $Q$ can avert this unphysical issue.  Secondly, after evolving the system at lower $Q$, we can evaluate the peak amplitude of the final horizontally-averaged profile, and then scale this state to whatever peak amplitude we desire, essentially adjusting $Q_{\rm eff}$.  We can then take this as an initial condition, and run at the higher $Q$ (the canonical $Q$ that we must use when a tube is eventually present in the next section, $Q=2\times 10^{8}$) until it relaxes into the pumped state for that value.  This process is equivalent to running at different amplitudes of the initial layer at fixed $Q$ (and therefore different $Q_{\rm eff}$) but allows us to have faster access with less numerical issues to a pumped state with a controllable peak amplitude of the profile.  Essentially, we can now realize any ratio of $Q_{eff}:Q$ and therefore any relative strength of the background field compared to that of the tube in the ultimate simulations to follow in the next section.

\begin{figure}
    \centering
    \includegraphics[width=\columnwidth,height=21 CM]{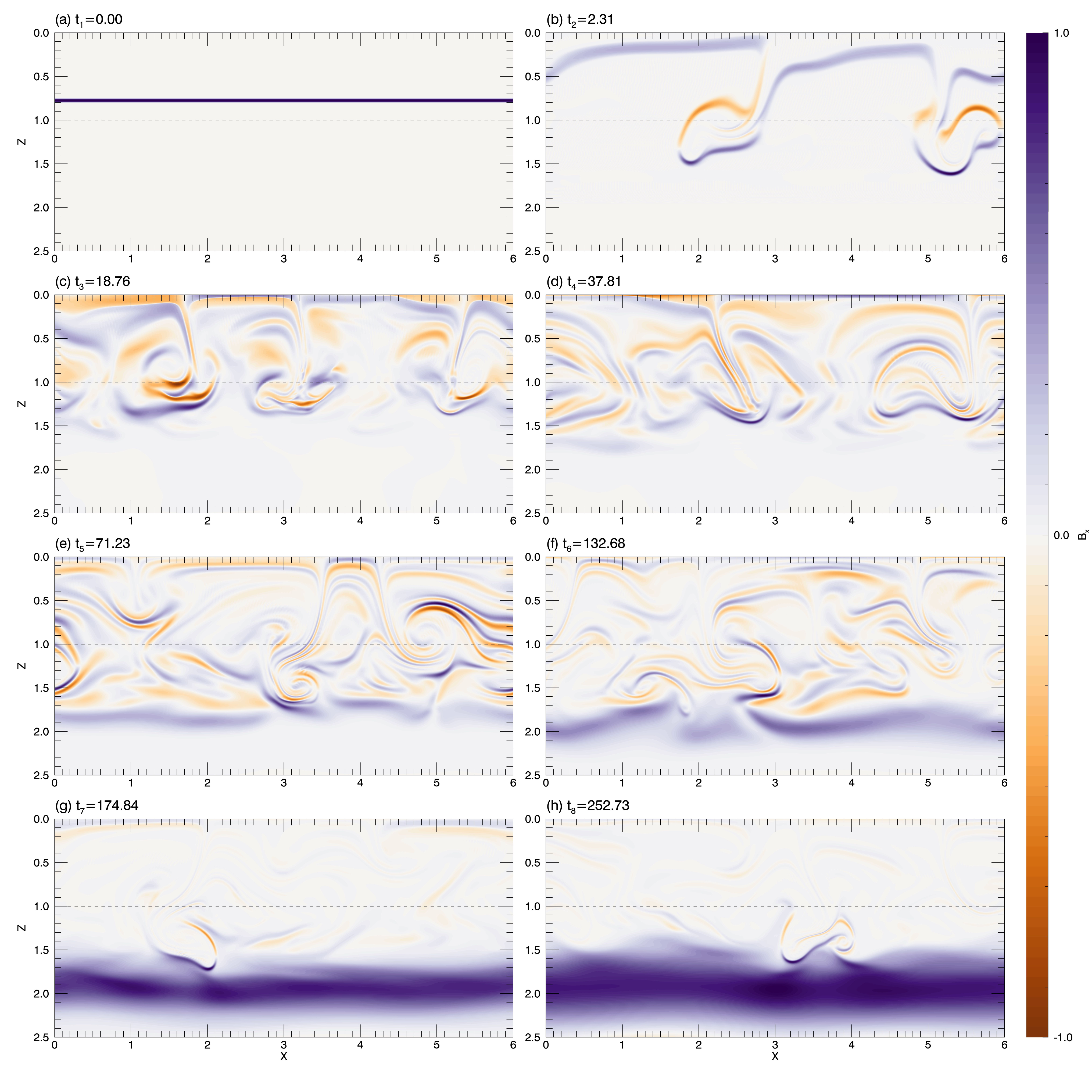}
    \caption{Snapshots of normalized horizontal magnetic field, $B_{x}$, as a function of time. (a)-(f) have $Q=4\times 10^{6}$ whereas (g)-(h) have $Q=2\times 10^{8}$.}
    \label{fig:pumped_bx_S7}
\end{figure}

\begin{figure}
    \centering
    \includegraphics[width=\columnwidth, height=7cm]{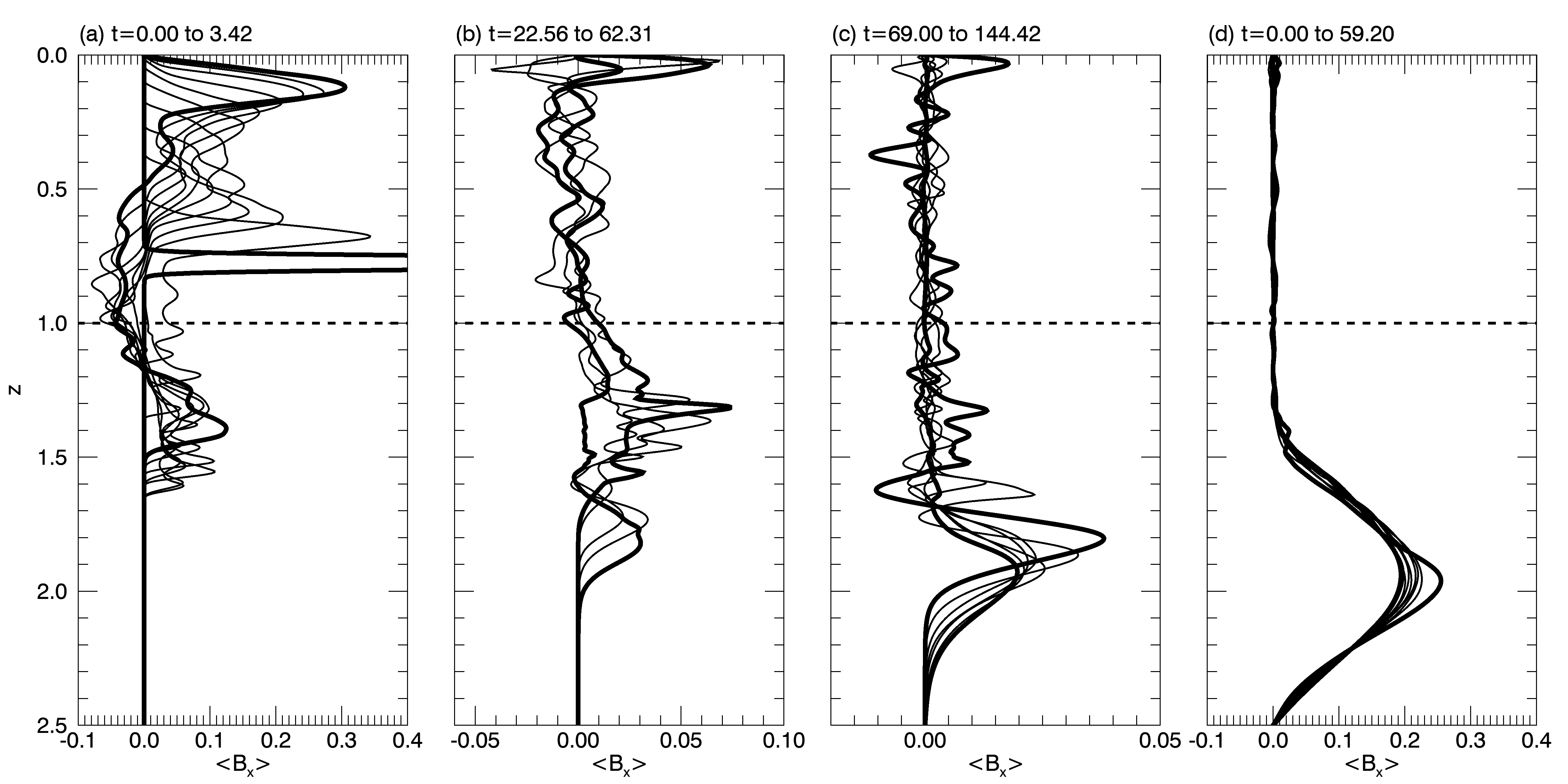}
    \caption{Time- and horizontally-averaged pumped horizontal field as a function of height (z) for the case with $S=7$. Initial and final profiles are plotted with a thicker line in each of the subplots. (a) shows the initial pumping of the imposed magnetic layer. The final profile of the pumped field is multiplied by a factor of 20 and its evolution is shown in (d).}
    \label{fig:Pumped_S7_AllCombined}
\end{figure}

We illustrate this whole process with Figures  \ref{fig:pumped_bx_S7} and \ref{fig:Pumped_S7_AllCombined}.   Figure  \ref{fig:pumped_bx_S7} gives a good impression of the magnetic pumping process by showing snapshots of the horizontal magnetic field $B_x$ (normalized by its maximum in any frame) as a function of time in the simulation at the canonical parameters. The initial magnetic layer can clearly be seen in Figure \ref{fig:pumped_bx_S7}a.  Panel b exhibits the early evolution of the layer, showing that the dynamics are dominated by the competing effects of magnetic buoyancy and downward transport of flux caused by advection in the convective plumes. Some sections of the layer rise by the combined effects of the magnetic buoyancy of the layer and upflow advection, whereas some sections get dragged towards and into the stable layer in strong downflows where advection overcomes the buoyancy. 
After this initial adjustment, the advective churning of the field takes over, as can be seen in panels c-d, and the small-scale turbulent interactions with the magnetic field dominate, leading to the turbulent transport mechanism of magnetic pumping.  The overshooting convective motions therefore transport magnetic flux to the stable region, as seen in Figure \ref{fig:pumped_bx_S7}e-f. The pumped fields slowly accumulate at the edge of the overshoot region, where their residual magnetic buoyancy is balanced by the pumping transport. 

In panels g-h of Figure \ref{fig:pumped_bx_S7}, we show an example of the adjustment of the peak amplitude of  pumped field. 
We evaluate the peak strength of the current pumped layer by calculating a time average of the horizontal field. Due to the influence of magnetic diffusion and vertical boundary conditions, the peak strength of the pumped layer has diminished considerably from the initial value of unity. We re-scale the magnetic fields in the whole simulation domain by a chosen scaling factor. If required, we can also adjust the $Q$ of the simulation at this point to the canonical value of $Q=2\times 10^{8}$. We then further evolve the new system over many convective turnover times, treating the scaled magnetic field as a new initial condition. This allows the system to relax to a new dynamically-consistent pumped magnetic layer at the desired $Q$, as shown in Figure \ref{fig:pumped_bx_S7}g-h.

It should be noted that even though the magnetic field in Figure \ref{fig:pumped_bx_S7}h (for example) looks fairly smooth (thanks to the considerably higher value now of the peak in the pumped layer), there are still  considerable local variations if we check the vertical profiles of the layer at different horizontal locations. In Papers 1 and 2, the artificially-imposed background field was uniform in the horizontal direction, whereas here, the self-consistently generated pumped field state is not.  In order to make contact between the two different works, we here calculate (and use for comparison) the horizontal-average of the horizontal magnetic field $B_x$. In Figure \ref{fig:Pumped_S7_AllCombined} we show the time evolution of the horizontally-averaged field at various stages in the simulation shown in the previous figure.  For each time portion (shown as a separate panel), the initial and final state is shown as a thicker line, and the thinner lines show regular time intervals between these end points.
Panel a again shows the initial dynamics where the average magnetic field profile evolves away from the initial layer located at $0.75 \leq z \leq 0.80$.  Initially, it can be seen that the profile relaxes in shape and amplitude somewhat, and moves upwards due to magnetic buoyancy and starts to accumulate at the top.  At the same time, some of the average field gets moved towards the lower stable layer, creating a second accumulation peaked at around $z=1.4$. In panel b, later in time, we see that on average, the field starts to be evacuated from the upper boundary and the interior of the convection zone.  The peak around $z=1.4$ becomes (relatively) more prominent, but also a second peak around $z=1.8$ starts to form, just below the overshoot region.  This latter peak is the progenitor of the pumped layer, as can be seen in panel c, where, as time progresses, this peak becomes the dominant dynamical feature in the average field.  

Panel d shows how we manipulate the amplitude of the pumped field to achieve a desired $Q_{eff}$ at a certain $Q$.  Here, we have taken the end result of the earlier calculations, and rescaled the pumped field by a chosen factor (a factor of 20 in this case) and continued the evolution.  It can be seen that this achieves a stable profile of pumped field on average, with a peak amplitude of about $0.2$.  This type of state is the desired initial condition for the calculations of the next section where a flux tube with peak field amplitude of unity is also added at the same $Q$.  Such a profile provides the desired ratio of amplitudes, where the background field is at about peak amplitude of 20\% of the tube peak amplitude.  It should be noted that pumped profile can still evolve further and run down due to the magnetic boundary conditions.  However, this evolution is in general very slow compared to other dynamics that we are interested in (such as the rise time of a flux tube) and therefore can be considered essentially steady.  As can be seen in panel d, the evolution of the profile is particularly slow at this point because it has evolved to a state where there are no gradients of the (average) field at the boundary and therefore no flux loss from the domain (especially at higher $S$ than is shown in this case: see later) and the profile becomes almost  piecewise linear, thereby engendering very little diffusion (except at the junctions of the piecewise profile).

\begin{figure}
    \centering
    \includegraphics[width=10cm, height=12cm]{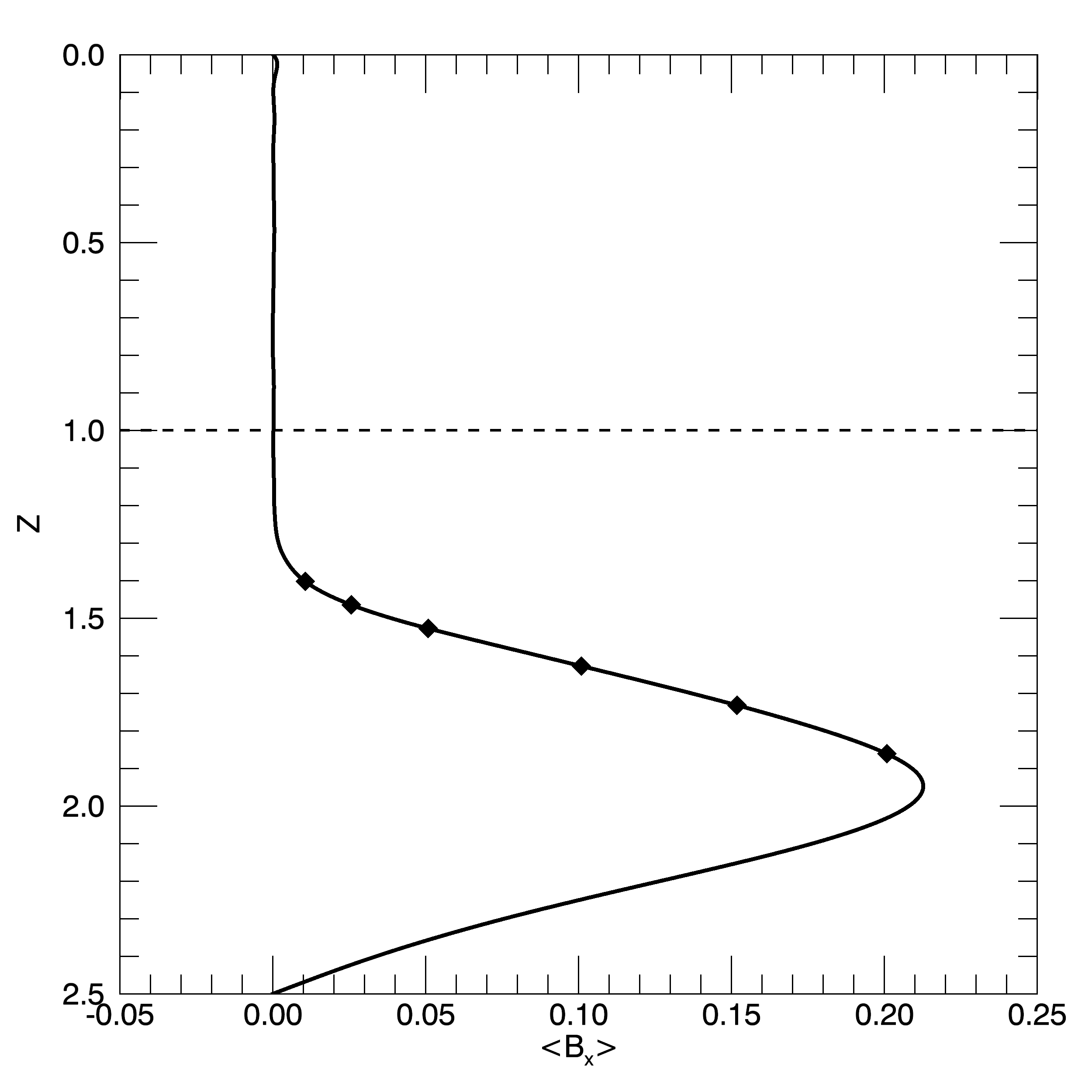}
    \caption{Time averaged pumped horizontal field as a function of height (z) for the case with $S=7$. Black diamond markers indicate the different z-coordinates centred at which flux tubes are introduced. These z-coordinates correspond to $1,~2.5,~5,~10,~15$ and $20\%$ strength of the pumped horizontal field as compared to the flux tube strength. The dashed line marks the transition between the convection zone and the radiative zone.}
    \label{fig:AveragePumpedField_S7_FTMarked}
\end{figure}

Figure \ref{fig:AveragePumpedField_S7_FTMarked} shows the final time- and horizontally-averaged state (averaged over many turnover times) of the pumped field as a function of height at the canonical parameters with $Q=2\times 10^{8}$.  This is the state that we will use for the next section. The peak amplitude of this pumped layer is about $0.21$, or $21\%$, compared to the unit strength flux tube. The filled-diamond markers are placed to indicate the locations where the average field strength is $1,~2.5,~5,~10,~15,$ and $20\%$ on the upslope section of the pumped layer, locations that we will use for initial conditions shortly.

\subsection{Rise of Flux tubes in the presence of a Background Field} \label{sec:fluxtube_backgroundfield}

We are now equipped to address the main point of this paper.  This section studies the dynamics of the rise of magnetic flux tubes in the presence of convection and a background magnetic field, and, in particular, examines bias introduced by the alignment of the twist of the tube with the background field. 

As initial conditions, we take the end point of the solutions arrived at in Section \ref{sec:Pumping_conv}, consisting of  statistically steady-state convection with a self-consistent magnetic field in the form of a pumped layer in the stable zone. To this state, we introduce a flux tube in the same manner as we did in \S \ref{sec:fluxtube_conv} where no background field was present. The total magnetic field is thus given by

\begin{equation} 
    \mathbf{B} = 
    (B_x,B_y,B_z) =
    \mathbf{B}_{tube}+
    \mathbf{B}_{pumped}
    \label{eq:fluxtube_pumpedfield}
\end{equation}
where $\mathbf{B}_{tube}$ is the field of a tube as in equation (\ref{eq:fluxtube}) and $\mathbf{B}_{pumped}$ is the field resultant from the end of the simulations in the previous section, \S \ref{sec:Pumping_conv}. 
In this section, we consider flux tubes with both positive (anticlockwise) and negative (clockwise) twist, $q$, but still with a fixed magnitude $|q|=0.5$. The other parameters for now remain as the canonical set: $Ra=4\times 10^{4}$, $Pr=0.1$, $\zeta=0.001, S=7$ and $Q=2\times 10^{8}$. 

There are a few important points to note at this stage. Firstly, introducing the field of a flux tube, $\mathbf{B}_{ tube}$, again requires a thermodynamic adjustment in the background stratification for consistency, as was the case when there was no large-scale background magnetic field already present.  As before, we choose to adjust the density whilst maintaining the current temperature in the tube, and this again imparts an initial magnetic buoyancy to the tube. The pumped layer, $\mathbf{B}_{ pumped}$, is already dynamically consistent with the remaining thermodynamics and no further adjustment is necessary for this component.

Secondly, as determined in detail in Paper 2, the relative strengths of the twist of the tube and the background field are the key determining factors in the differential evolution of differently twisted flux tubes.  Here, we have fixed the strength of the tube (via its unit maximum axial field strength and fixed twist value, $|q|$) and are left with two controls over the relative strength of the background field.  One control is that we can adjust the peak amplitude of the pumped layer as described in the previous section.  The other control is that, even for a given peak amplitude in the initial condition for $\mathbf{B}_{pumped}$, we can adjust the depth at which the flux tube is introduced, thereby determining both the  background field value that the tube experiences initially and the amount of pumped background field the tube experiences during transit upwards. 
Our strategy here initially is therefore to select a case from our work of the previous section where the peak (time- and horizontally-averaged) amplitude of  $\mathbf{B}_{pumped}$ lies in realm where Paper 2 predicts that a selection mechanism may occur.  We then select particular depths at which the flux tube is introduced into the pumped magnetic layer in order to vary the precise ratio of the twist field in the tube to the averaged pumped field strength; these depths are the diamond markers in Fig. \ref{fig:AveragePumpedField_S7_FTMarked}.  We investigate the details of varying the peak amplitude and related local effects later.

We initially choose all the locations for the introduction of the flux tube to be on the upper side of the pumped layer. We do this because we assume that flux tubes are likely to be formed by magnetic buoyancy instabilities, which require that the magnetic field increases sufficiently strongly downwards, therefore the most likely initiation of tubes is on the upper side of the pumped layer.  We note that there is an intermediary process (velocity shear) that connects the horizontal $B_x$ field to the horizontal $B_y$ field that is the source of the tubes.  This process might alter the expectation of where tubes are produced.  Our current work does not model this process and so there is an element of uncertainty in this matter.  However, we also argue that, while flux tubes may perhaps form below the peak of the pumped magnetic layer, the eventual rise of such tubes is more likely to fail completely due to the greater extent of strong magnetic field that it has to rise through, and that failed cases are less interesting to us initially.
We adopt the notation that, for example, $B_{s}=0.10$ corresponds to an averaged pumped background field strength at the depth of the center of the tube that is $10\%$ of the unit peak strength of the flux tube (in both axial and, more importantly, azimuthal field).  Locating the flux tube at the vertical position corresponding to this strength in the average pumped field is now one of our key control parameters.

\begin{figure}
    \centering
    \includegraphics[width=\columnwidth,height=21cm]{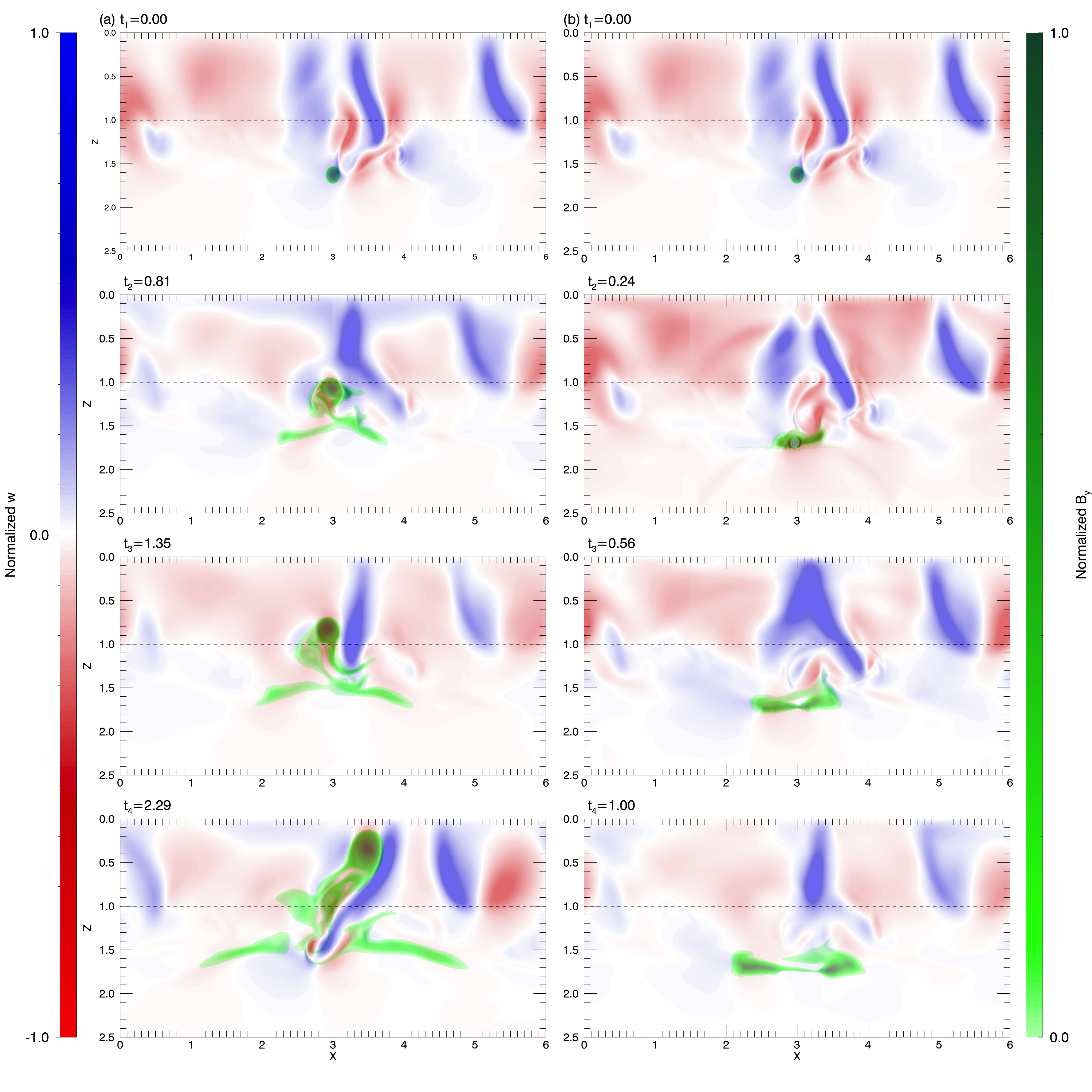}
    \caption{Intensity plots of normalized $B_{y}$ (shown in green tones)  overplotted with normalized vertical velocity, $w$ (shown in red and blue tones for upflows and downflows, respectively), for (a) $q=0.5$ and (b) $q=-0.5$ at four different times (as shown) in the simulation evolution. The (tine- and horizontally-) averaged pumped background field strength, $B_{s}$, at the center of flux tube initially in both cases is $0.10$, i.e., $10\%$ of the initial strength of the flux tube.}
    \label{fig:S7_TP05_TN05_X3_By}
\end{figure}

We now exhibit an example of the main result of this paper. We initialise two simulations as described above, but one includes a positively-twisted ($q=+0.5$) tube whereas the other has a negatively-twisted ($q=-0.5$) tube;  all other aspects are  identical.  Figure \ref{fig:S7_TP05_TN05_X3_By} shows intensity plots of normalized axial field, $B_{y}$ (green color scale), overplotted with normalized vertical velocity, $w$ (red-blue color scale), for both a positively-twisted (panel a, left column) and negatively-twisted (panel b, right column) flux tube for four different times in the evolution of the system. Here the red and the blue color represents upwards and downwards vertical velocity respectively. For this case, the flux tube in both cases is introduced at a depth such that $B_{s}=0.10$ as explained above. Figures \ref{fig:S7_TP05_TN05_X3_By}a and b therefore look identical, with the flux tube embedded between two convective downflow plumes. However, the dynamics subsequently evolve substantially differently in the two cases. 

The flux tube with a positive twist rises initially due to its imposed magnetic buoyancy, as in the case with no background field. However, the rise this time is not entirely coherent as the flux tube is rising through the pumped horizontal background field. Instead, as the tube rise, it experiences some axial flux loss along the background field $B_x$ accompanied by substantial disruption of the vortices that form in the isolated tube case.  This can be seen qualitatively but clearly in Figure \ref{fig:S7_TP05_TN05_X3_By}a at $t_{2}=0.81$. These effects, due to the presence of the background field, were also seen and described in Papers 1 and 2.  Furthermore, the flux tube in this case is located between two merging downflows at these early stages in the plots.  However, even with all these relatively unfavorable circumstances for its rise, the flux tube successfully enters the convectively unstable layer (Fig. \ref{fig:S7_TP05_TN05_X3_By}a at $t_{3}=1.35$) and eventually rises to the top in a weak upflow next to the strong merged downflow (Fig. \ref{fig:S7_TP05_TN05_X3_By}a at $t_{4}=2.29$). Note that the effect of convective motions on the rise is immediately evident as the flux tube does not rise symmetrically (about the line $x=3$) and so these simulations are clearly distinct from those of Papers 1 and 2.  

In stark contrast, the dynamics of the flux tube with a negative twist are entirely different. Figure \ref{fig:S7_TP05_TN05_X3_By}b at $t_{2}=0.24$ shows that, in the presence of the background field, the flux tube does not even begin to rise through the overshooting  convective background. The flux tube structure simply gets distorted quickly and  breaks apart, with axial flux rapidly drained along the $B_x$ fieldlines (see Fig. \ref{fig:S7_TP05_TN05_X3_By}b at $t_{3}$ and $t_{4}$).

At this point, we have already arrived at an important conclusion that parallels those of Papers 1 and 2; the flux tube dynamics in the presence of a background field are very different than cases where there is no background field present.  With no background field, tubes of both twists rise very similarly, with the dynamics dominated by their identical magnetic buoyancy (see \S \ref{sec:fluxtube_conv}), whereas, here, in the presence of background fields, positively- and negatively-twisted tubes evolve very differently. In these new simulations, these conclusions are reached despite the presence of convection and a more self-consistent background field.  
Put in a slightly different way, we have shown that a positively-twisted flux tube is more likely to rise than a negatively-twisted tube even in the presence of convection and a self-consistently pumped background field (at these parameters). 

\begin{figure}
    \centering
    \includegraphics[width=9cm, height=10cm]{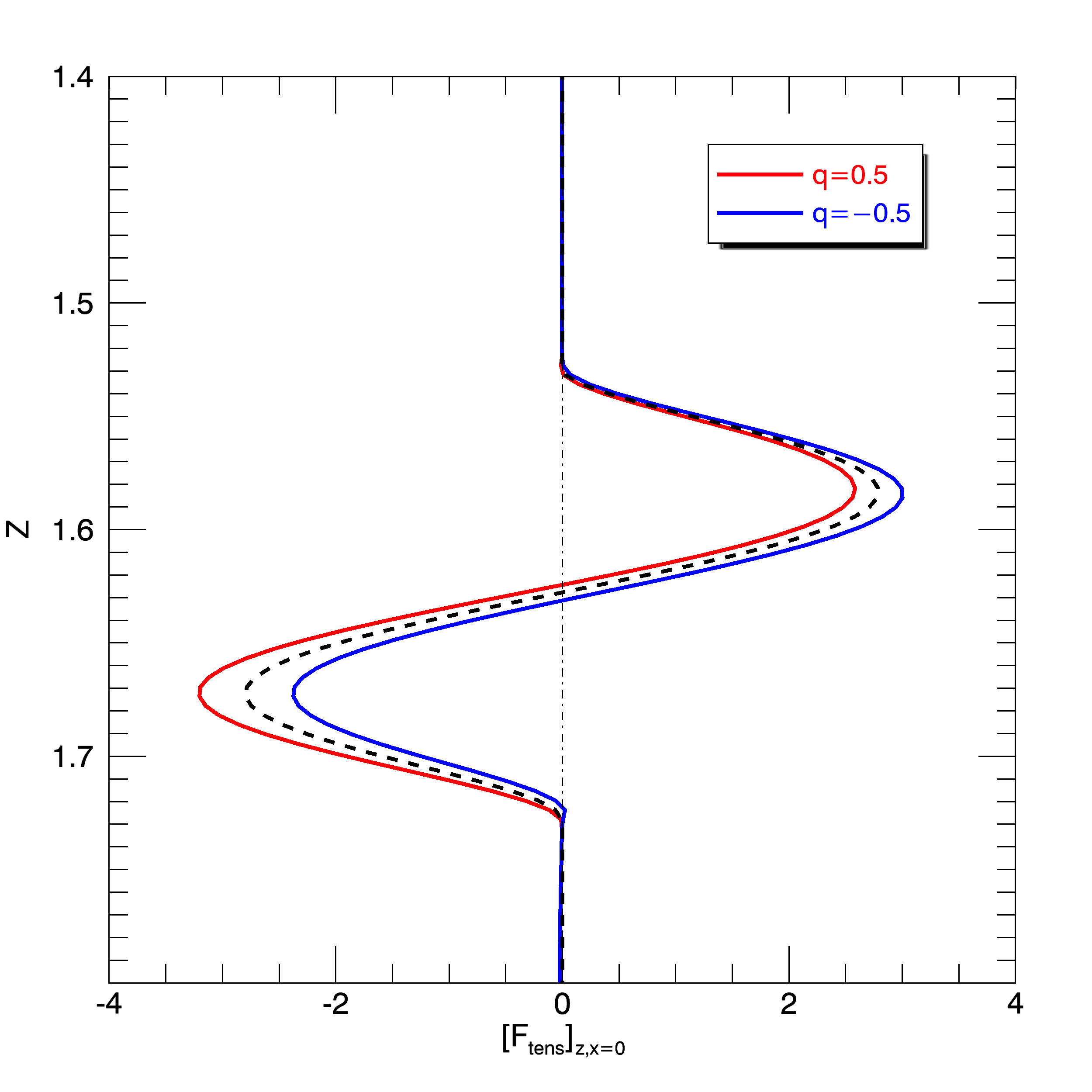}
    \caption{Plots from the canonical case at $S=7$ (from \S  \ref{sec:fluxtube_backgroundfield}) showing profiles of the tension force acting in the $z$-direction, $F_{tens,z}=(B_{x} \partial_{x} + B_{z} \partial_{z})B_{z}$, evaluated at the middle of the tube, $x=x_c=3.0$, in the initial conditions, $t=0$, for both positively-twisted ($q=0.5$) and negatively-twisted ($q=-0.5$) tubes. The dashed line shows the equivalent  profile for the tube from the simulation in \S \ref{sec:fluxtube_conv} where no pumped background field was present. 
    \label{fig:S7_TP05_TN05_VerticalTension_LineCut}}
\end{figure}

Since the magnetic buoyancy forces exerted by both signs of twist are nearly identical, we attribute these effects to magnetic tension effects, as described in Papers 1 and 2, even though this system is more complex.  We verify this conclusion quickly here via Figure \ref{fig:S7_TP05_TN05_VerticalTension_LineCut} and investigate this in more detail later.  Figure \ref{fig:S7_TP05_TN05_VerticalTension_LineCut} shows the initial ($t=0$) vertical profile of the vertical tension force $F_{tens,z}=(B_{x} \partial_{x} + B_{z} \partial_{z})B_{z}$ through the center of the tube ($x=3$) for each of three cases: the control case with no background field and the two cases with background field and opposite signs of the twist, $q = \pm 0.5$.   These results correspond to the state in the first panel of Figure \ref{fig:tube_convection_s7} and the top panels of columns a and b in Figure \ref{fig:S7_TP05_TN05_X3_By}.  These  profiles confirm that, compared to the case where no background field is present (dashed line), the positively-twisted tube has enhanced negative (upward) tension in the lower half of the tube and decreased positive (downward) tension in the upper half, leading to a net positive (upward) tension force, whereas the negatively-twisted tube has the opposite bias (enhanced positive tension in the upper half of the structure, decreased negative tension in the lower half, leading to a net downward force).  For the positively-twisted tube then, the net tension force acts in concert with the buoyancy force, helping the rise of the tube, whereas for the negatively-twisted tube, tension acts against buoyancy, decreasing the chances of rise.  Note that this is the case here despite the fact that the local background pumped field that the tube experiences may have significant variations from the mean profile at this particular location. This issue is examined in more detail shortly (in \S \ref{sec:fluxtube_backgroundfield_MC}).

\subsection{Dependence on $B_s$ or vertical location}
\label{sec:fluxtube_backgroundfield_Bs}

Having given an initial example of the biased behavior resulting from the interaction of twisted tubes and background field, we now explore the dependence of this selection mechanism on the relative strength of the background field by varying the initial vertical location of the flux tube and examining the subsequent rise dynamics. Such a dependence was found in the simpler context of Papers 1 and 2.  We explore a broad range of $B_{s}$ in the canonical case, ranging from relatively very weak, $B_{s}=0.01$ (relative to the unit amplitude of the tube fields), to fairly strong, $B_{s}=0.20$. 
The vertical locations corresponding to the full set of pumped background field strengths examined are marked in Figure \ref{fig:AveragePumpedField_S7_FTMarked}.
The parameters for all of these simulations remain the same at the canonical values: $Ra=4\times 10^{4},~Pr=0.1,~\zeta=0.001,~Q=2\times 10^{8},$ and $S=7$.

\begin{figure}
    \centering
    \includegraphics[width=\columnwidth,height=9cm]{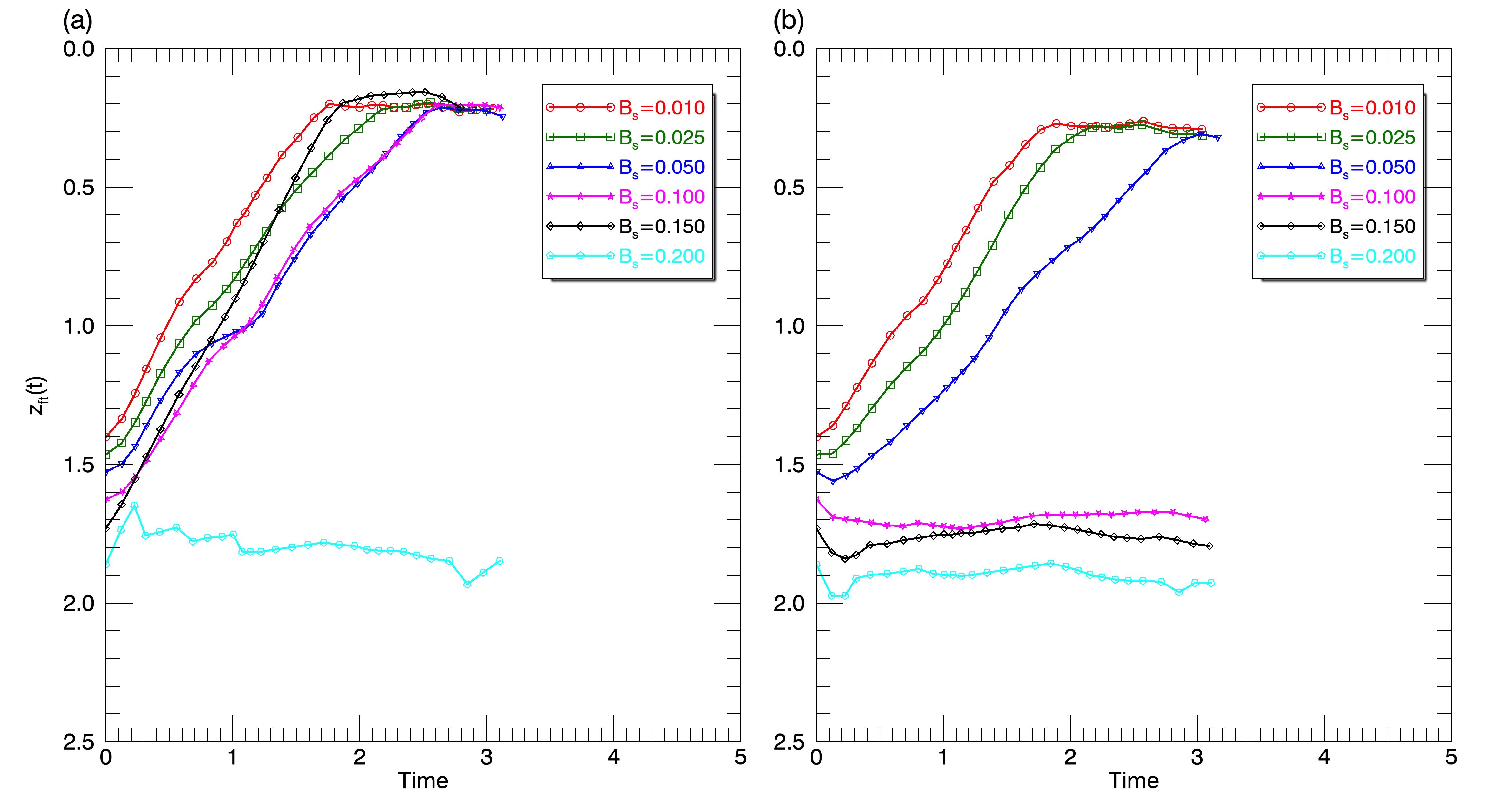}
    \caption{Plot of $z_{ft}$ as a function of time for (a) $q=0.5$ and (b) $q=-0.5$. Each subplot shows five cases performed at  different $B_{s}$:  $B_{s}=0.01,0.025,0.05,0.10,0.15,~0.20$.}
    \label{fig:TrackBy_S7_TP05_TN05}
\end{figure}

Rather than qualitatively analyzing intensity plots to determine whether the rise of a flux tube was  successful or unsuccessful, in Figure \ref{fig:TrackBy_S7_TP05_TN05} we return to plots of $z_{ft}$ that track the location of the peak field and therefore the location of the tube (at least while it remains coherent) as a function of time for the various cases. 
This figure shows $z_{ft}$ as a function of time for positively-twisted flux tubes in panel a and negatively-twisted flux tubes in panel b for each of the six different locations corresponding to the six different relative background field strengths at those locations, namely  $B_{s}=0.01,~0.025,~0.05,~0.10,~0.15,~0.20$, at the canonical parameters. It can immediately be confirmed that the initial $z_{ft}$ location, $z_{ft}(t=0)$, for each of these values is different, and corresponds to the appropriate point in Figure \ref{fig:AveragePumpedField_S7_FTMarked}.  

Figure \ref{fig:TrackBy_S7_TP05_TN05}a shows that positively-twisted flux tubes rise through the stable layer into the convectively-unstable layer and reach the top of the domain for all cases surveyed except $B_{s}=0.20$.  It appears that background field strengths of around $B_s>0.15$ are sufficient to prevent the rise of a positively-twisted tube.
On the other hand, Figure \ref{fig:TrackBy_S7_TP05_TN05}b shows that negatively-twisted flux tubes rise when embedded in  relatively weak background field strengths ($B_{s}=0.01,~0.025,0.05$) but do not rise for somewhat stronger background field strengths ($B_{s}=0.10,~0.15,~0.20$).  

These simple plots verify the conclusions of Papers 1 and 2 in the more complex simulations performed here.  That is, at low background field strengths ($B_s \leq 0.05$ for this $|q|$), tubes with both signs of twist can rise uninhibited.  At relatively high background field strengths ($B_s \geq 0.2$ for this $|q|$), the background field completely quenches the rise of tubes with either sign of twist.  Somewhere between these upper and lower limits, here (approximately) $0.05 < B_s < 0.2$, there exists a region of parameter space that we call the Selective Rise Regime (or SRR), where positively-twisted tubes rise, and negatively-twisted tubes do not (for a positively-oriented background field).

\subsection{Dependence on horizontal location and initial timestep}
\label{sec:fluxtube_backgroundfield_MC}

The previous two sections have established  the role of the {\it time- and horizontally-averaged} pumped background field in creating an asymmetry between the rise of differently twisted flux tubes  via examples from a canonical case.  It is essential to note that the pumped magnetic layer can have significant local temporal and spatial variations around this averaged state, due to the turbulent nature of the overshooting convection. These variations can  potentially affect the initial localized background state that the flux tube experiences quite substantially.  One could argue that the results that were presented earlier were perhaps not necessarily representative, since they may be a result of a particularly  favorable (or unfavorable) fluctuation in the initial conditions.  Indeed, as seen in the SHHR observations and as explored in Paper 2, we might actually expect substantial variations in the resultant dynamics, including violations to any ``rules" governing the behavior, even under the same general dynamical conditions.  
Therefore, to establish the previous single-example results of \S \ref{sec:fluxtube_backgroundfield} and \S \ref{sec:fluxtube_backgroundfield_Bs} in a more statistical sense, and to better understand the implications of initial conditions, we carry out a Monte Carlo-type (MC) study that examines 
two sources of local variations in the  initial background conditions that can potentially affect the rise dynamics of the flux tube.  
Firstly, while the averaged pumped field reaches a reasonably well-established statistically-steady profile, the overshooting convection is highly time-dependent, and the associated local thermodynamic and magnetic state are similarly so.  Therefore, the state at the particular time that we choose for the initial conditions in which we embed the flux tube may deviate substantially from the time average. 
Selecting the background state at different times from the magnetic layer pumping simulations (from Section \ref{sec:Pumping_conv})
can significantly change the  conditions the flux tube experiences initially. 
Secondly, even at a particular time and a particular vertical location, there may be substantial variations in the background dynamics in the horizontal.  Again, local conditions at the point of insertion of the tube can be significantly different from the horizontal mean, deeming the latter potentially unrepresentative for a particular simulation.  These two sources may be essentially equivalent in creating random variations, but we explore both here.  We note that the more systematic variation of the dynamics with the vertical location of the tube has already been discussed in the previous section. 

The simulations in \S \ref{sec:fluxtube_backgroundfield} and \S \ref{sec:fluxtube_backgroundfield_Bs} all started with the flux tube located at the middle of the horizontal domain ($x_{c}=3.0$) and used a particular timestep ($t=62.01$) from the simulation in \S \ref{sec:Pumping_conv} (shown as Figure \ref{fig:Pumped_S7_AllCombined}d). 
We now explore the consequences of varying these choices of time and horizontal location. At our canonical parameters, for each nominal background field strength $B_s$ (i.e. vertical location, gleaned from the time- and horizontally-averaged field), we consider three randomly-chosen different timesteps from the later stages of the simulation in \S \ref{sec:Pumping_conv}.  Using each of these three pumped field states as initial conditions, we run a series of simulations where we add single flux tubes placed at one of five different horizontal locations (chosen non-randomly as $x=1,2,3,4,5$). Every one of these simulations is run with both positively-twisted and negatively-twisted tubes.  Overall, this amounts to $15$ separate cases of flux tube dynamics for each sign of the twisted tube that (somewhat randomly) sample different background dynamics (all at the canonical parameters).

\begin{figure}
    \centering
    \includegraphics[width=16 cm,height=21 cm]{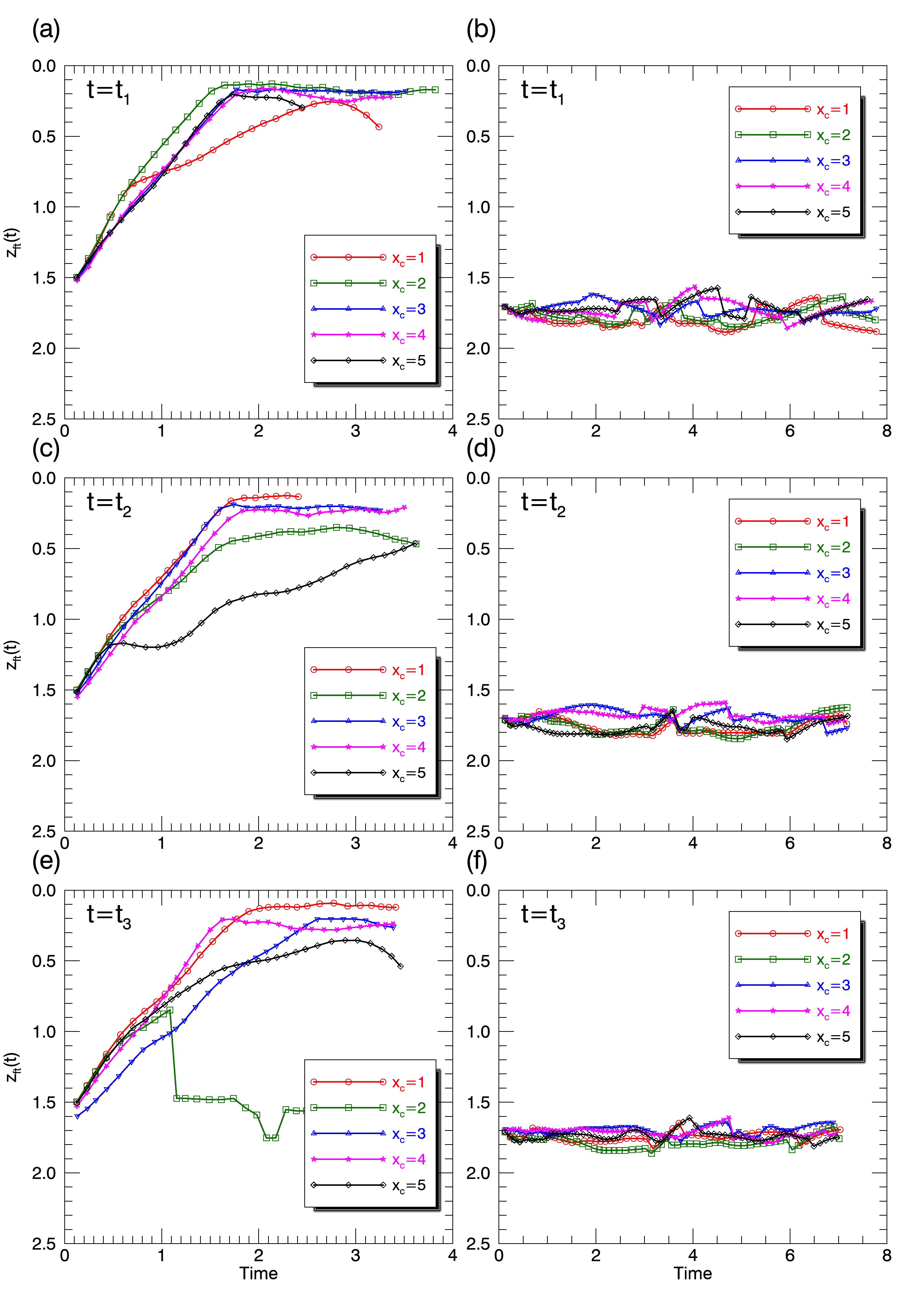}
    \caption{Plots of $z_{ft}$ as a function of time showing the successful or unsuccessful rise of the flux tube from a set of MC simulations incorporating a pumped background field of (average) strength, $B_{s}=0.10$. The first column (panels a, c and e) shows the cases with $q=0.5$ whereas the second column (panels b, d and f) has $q=-0.5$. Each of the three rows represent different initial conditions chosen from the results of Section \ref{sec:Pumping_conv} at times  $t_{1}=35.62$, $t_{2}=48.78$, and $t_{3}=62.01$. Each subplot exhibits the evolution for five different initial horizontal locations of the flux tube, $x_{c}$.}
    \label{fig:S7_BS10_TrackBy}
\end{figure}

Figure \ref{fig:S7_BS10_TrackBy} shows our measure of the location of the tube, $z_{ft}$, as a function of time, from the MC  simulations utilising the initial tube insertion vertical level nominally corresponding to $B_{s}=0.10$. Panels a, c, and e are for positively-twisted flux tubes at the three chosen times, 
$t_{1}=35.62$, $t_{2}=48.78$, and $t_{3}=62.01$, taken from the canonical pumping simulation exhibited in Figure \ref{fig:Pumped_S7_AllCombined}.
Panels b, d, and f exhibit simulations with exactly the same initial conditions except that a negatively-twisted flux tube is embedded initially instead of a postively-twisted tube. Each panel in Figure \ref{fig:S7_BS10_TrackBy} contains five traces of  $z_{ft}$ corresponding to the five different horizontal locations of the inserted flux tube, 
$x=1,2,3,4,5$.  

Some general conclusions can quickly be reached from this figure.  Studying any individual panel shows that there can be significant variations in the rise characteristics due to the horizontal positioning of the flux tube. For example, Figure \ref{fig:S7_BS10_TrackBy}a, 
using the $t=t_1$ timestep as an initial condition, 
shows that 
the flux tube embedded at $x_{c}=1.0$ takes a relatively long time to reach the top of the convective layer, whereas the flux tube at $x_{c}=2.0$ reaches the top much faster. Other horizontal locations lead to very similar rise times to each other, with this rise time intermediate between the fastest and slowest. 
Looking at the other timesteps in the left column, we see very similar rise behaviour (to both each other and to the majority of locations in timestep $t=t_1$) from most locations in Figure \ref{fig:S7_BS10_TrackBy}c,  although now the tube initiated from position $x_c=5$ rises significantly more slowly. 
The final initial timestep in the left column at $t=t_3$, shown in  \ref{fig:S7_BS10_TrackBy}e, again exhibits similar overall behaviour except that the flux tube initiated at $x_{c}=2.0$ initially rises but then is halted completely and the location of the maximum axial flux  jumps back to the overshoot layer.  If the full dynamics from this particular simulation are examined in detail, we find that the flux tube in this case encounters an especially strong downflow during its rise that completely destroys its integrity.
While the dynamics in individual cases may somewhat differ due to such fluctuations, the broader conclusion is that the vast majority of the cases initiated with a positively-twisted flux tube (left column of panels in the figure) lead to a successful rise of the flux tube. 
Similarly, from the right column of subplots in the figure, i.e. Figures \ref{fig:S7_BS10_TrackBy}b, d and f, we can quickly see that
the negatively-twisted flux tube cases even more consistently fail to rise; regardless of timestep or horizontal location of initiation, all of the negatively-twisted flux tubes remained in the overshoot layer.

Overall then, the MC set of simulations represented by Figure \ref{fig:S7_BS10_TrackBy} does indeed reinforce the idea discussed in Section \ref{sec:fluxtube_backgroundfield} that there are strongly  different dynamics depending on the sign of the twist of the flux tube.  One sign of the twist (positive in this case, with the orientation of the pumped background field fixed as the positive direction) is statistically far likelier to rise than the other.  There are variations in rise times and even violations to the rule due to the precise location of the origination of the flux tube in space and time and therefore the exact magnetic-convective conditions that the tube experiences, but statistically the result is clear.

\begin{figure}
    \centering
    \includegraphics[width=16 cm,height=22 cm]{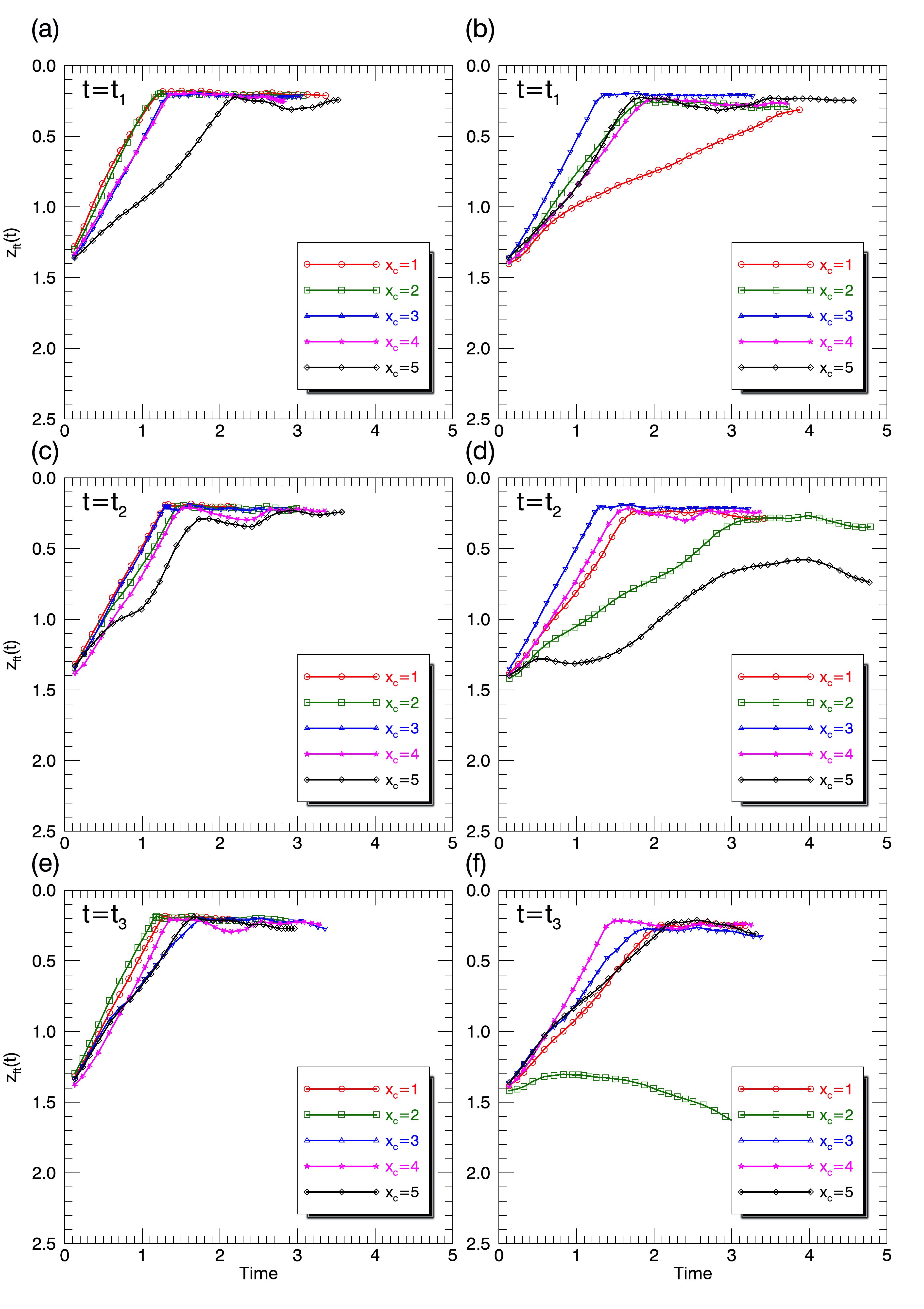}
    \caption{Same as Figure \ref{fig:S7_BS10_TrackBy} but for $B_{s}=0.01$.}
    \label{fig:S7_BS1_TrackBy}
\end{figure}

\begin{figure}
    \centering
    \includegraphics[width=16 cm,height=22 cm]{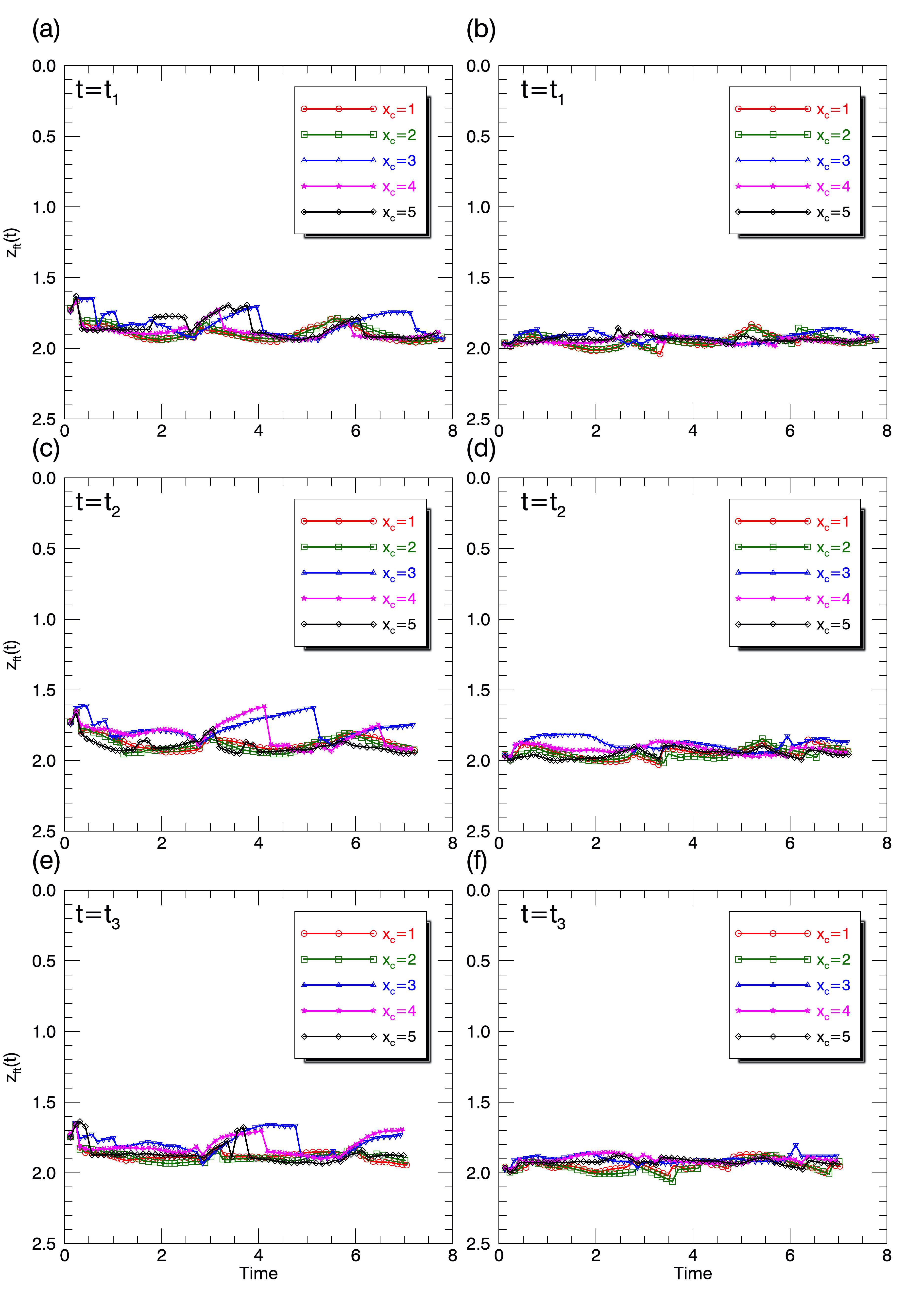}
    \caption{Same as Figure \ref{fig:S7_BS10_TrackBy} but for $B_{s}=0.20$.}
    \label{fig:S7_BS20_TrackBy}
\end{figure}

Thus far, we only examined one vertical location, corresponding to one moderate (averaged) pumped background field strength, $B_s=0.1$, in this more statistical MC manner.  We now also  confirm the single-example results of \S \ref{sec:fluxtube_backgroundfield_Bs} regarding other $B_s$ via similar MC-type statistical surveys.
Figures \ref{fig:S7_BS1_TrackBy} and \ref{fig:S7_BS20_TrackBy} are the same as Figure \ref{fig:S7_BS10_TrackBy} but for a relatively weak background field,  $B_{s}=0.01$, and a relatively strong background field, $B_s=0.20$, respectively. 
Figure \ref{fig:S7_BS1_TrackBy} shows that most of the flux tubes initiated in the presence of a  relatively weak  background field at $B_{s}=0.01$ rise successfully  through the stable layer into the convective layer and reach the top of the domain.  Of course, some variations in the background state do affect the rise of individual flux tubes.  For example, the tube initiated at $x=5$ (black markers) in panel a rises more slowly than average. However, in general, all of the positively-twisted tubes in panels a, c, and e rise relatively consistently.  In this weak background field case, we also find that the  negatively-twisted tubes in panels b, d and f mostly rise as well, although there is greater variation for this sign of the twist.
For example, the negatively-twisted tubes initiated at $x=1$ in panel b (red marker) and the tubes initiated at $x=2,5$ in panel d (green and black markers) rise much more slowly, and the tube at $x=2$ in panel f does not  rise at all.  That more obvious  variations for the negatively-twisted flux tubes are more likely was predicted by the work in Papers 1 and 2 (in the absence of overshooting convection), where it was explained that negatively-twisted  tubes add a tension effect that reduces the likelihood of rise, and possibly even quenches the rise if the state is close to marginal buoyancy, whereas the existing rise is only enhanced in positively-twisted tubes.  

Figure \ref{fig:S7_BS20_TrackBy} exhibits the MC runs for a relatively strong background field, $B_s=0.20$.  This figure simply shows that all the cases studied fail to rise, as was predicted in Paper 2 (for dynamics in the absence of convection).  Variations in the dynamics of these cases are minimal since, with the immediate defeat of the rise by the strong background field, the flux tubes do not interact with overshooting convection much at all.  This makes the predictions of Papers 1 and 2 more likely to be valid.

This section of the current paper  confirms statistically that: (a) at moderate background field strength, there is a strong difference in the dynamics between differently twisted tubes; (b) weak background field allows both signs of twisted tubes to rise in general, although violations to this part of the rule are more likely for negatively-twisted tubes (for a positively-oriented background field); and that, (c) relatively strong background field quenches the rise of both signs of twisted tubes.
Paper 2 discussed in detail the possible mechanisms that can lead, in the absence of convection, to variations and violations in the selection mechanism that might contribute to the observational scatter in the SHHR.  In the presence of convection and a  dynamically-formed pumped field, variations in the  background state experienced by the flux tube can clearly be a significant additional source of scatter.

\begin{figure}
    \centering
    \includegraphics[width=\textwidth,height=9cm]{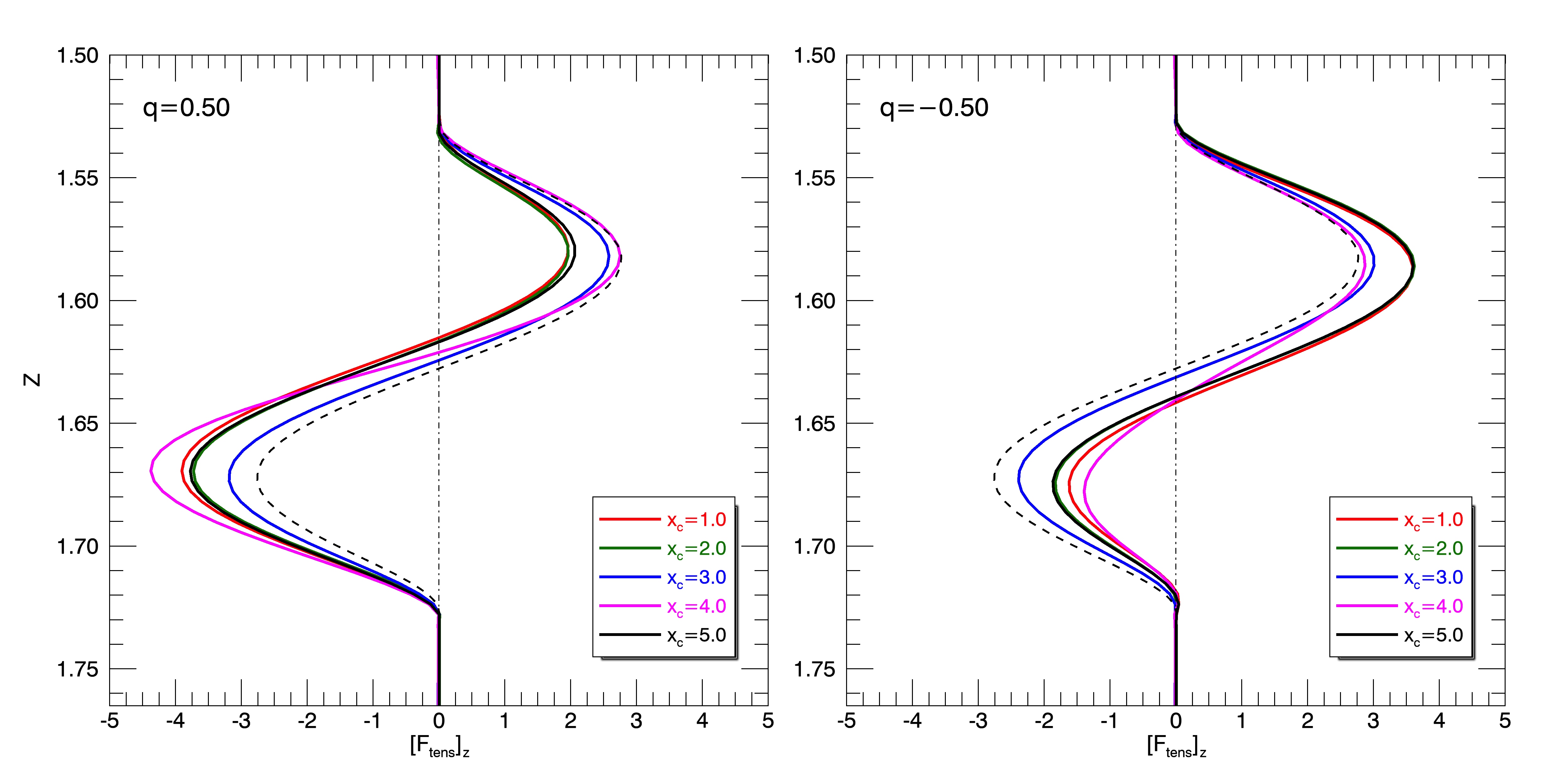}
    \caption{Plots of the tension force acting in the $z$-direction, $F_{tens,z}=(B_{x} \partial_{x} + B_{z} \partial_{z})B_{z}$, from the MC data in \S \ref{sec:fluxtube_backgroundfield_MC}. The line cuts are evaluated from the initial conditions of each simulation ($t=0.0$) and at the center ($x=x_c$) of each tube, for both positively-twisted ($q=0.5$) and negatively-twisted ($q=-0.5$) tubes. The dashed line shows the profile from the symmetric tube of \S  \ref{sec:fluxtube_conv} where no pumped background field is present.  The data in this figure corresponds to that in Figure \ref{fig:S7_BS10_TrackBy}e and f. }
    \label{fig:S7_BS10_TP05_TN05_T3_TensionLinecut_N3}
\end{figure}

\subsection{Tension forces}\label{sec:tension_S7}

That these more complex simulations seem to demonstrate statistically a selection mechanism akin to the one suggested in Papers 1 and 2 now seems well-established, at least at the canonical parameters.  The quick check of the tension force for one realization  exhibited as Figure
\ref{fig:S7_TP05_TN05_VerticalTension_LineCut} showed that the explanation in terms of tension forces from Papers 1 and 2 may indeed carry over to the more complicated dynamics here.  We now confirm this by plotting tension forces for some of the MC simulations to show that this effect is universal.  Figure \ref{fig:S7_BS10_TP05_TN05_T3_TensionLinecut_N3} shows profiles of the vertical tension force, $F_{tens,z}$, along the centerlines of the tubes in the initial conditions for some of the MC cases previously exhibited in Figure \ref{fig:S7_BS10_TrackBy}e and f.  These cases are at the canonical parameters, using timestep $t=t_3$ as an initial condition, and the left and right  panels (of both figures) show the results for positively-twisted and negatively-twisted initial tube conditions respectively.  Figure \ref{fig:S7_BS10_TP05_TN05_T3_TensionLinecut_N3} clearly shows that, for all the cases with a positive twist, there is a net shift of the tension profile to the left, towards more negative (upwards) net force values (when compared with the dashed line, which represents the symmetrical tension profile of a tube with no background field present from Section \ref{sec:fluxtube_conv}); for all cases with a negative twist, the shift is towards the right or more positive (downwards) values.  These results are all commensurate with Figure \ref{fig:S7_TP05_TN05_VerticalTension_LineCut} and the ideas of Paper 1 and 2.  This particular set of results from the MC  simulations also shows however that there are significant variations in the profiles, and the fact that these varying profiles can help explain the variations in Figure \ref{fig:S7_BS10_TrackBy} is supportive of the argument that these tension effects are causal.  For example, the positively-twisted tube placed at location $x=3.0$ (blue lines) has the slowest (successful) rise in Figure \ref{fig:S7_BS10_TrackBy}e and can be seen here to have the weakest deviation of the tension profile from that of the symmetric flux tube.  These plots also confirm that some of the drastic variations in Figure \ref{fig:S7_BS10_TrackBy} that occur later in the rise are due to advective effects of the convection.  For example, the positively-twisted tube placed at $x=2.0$ (green lines) has a substantial bias in the tension force and initially rises strongly due to this, but, on entering the convection zone, meets a convective downflow that overcomes the buoyancy and tension effects thereby disrupting its trajectory dramatically.  It now seems convincingly proven that the selection mechanism does indeed operate due to tension effects and certainly lends a bias to the rise statistics, but also that the convection provides an additional source of significant variation. 

\subsection{Variation with S}\label{sec:vary_S}

The previous simulations have all been carried out at a canonical set of parameters including a stiffness parameter of $S=7$.  We now examine the implications of varying the stiffness, $S$.  In our model, changing $S$ changes the polytropic initial stratification of the lower stable layer, with increasing $S$ leading to increased buoyancy deceleration in the stable layer. 
This leads to a reduction of the extent of overshooting convection in the stable region, as was mentioned earlier and was shown in Figure \ref{fig:ke_avg_svar} \citep[and as was investigated in detail in the past in, for example,][]{Hurlburt:etal:1994, Brummell:etal:2002}.  
If a pumping simulation is performed by adding a magnetic layer as an initial condition to such simulations, as in \S \ref{sec:Pumping_conv}, then the depth to which the magnetic layer is pumped is also reduced at higher $S$ for layer strengths that are relatively passive (high plasma $\beta$), since the pumping depth is then strongly related to the extent of overshooting.  This was investigated in detail in \cite{Tobias:etal:2001}.
As the form of the pumped background field is an essential component of the initial conditions in which we embed the flux tube for the main simulations of this paper, and, as already found above, plays a crucial role in the subsequent evolution of a flux tubes, it is essential to ascertain the role of $S$.

\begin{figure}
    \centering
    \includegraphics[width=\columnwidth,height=9.5cm]{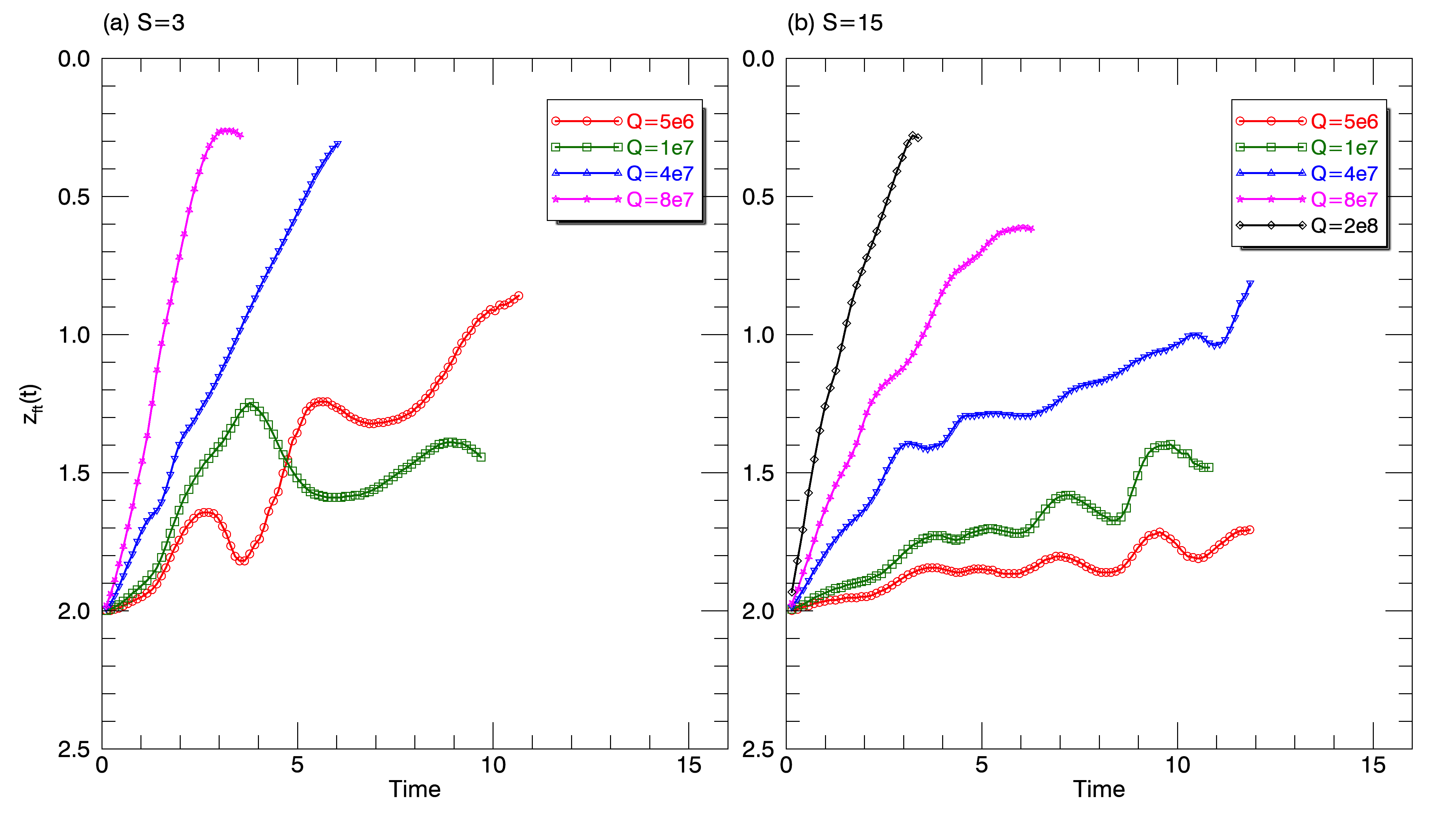}
    \caption{Same as Figure \ref{fig:S7_TP05_NoBG} but for (a) $S=3$ and (b) $S=15$.}
    \label{fig:S3_S15_TrackBy_VaryQ}
\end{figure}

In this section, we carry out a suite of MC simulations, similar to those in \S \ref{sec:fluxtube_backgroundfield_MC}, but now at $S=3$ and $S=15$ instead of the former $S=7$. We keep all the other parameters at the canonical values:  $Ra=4 \times 10^4, Pr =0.1, \zeta=0.001$.  A number of preliminary simulations must be completed to create a meaningful initial condition, as was done for the $S=7$ canonical case earlier. We first take the simulations of purely hydrodynamic overshooting convection from \S \ref{sec:conv_2D} that were performed at $S=3$ and $S=15$.  
As in \S \ref{sec:fluxtube_conv}, we then perform a number of simulations at each $S$ where we add an {\it isolated} flux tube to the convection at various different $Q$ values to determine the necessary value of $Q$ that dictates that an {\it isolated} flux tube will rise through the convection.  Figure \ref{fig:S3_S15_TrackBy_VaryQ} shows plots of $z_{ft}$ that compactly characteristize the rise dynamics for the new $S$ values (equivalent to Figure \ref{fig:S7_TP05_NoBG} for the $S=7$ case).  
When choosing $Q$ for $S=7$ before, we chose the case at $Q=2 \times 10^8$ that exhibited a very clear rise.  Here, we choose $Q = 4 \times 10^7$ (blue markers) that is a definite rise for $S=3$ but is closer to a marginal rise for $S=15$.  Being close to the marginal state is not necessarily a bad thing, as then the effects of tension can be highlighted, if they exist, by clearly creating distinctive rise or no-rise realizations.

\begin{figure}
    \centering
    \includegraphics[width=\columnwidth,height=5cm]{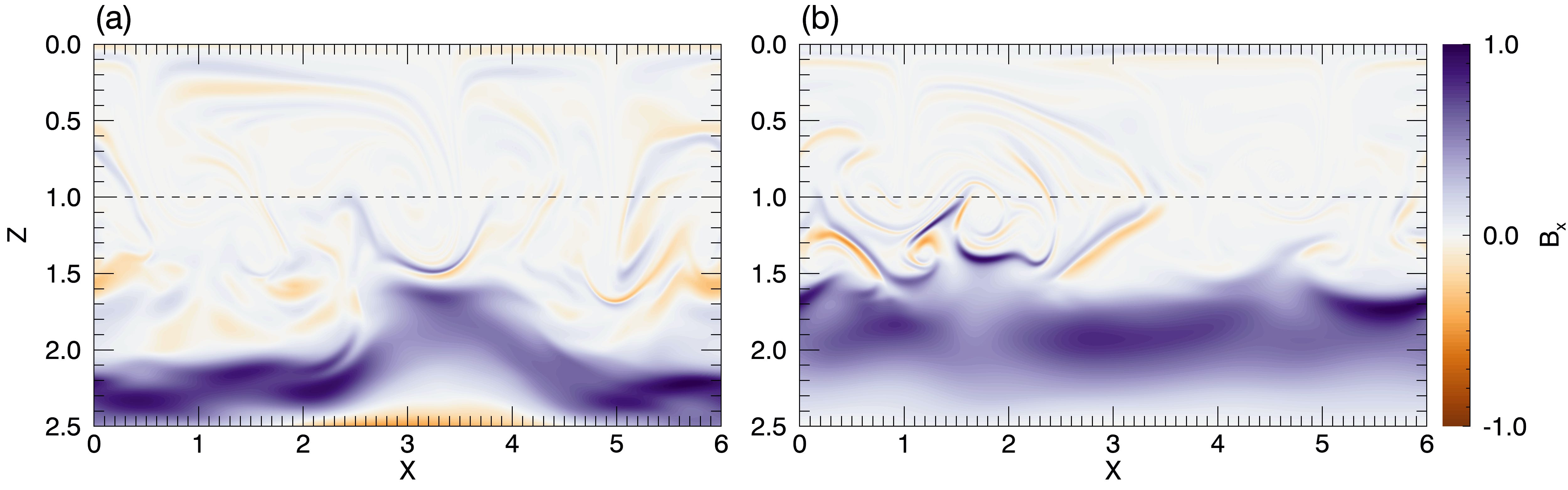}
    \caption{Snapshots of the normalized horizontal magnetic field, $B_{x}$, for (a) $S=3$ and (b) $S=15$, at an arbitrary late time (denoted as $t=t_{1}$) in each simulation. Note that $t_{1}$ represents different times in both the subplots. Here, the parameters are the canonical set except that $S$ is as stated and $Q=4\times 10^{7}$.}
    \label{fig:Pumping_S3_S15_Combined}
\end{figure}

\begin{figure}
    \centering
    \includegraphics[width=10cm,height=12cm]{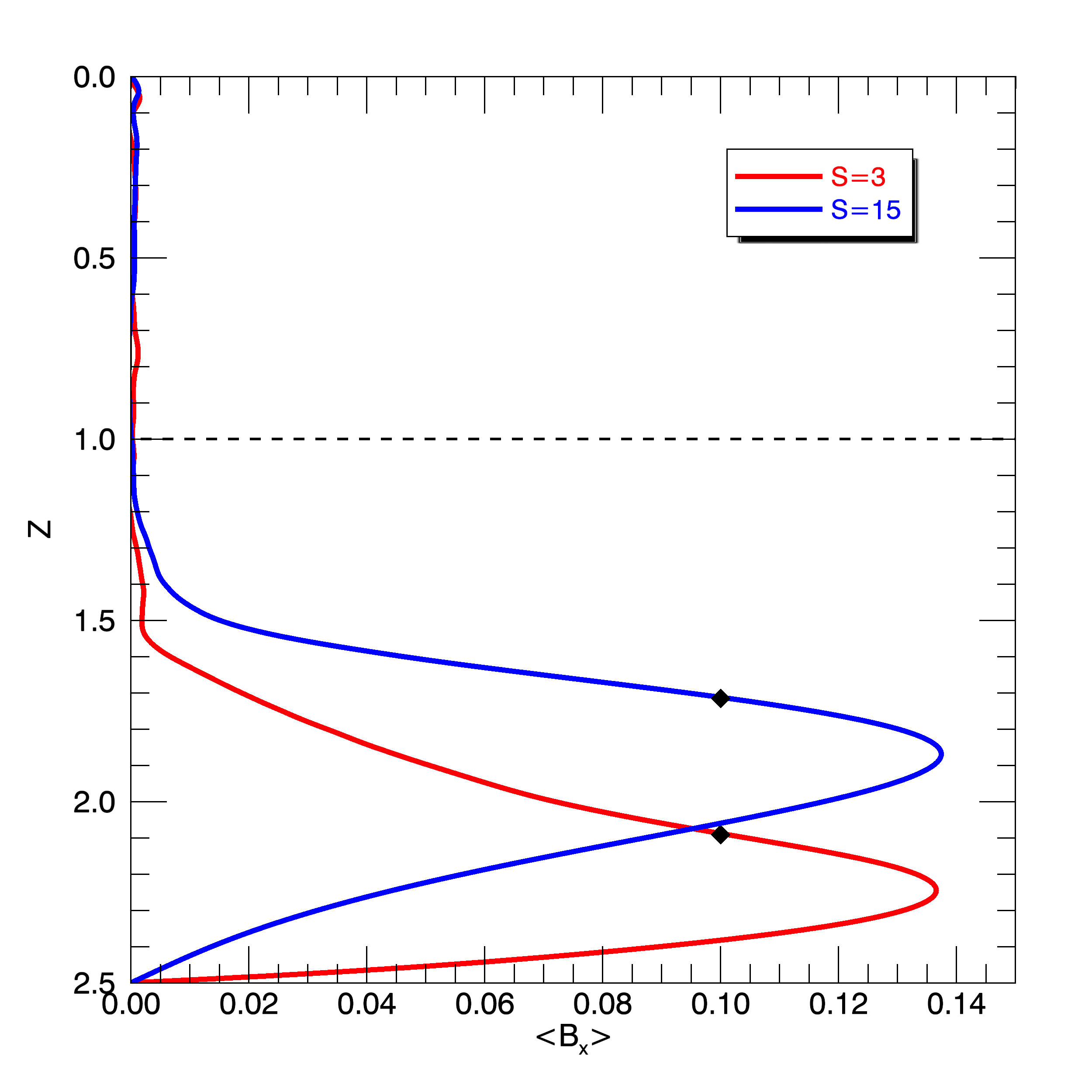}
    \caption{Time- and horizontally-averaged pumped field as a function of $z$ for $S=3$ and $15$. Diamond markers indicate the vertical location corresponding to $B_{s}=0.10$.}
    \label{fig:PumpedField_S3_S15_Bs10Marked}
\end{figure}

The next step in preparing the ultimate initial conditions is to create a pumped layer. We again take the statistically-stationary state of the purely hydrodynamic turbulent convection calculations (without a flux tube) and introduce a magnetic layer in the same manner as was done for $S=7$ in Section  \ref{sec:Pumping_conv} but now for the two new $S$ values. 
In each of these cases, we evolve the layer for a significant amount of time and, again, the overshooting convection pumps the magnetic layer into the stable region, albeit with significant differences. 
Figure \ref{fig:Pumping_S3_S15_Combined} shows a snapshot of the pumped state in the late stages of the simulation for each $S$ (equivalent to Figure \ref{fig:pumped_bx_S7}) and Figure \ref{fig:PumpedField_S3_S15_Bs10Marked} shows the time- and horizontal-average of this state (equivalent to Figure \ref{fig:AveragePumpedField_S7_FTMarked}).
It can be seen from these figures that the depth at which the pumped layer is formed is different, and the form of the average profile is somewhat adjusted. For the lower $S=3$ case, the extent of overshooting is greater than that for the earlier $S=7$ case, and strong fluctuations in the ``softer'' overshoot zone lead to a deeper pumped magnetic layer, that is possibly somewhat influenced by the lower boundary conditions, leading to a more asymmetric profile .  For the $S=15$ case, a slightly more compact pumped layer is generated, located higher up in the stable region due to the reduced overshooting in this ``stiffer'' case.

We now have all the prerequisite knowledge and initial conditions necessary to run the ultimate simulations at $S=3$ and $S=15$ involving turbulent overshooting convection, its associated pumped field, plus a flux tube as an initial condition. We choose to run simulations only for an intermediate value  of the relative background field strength, $B_s=0.1$,  where selective dynamics are most likely.
This strength 
corresponds to the vertical locations, marked (as black diamonds) on Figure \ref{fig:PumpedField_S3_S15_Bs10Marked} for each case.
We again run a small MC-like suite of new  simulations for each $S$ starting from a random late timestep of the appropriate pumping simulations (which we denote arbitrarily by $t=t_1$), initiating flux tubes at the desired vertical location and at one of the five different horizontal locations for each (as previously in \S {\ref{sec:fluxtube_backgroundfield_MC}}).

\begin{figure}
    \centering
    \includegraphics[width=\columnwidth,height=9cm]{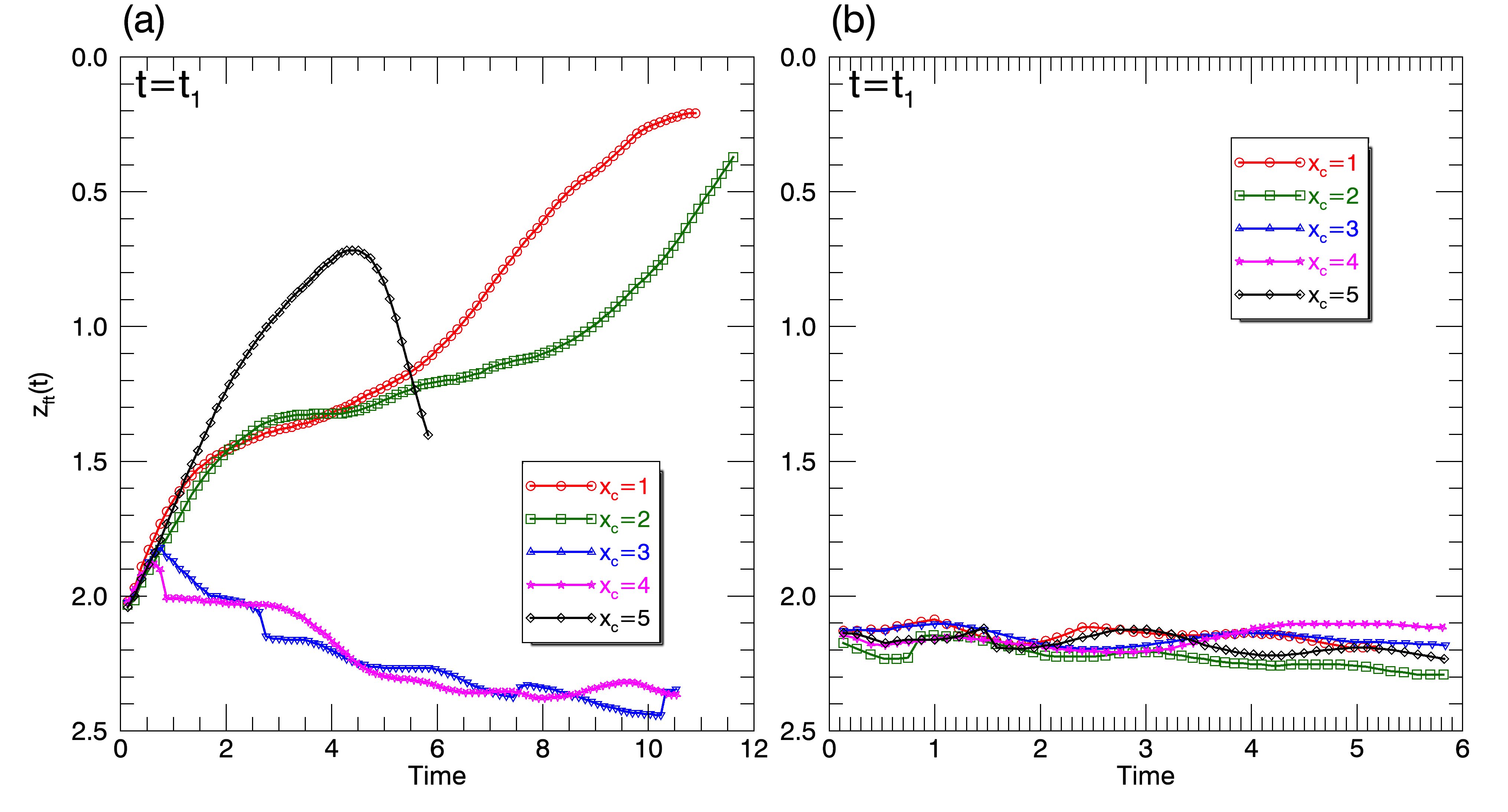}
    \caption{The tracer $z_{ft}$ for $S=3, B_s=0.1$ and  $Q=4\times 10^{7}$ plotted as a function of time, for various  different horizontal initial starting locations of the tube, $x_{c}=1,2,3,4,5$.  Panel (a) shows positively-twisted tubes, $q=0.5$, and panel (b) shows negatively-twisted tubes, $q=-0.5$. Here, $t=t_{1}$ indicates the time chosen arbitrarily from the associated pumping simulation at each $S$.}
    \label{fig:S3_NT163000_BS10_TP05_TN05_XVar_TrackBy}
\end{figure}

Figure \ref{fig:S3_NT163000_BS10_TP05_TN05_XVar_TrackBy} is similar to Figure \ref{fig:S7_BS10_TrackBy} except that $S=3$ and $Q=4\times 10^{7}$.
Figure \ref{fig:S3_NT163000_BS10_TP05_TN05_XVar_TrackBy}a shows the traces of $z_{ft}$ for the five different horizontal locations of a positively-twisted flux tube. We notice that the dynamics of each case are markedly different and reveal some interesting characteristics of the impact of overshooting convection on the rise of the tube. 
The tubes initiated from  $x=1$ and $2$ (red and green markers) rise through the stable and convective layer and reach its top, but the rise is slow compared to any of the cases studied in the previous section for $S=7$, 
taking about twice as much time to reach the top of the convective layer compared to any of the previously discussed successful rise cases. There are significant reasons for this. First, the initial $Q$ for this case is weaker than any of the previous simulations, which results in lower initial buoyancy imparted to the tube. Second, the deep overshooting convective motions in this case influence the directness of the rise of the tube, making its path to the top of the convective layer more convoluted, thus taking more time.
Additionally, the flux tubes are rising from a vertical location deeper in the stable layer, so the distance they have to cover during their rise is comparatively longer.  The second reason seems perhaps to be the most influential, since the rise speed drops again  significantly as the tubes encounter overshooting of significant strength at around $z \sim 1.5$.

Next, from Figure \ref{fig:S3_NT163000_BS10_TP05_TN05_XVar_TrackBy}a, we notice somewhat surprisingly that flux tubes that were initiated near the center of the horizontal domain ($x=3$ and $4$) have similar $z_{ft}$ profiles, each of them showing that the rise is unsuccessful. The influence of deep overshooting convection and its impact on the pumped layer is dramatic here. Figure \ref{fig:Pumping_S3_S15_Combined}a shows the background pumped field (at the arbitrary time $t=t_1$) that forms the initial condition in which the flux tube has been embedded. Due to the deep overshooting motions, the pumped layer gets locally lifted at horizontal locations roughly between $x=2$ and $4.5$.  This increases the effective thickness of the pumped layer through which the flux tube has to rise when it is inserted at the $B_s=0.1$ vertical location deduced from the average.  The local vertical profile of $B_x$ at this horizontal location is substantially different from that average. This hampers the rise from these two locations sufficiently to make them actually fail.  

The tube located at $x=5$ also has an interesting trajectory. In this case, we see that the flux tube rises quickly through the stable layer and enters the convective layer. However, it soon interacts with a downward convective plume and is strongly advected downwards back into the stable layer, resulting ultimately in an unsuccessful rise. 

From the limited set of cases explored here, at the very same parameters we have two examples of the successful rise of flux tube, another case where the tube starts to rise but eventually fails, and two more simulations where the tubes fail to rise dramatically. These are intriguing results that show how local convective fluctuations can impact the overall dynamics of the flux tube in these simulations. The fluctuations in the overshoot layer in this $S=3$ case are clearly substantial.  This may potentially translate to the solar context, possibly contributing to the SHHR scatter.  

In contrast, Figure \ref{fig:S3_NT163000_BS10_TP05_TN05_XVar_TrackBy}b shows identical simulations to those discussed for panel a but where negatively-twisted flux tubes are embedded in the initial condition instead of positively-twisted tubes. Here, very clearly none of the tubes rise.  The retarding effect of the negative twist is strong enough that no fluctuations in the overshooting or pumped layer are sufficient to enable rise.  Once again, a broad conclusion is that there is clearly a significant difference between postively- and negatively-twisted tubes in this $S=3$ case, commensurate with what was observed in the $S=7$ case, and the ideas of Papers 1 and 2.

\begin{figure}
    \centering
    \includegraphics[width=\columnwidth,height=9cm]{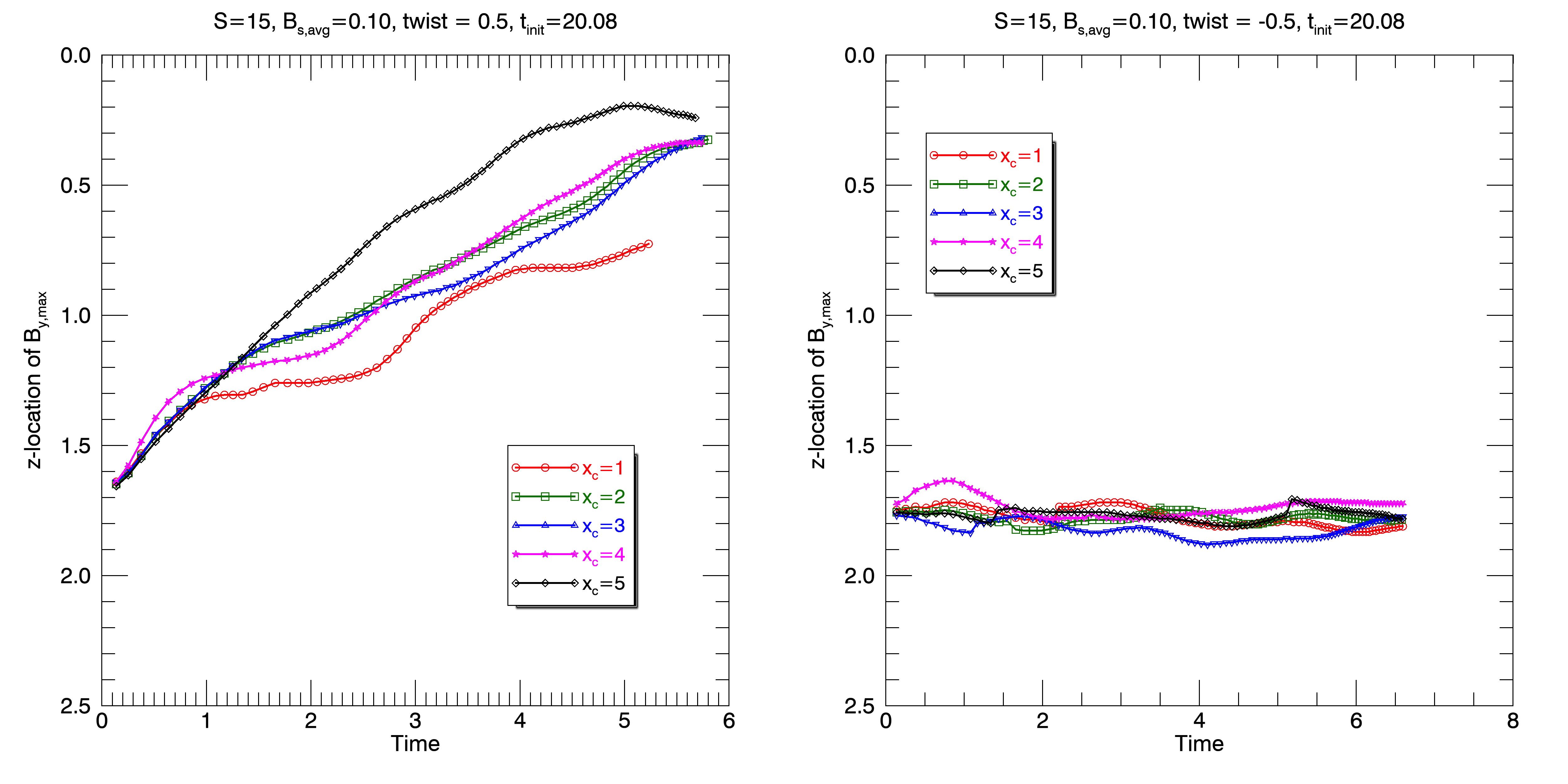}
    \caption{Same as Figure \ref{fig:S3_NT163000_BS10_TP05_TN05_XVar_TrackBy} but for $S=15$.}
    \label{fig:S15_NT176000_BS10_TP05_TN05_XVar_TrackBy}
\end{figure}

Examining the case with a relatively stiffer stable region at $S=15$, the results are far more consistent between realizations and again  commensurate with the selection mechanism ideas. 
Figure \ref{fig:S15_NT176000_BS10_TP05_TN05_XVar_TrackBy} shows the trajectories $z_{ft}$ for positively-twisted (panel a) and negatively-twisted flux tubes (panel b). Each panel again shows the five cases corresponding to the different horizontal locations of the flux tube.  Figure \ref{fig:S15_NT176000_BS10_TP05_TN05_XVar_TrackBy}a shows that each individual case is slightly different due to the local background variations that the tube experiences. However, in this case, all the flux tubes regardless of  horizontal location successfully rise through the stable layer into the convective layer and eventually reach the upper boundary. 
In contrast, once again, Figure \ref{fig:S15_NT176000_BS10_TP05_TN05_XVar_TrackBy}b shows that none of the negatively-twisted tubes rise.  The greater regularity here can be attributed to the closer correlation between the time-average field and the local conditions that the tubes experience.  Figure \ref{fig:Pumping_S3_S15_Combined} gives some visual confirmation that the overshooting convection motions only weakly influence the pumped field layer in the stiffer $S=15$ case.

\begin{figure}
    \centering
    \includegraphics[width=\columnwidth,height=9cm]{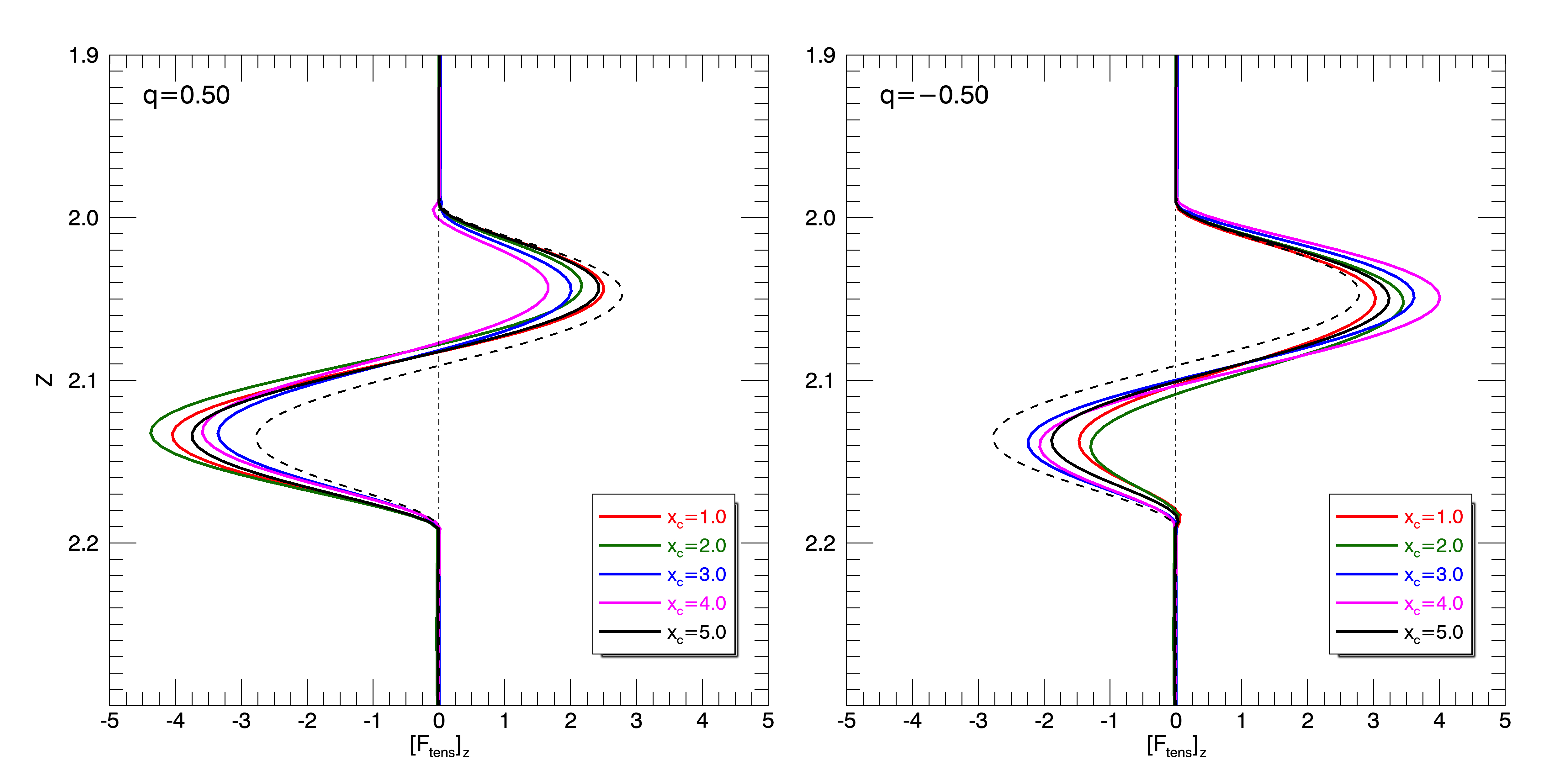}
    \caption{Plots of the tension force acting in the z-direction, $F_{tens,z}=(B_{x} \partial_{x} + B_{z} \partial_{z})B_{z}$ on line cuts through the middle of each  tube ($X=x_c$) in the initial conditions ($t=0$) for both signs of $q$: $q=0.5,~-0.5$. This data corresponds to the $S=3$ MC cases shown in  Figure \ref{fig:S3_NT163000_BS10_TP05_TN05_XVar_TrackBy}a and b.}
    \label{fig:S3_BS10_TP05_TN05_T1_TensionLinecut_N3}
\end{figure}

\begin{figure}
    \centering
    \includegraphics[width=\columnwidth,height=9cm]{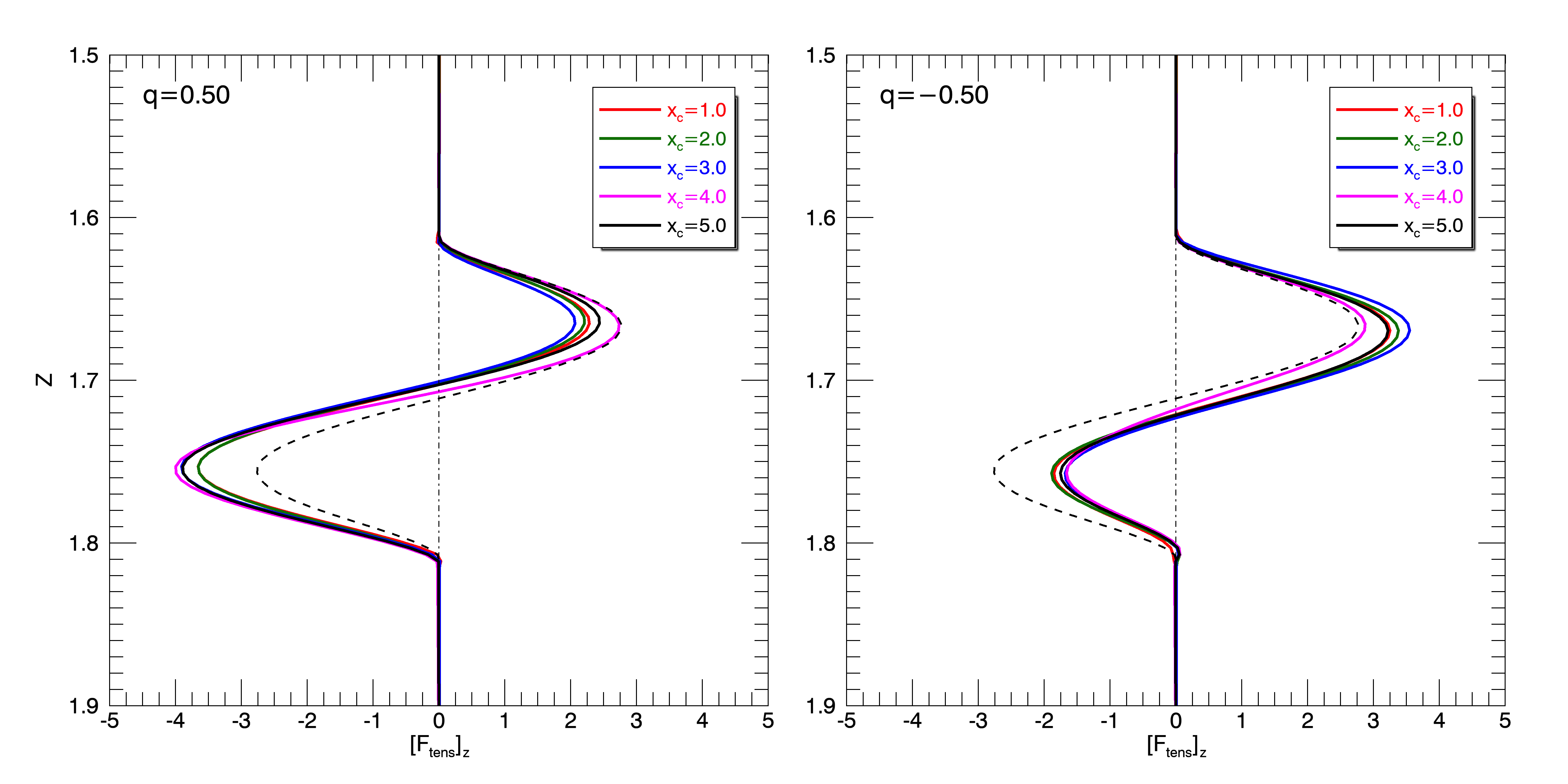}
    \caption{Same as Figure \ref{fig:S3_BS10_TP05_TN05_T1_TensionLinecut_N3} but for the MC cases at $S=15$ that correspond to data shown in Figure \ref{fig:S15_NT176000_BS10_TP05_TN05_XVar_TrackBy}.}
    \label{fig:S15_BS10_TP05_TN05_T1_TensionLinecut_N3}
\end{figure}

Again, we can check that the ideas of Papers 1 and 2 regarding tension forces are indeed responsible for these effects.  Figures \ref{fig:S3_BS10_TP05_TN05_T1_TensionLinecut_N3} and \ref{fig:S15_BS10_TP05_TN05_T1_TensionLinecut_N3} show vertical profiles of the centerline tube vertical magnetic tension force, $F_{tens,z}$, for the initial conditions of the MC simulation for $S=3$ and $S=15$ respectively.  In all these cases, the expected tension biases do indeed occur, with positively-twisted tubes contributing a rise-enhancing force, and negatively-twisted tubes creating a net tension in the tube that opposes buoyancy.  Again, the MC sets show variations which can account for the individual variations in rise times.  It should be noted once more that the variation for $S=3$ is far greater than that for $S=15$, leading to some of the less common effects after the initial rise seen in Figure \ref{fig:S3_NT163000_BS10_TP05_TN05_XVar_TrackBy}a, for example.

In this section, we have shown how changes in the stiffness of the stable region impact the overall rise dynamics of flux tubes. There exist some significant differences in overshooting, pumping convection, such as the variation of the overshoot depth, and therefore the depth and form of the pumped field profile (which were previously known).  Here, we have found that these can have a fairly substantial impact on the rise dynamics of individual tubes, but overall, the selection mechanism still operates to enhance the statistical likelihood of the rise of tubes with one sign of twist (here positive for a positively-directed background field) over the other (here negative).  The more significant variations in the overshooting and pumped layer in less stiff layers (e.g. $S=3$) lead to greater variability in the agreement with this average result for lower $S$ cases.  Higher $S$ cases conform to the selection rules more uniformly.

\section{Discussion and Conclusions}
\label{sec:discussion_conclusions}

In this paper, via a series of numerical simulations, we have clearly established an asymmetry between the buoyant rise characteristics of oppositely twisted flux tubes when in the presence of overshooting convection and a dynamically-organized large-scale background field. Specifically, we find that a positively-twisted (anticlockwise) flux tube is more likely to rise than a negatively-twisted (clockwise) tube when the rise is in the presence of a pumped field that is, on average, oriented horizontally in the positive direction.  By symmetry, if the horizontal background field were reversed to be in the negative direction, the rise bias would also be reversed, then preferring negatively-twisted tubes.  This result affirms the existence and applicability of a selection mechanism based on tension effects akin to the non-convective effect described in Papers 1 and 2. This mechanism operates because of the relative orientations of the azimuthal interior field of the tube and the horizontal background field.  For a positively-oriented background field, the azimuthal field of a positively-twisted flux tube is enhanced on the lower side and reduced on the upper side, leading to a net tension force that acts upwards in concert with buoyancy, encouraging rise.  For negatively-twisted flux tubes, the resultant net tension force is downwards, acting against buoyant rise.  As explained in Papers 1 and 2, this selection mechanism, when related to the solar context, is in agreement with the solar hemispherical helicity rule(s).

We find that the effect of convection on the {\it initial} rise dynamics of the flux tubes is weak.  The reason for this is mainly because the initial rise characteristics are dictated by magnetic buoyancy in conjunction with the tension effects described above, and the fact that the pumped layer peaks where the overshooting convective power drops towards zero.  This means that the likely location of tube formation and therefore the selection mechanism only happens in a region of weak convection.  Because of this, examining the tension forces of the initial conditions is a strong predictor of at least the initial rise (or lack of rise) of any tube.  

Convective effects are not necessarily insignificant overall though.  In order to assess this issue, we performed detailed sets of MC simulations to establish the statistical relevance of the selection mechanism described. We studied the effect of temporal and spatial variations in the convection and its associated pumped background field on the rise characteristics of tubes. 
The overshooting process is quite intermittent, with some strong events that can lead to significant perturbations in the local environment that influences the tube's rise dynamics.  These dynamical perturbations can be strong deviations of the local pumped field away from the average, or strong advective effects, and, if affecting the initial conditions, these can lead to less predictable outcomes.  Furthermore, even if the tube initially rises, as it enters more vigorous convection in the overshoot zone and the actual convection zone, it can be subject to advective forces that can significantly affect the rise.  The presence of convection is a significant source of variation in the  {\it overall} rise dynamics of the tube.  However, despite these statistical fluctuations, the ultimate conclusions are the same:  that there is a definite statistical difference in the dynamics depending on the twist of the flux tube (for fixed orientation of the average background field), and that the rise of one (positive, in the case exhibited) is preferred.

This work also confirms other elements of the predictions of  Papers 1 and 2 under the influence of the more complex dynamics examined here.  For example, Papers 1 and 2 predict that there is a lower threshold in the strength of the background field below tubes with both types of twist will rise, and another upper threshold above which both types of tubes will fail to rise, thereby confining the selective dynamics to a region in parameter space that Papers 1 and 2 deem the Selective Rise Regime (SRR).  Here, we statistically confirm the presence of this region in the convective dynamics.  

The role of the stiffness parameter $S$ was also investigated.  We performed a number of MC suites of simulations and a statistical analysis of the presence of the selection mechanism for $S=3,~7$ and $15$, covering the range from a weakly-stable lower region to a fairly strongly-stable lower region.   For lower $S$, the motions overshoot a substantial fraction of the convection zone depth into the stable lower layer, thereby creating a much deeper and somewhat wider pumped layer, whereas for higher $S$ the overshoot and pumped layer are shallower and perhaps more compact.  
Regardless, the selection mechanism appears to operate consistently at all $S$.  At lower $S$, due to the greater degree of fluctuating dynamics that the tube must encounter during any attempted rise, the variations in the rise dynamics are greater than at higher $S$, where agreement with the selection rules is more uniform. 

It is an interesting question as to what is the correct $S$ for the solar context.  Early works \citep{Milne:1932,Eddington:1938} and some more recent unpublished discussions (see https://astronomy.stackexchange.com/questions/16633/does-eddingtons-variable-polytropic-index-better-fit-data-from-the-standard-sol) on fitting polytropic models to solar data put the adiabatic index of the stable region, $m_2$, in the range $m_2=3-4$.  This corresponds to $S=3-5$ here.  However, many researchers quote an $S$ that is much higher, usually based on the fact that, in the turbulent stationary state, convection relaxes the upper layer to an adiabatic state away from any boundaries.  This would imply $m_1=1.5$ and therefore an effective ``turbulent" $S$ of infinity in our definition.  We think this logic is a bit misleading, but have nonetheless erred on the side of including simulations at higher $S$ to accommodate researchers who think this way.  Regardless, the $S=15$ case is very similar to the $S=7$ in all respects. Perhaps the more important distinction is that our simulations are not at high enough $Ra$ to be fully representative of the solar context. 
Note that increasing $S$ also increases the plasma $\beta$ of the dynamics at fixed $Q$.
We have considered flux tubes of fixed strength, both in the amount of twist and the axial field strength, and thereby the flux tube has a fixed (initial) magnetic pressure, dictated by the chosen $Q$ (or $\alpha$ to be more precise, but the other factors in $\alpha$ are held fixed in these simulations).  For larger $S$, this implies a larger plasma $\beta$ since the gas pressure is increased at fixed magnetic pressure.
This could be compensated for via a larger $Q$.  However, note that the choice of $Q$ cannot be arbitrarily large when inserting a magnetic structure (tube or layer) into the stratification since the thermodynamic quantities need to be adjusted to preserve total pressure balance;  this sets an upper limit.

Of course, there are many caveats to our results.  Perhaps the foremost issue is that the flux tubes we examine are not self-consistently generated from the background magnetic field.  Our aim has been to examine volume-filling fields where flux tubes are concentrations in the field.  The work presented here is a very crude attempt at this.  A more realistic approach, perhaps, would be to incorporate a shear flow below the convection zone, as is the case in the solar tachocline, and thereby to create a layer of $B_y$ or structures in $B_y$ by  stretching action of this shear acting on the pumped field, $B_x$.  The hope would be that these structures could become unstable to magnetic buoyancy instabilities and create rising self-consistent magnetic structures that are then truly magnetic concentrations in volume-filling fields. Such simulations in the absence of convection  \citep[see e.g.][]{Vasil:Brummell:2008} and (to a certain extent, in a slightly different context) in the presence of convection \citep{Guerrero:Kapyla:2011} have been performed before, but a comprehensive study of this particular setup  is future work.

In terms of parameters, our work could be critiqued in a number of ways.  The Rayleigh number is orders of magnitude below stellar values and the Prandtl number is orders of magnitude too large, and therefore our simulations do not operate at anything like astrophysically realistic levels of turbulence.  These limitations are numerical. We could have done better values for these parameters using higher resolution in these 2D studies if we had possessed greater patience.  These simulations, however, if nothing else, serve as a first insight into mildly turbulent convective effects.  The magnetic Prandtl number, $Pm=\sigma/\zeta$, which in our case has value  $Pm=100$, is also not astrophysical, since the viscosity should be smaller than the resistivity for astrophysical objects, i.e. $Pm<1$.  We worked with low resistivity to avoid issues of the magnetic flux tubes diffusing too quickly, but our high viscosity leads to $Pm>1$ as well as the low $Ra$ mentioned above.  While having a low $Pm$ would be ideal, the role of $Pm$ is probably more critical in dynamo simulations.

It is perhaps  interesting to question what might happen at higher levels of turbulence, and this brings up the complex dependence of the necessary $Q$ for a rise simulation on the other non-dimensional parameters.  To overcome the increased convective kinetic energy in a high $Ra$ and low $Pr$ scenario, a high $Q$ value is required for the flux tube to rise successfully through the flows. Moreover, we expect vigorous convection to transport (pump) magnetic fields from the convection zone toward its base and into the stable region more efficiently. Efficient pumping of magnetic fields is expected to create a relatively strong pumped background magnetic field layer, which woiuld also require a higher effective $Q$ value for the flux tube to rise through it successfully, since more magnetic buoyancy would be needed to overcome the greater tension in the overlying field. This dependence, where very strong tubes might be required for transit through highly turbulent convection zones, is somewhat unnerving.  However, some of this discomfort is perhaps a result of  the fact that our flux tubes are not self-consistently generated from the background field.  If toroidal field ($B_y$) structures were to be created by the stretching action of velocity shear, we might expect them to be  substantially stronger than any $B_x$ that is present.  This fact, and the fact that $B_x$ is pumped and therefore must be weak in some energetic sense compared to the convection, might guarantee that structures can always survive transit if the shear is sufficiently strong.  Alternatively, there may some issue of scale, whereby small-scale magnetic structures can transit relatively easily (see later). 
It is essential to note that none of this directly influences the significant result of the paper (the existence of the selection mechanism) since this merely depends on the relative strengths of the azimuthal field in the tube compared to the background field.  If background field strengths change, an adjusted  range of $q$ still defines a Selection Rise Region (and vice versa).  If we assume that a wide spectrum of twists is generated in the process of forming the structures, then this adjustment may always be allowable, although there are a few physical limitations on the size of $q$, as described in Paper 2.  

The non-astrophysical $Pm$ that we used brings up the more general question of timescales.  With the magnetic boundary conditions that we have chosen, horizontal magnetic flux is not conserved in the domain and the pumping simulations can potentially lose their  initial $B_x$ and run down.  However, when we achieve a pumped state, the majority of the horizontal field becomes isolated from the boundaries (in most cases), and the average profile becomes almost piecewise linear, and therefore only diffuses and runs down very slowly, in the sense that the rundown is a lot slower than the timescales of the other dynamics that we are interested in.  For example, the timescale of the rise is set via $Q$ and we make sure that this is such that the rise is relatively quick compared to the rundown timescale.  Our simulations are once again not perfectly astrophysical but we believe that the ordering of timescales is the relevant one, with the rise being slow compared to convective overturning, but fast compared to any significant evolution of the large-scale background field.  If we use other boundary conditions, magnetic flux could be conserved, but field can then accumulate temporarily at the upper boundary during the initial evolution, causing unphysical numerical issues.

Similarly, we attempt to get the ordering of length scales correct. The mildness of the turbulence means that an extended inertial cascade of dynamical scales is not realized.  However, the most influential scale issue is probably the relative size of the flux tube compared to the deeper convective dominant scales.  Unfortunately there is no direct solar data on either of these scales.  We might expect the scales of buoyancy instabilities to be of the order of the depth of the shear (the tachocline), and the strong stratification suggests that the scale of the flux tubes in the deep interior should be perhaps smaller than those observed to emerge at the solar surface. Therefore, it seems likely that flux tubes at initiation are smaller in width than some characteristic overturning scale of the convection.  The flux tubes in our simulations obey this ordering. We suspect, if anything, our flux tubes are likely far larger than they should be, since we choose a scale smaller than the main turnover scale but large enough to be well-resolved by our numerical grid resolution.

Finally, it is important to acknowledge that the work described in this paper has been carried out in 2D as an initial investigation, and, of course, 3D simulations would be more realistic. The 3D simulations are already underway and will be reported soon, but it is worth noting here some differences that might be anticipated.
\cite{Brummell:etal:2002} and 
\cite{Tobias:etal:1998, Tobias:etal:2001} carried out 3D numerical simulations investigating  overshooting convection and the transport (pumping) of magnetic fields respectively, and there are some notable differences between 2D and 3D versions of these processes. In particular, the overshooting depth of convection in 2D is significantly larger than in 3D. This is due to the more coherent and space-filling nature of 2D convection.  Convection in 2D even at high Reynolds numbers is dominated by large, more cellular overturning motions.  In 3D, the downwards motions are characterized by much more localized plumes.  The sparse nature of 3D convection leads to reduced mixing below the convection zone, and therefore reduced overshoot.
This further means that the pumping mechanism in 2D calculations is far more efficient than in 3D and may even be more akin to flux expulsion \citep{Weiss:1966} than a $\gamma$-effect \citep{Ossendrijver:etal:2002} from a mean-field perspective. We find in 2D that the average pumped field is virtually absent in the convective layer; this is not the case in three-dimensional calculations where the overshooting convection transfers the majority of magnetic fields from the convective layer to the region below its base but does not entirely evacuate the convection zone.  As a corollary, the pumped layer in 2D seems to sit firmly below the overshoot region, whereas in 3D the two seem to overlap more substantially.  In terms of the rise of magnetic structures then, we might expect that tubes initiate from a slightly different location in 3D, and might experience more interaction with the background field and the convective flows both at initiation and during transit through the convective layer. With regards to the selection mechanism, this suggests that the relative strengths of the background field that prevent a positive and a negative twisted flux tube from rising may be somewhat different than those found to be required here, with lower values possibly needed to quench a rise.   It might be easier to prevent a tube from rising for these reasons, but the presence of less efficient pumping may counteract this issue. While these are important details and certainly require further investigation, it is not likely that such issues will impact the broader conclusion of this paper, i.e., that there is an asymmetry in the rise of oppositely-twisted flux tubes.  

Overall, we have clearly established a case for the selection mechanism of Papers 1 and 2 being relevant to the rise of magnetic structures in the more complex situation involving a convection zone.  The earlier papers established the connection between this mechanism and the SHHR, and it appears that this mechanism could be a contributor to the bias observed as these rules.  There are other mechanisms that could also certainly  contribute to the bias, such as the $\Sigma$ mechanism suggested by \cite{Longcope:etal:2003}, and it is an interesting future question as to what the contribution of each mechanism is to the total helicity budget.  It should be noted that the mechanism presented here has some sound theoretical arguments for the statistical error in the SHHR.  If the mechanism presented here then were to be a substantial  contributor, the observed SHHR error could perhaps even be used to infer some of the likely magnetic conditions at the base of the convection zone via this causal relationship,  providing a very useful probe into the deep solar interior.


\begin{acknowledgments}
This work was initiated during the 2021 Kavli Summer Program in Astrophysics, hosted by the Max Planck Institute for Solar System Research and funded by the Kavli Foundation, where CMP was a student and BM and NB were participants. This research was also  supported by the National Science Foundation, under grants NSF AST-1908010, and by the joint NASA-NSF Diversity, Realize, Integrate, Venture, Educate (DRIVE) Science Center (DSC) Phase 1 grant 80NSSC20K0602 for the Consequences Of Fields and Flows in the Interior and Exterior of the Sun (COFFIES) DSC (subcontract to University of California, Santa Cruz). The authors also acknowledge NSF XSEDE allocations for access to the Texas Advanced Computing Center (TACC \url{http://www.tacc.utexas.edu}) at The University of Texas at Austin for providing high-performance computing (HPC), visualization, and database resources that have contributed to the research results reported within this paper. We thank the anonymous referee for their time and helpful suggestions.
\end{acknowledgments}

\bibliography{main}{}
\bibliographystyle{aasjournal}



\end{document}